\documentclass[
 reprint,
 amsmath,amssymb,
 superscriptaddress,
 aps,
 prx,
 nofootinbib,
]{revtex4-2}

\usepackage{graphicx}
\usepackage{dcolumn}
\usepackage{bm}
\usepackage{physics2}
\usephysicsmodule{ab}
\usephysicsmodule{op.legacy}
\usephysicsmodule{braket}
\usephysicsmodule{nabla.legacy}
\usepackage{derivative,fixdif}
\usepackage{mathtools}
\usepackage[normalem]{ulem}

\usepackage{siunitx}
\usepackage{appendix}
\usepackage[colorlinks,linkcolor=blue,urlcolor=blue, citecolor=blue]{hyperref}
\usepackage[whole]{bxcjkjatype}
\usepackage[T1]{fontenc} 
\usepackage{comment}
\usepackage{amsfonts}
\usepackage{svg}
\usepackage{multirow}

\usepackage{newtxtext}  
\usepackage{newtxmath} 
\usepackage{enumitem}

\begin{document}

\title{Entanglement boosting: Low-volume logical Bell pair preparation for distributed fault-tolerant quantum computation}

\author{Shinichi Sunami}
\email{shinichi.sunami@nano-qt.com}
\altaffiliation{equal contribution}
\affiliation{Nanofiber Quantum Technologies, Inc. (NanoQT), 1-22-3 Nishiwaseda, Shinjuku-ku, Tokyo 169-0051, Japan.}
\affiliation{Clarendon Laboratory, University of Oxford, Oxford OX1 3PU, United Kingdom}
\author{Yutaka Hirano}
\altaffiliation{Equal contribution}
\affiliation{Nanofiber Quantum Technologies, Inc. (NanoQT), 1-22-3 Nishiwaseda, Shinjuku-ku, Tokyo 169-0051, Japan.}
\author{Toshihide Hinokuma}
\affiliation{Nanofiber Quantum Technologies, Inc. (NanoQT), 1-22-3 Nishiwaseda, Shinjuku-ku, Tokyo 169-0051, Japan.}
\author{Hayata Yamasaki}
\email{hayata.yamasaki@nano-qt.com}
\affiliation{Nanofiber Quantum Technologies, Inc. (NanoQT), 1-22-3 Nishiwaseda, Shinjuku-ku, Tokyo 169-0051, Japan.}
\affiliation{Department of Computer Science, Graduate School of Information Science and Technology, The University of Tokyo, 7-3-1 Hongo, Bunkyo-ku, Tokyo 113-8656, Japan}

\begin{abstract}
Distributed architecture is a promising route to scaling fault-tolerant quantum computing (FTQC) beyond the inherent limitations of single processors, for which high-fidelity logical Bell pairs need to be prepared from many noisy physical Bell pairs with high efficiency.
For practical implementation of distributed FTQC, logical Bell pair preparation must be designed not only for efficient Bell pair consumption but also for the spacetime volume of the protocol; however, entanglement distillation protocols have primarily focused on minimizing the consumption of Bell pairs, often resulting in protocols that require a substantial number of local operations.
A key challenge is to find an appropriate balance between these two contrasting features.
To resolve this issue, we introduce a metric for characterizing the practical cost of preparing high-fidelity logical Bell pairs, \emph{link-limited volume} (LLV), which is a circuit-volume metric incorporating, in a single quantity, both the cost of physical Bell pair consumption and the volume associated with local operations.
Guided by this metric, we propose the \emph{entanglement boosting} protocol that achieves efficient preparation of logical Bell pairs encoded in rotated surface code, with LLV reduced by orders of magnitude compared to prior state-of-the-art methods.
In this protocol, paralleling recent advances in magic state cultivation, we employ soft-information decoders and postselection to suppress the logical error rates of Bell pairs to practical levels, e.g.~less than $10^{-10}$ from 86 noisy physical Bell pairs at 1\% error, while all local operations are implementable within a spatial region of a single surface code patch with two-dimensional local connectivity.
This is a substantial reduction from other protocols, such as remote lattice surgery operations requiring nearly 1000 physical Bell pairs under the same setting.
To further augment the entanglement boosting, we also present a pipelined implementation of entanglement distillation using high-rate quantum error-correcting codes, enabling arbitrarily low logical error rates while also maintaining physically efficient implementations.
These results pave the way for the practical implementation of distributed FTQC, reinforcing the benefits of fast interconnect technologies and serving as a guiding principle for the efficient design of protocols and devices.
\end{abstract}

\maketitle

\section{Introduction}\label{sec:intro}

High-fidelity, maximally entangled qubit pairs (Bell pairs) prepared over a network are a fundamental resource for distributed quantum technologies, including quantum communication~\cite{Panayi2014,Azuma2023}, blind quantum computing~\cite{Fitzsimons2017}, quantum sensing~\cite{Gottesman2012}, and distributed fault-tolerant quantum computing (FTQC)~\cite{Monroe2014,Sunami2025}. 
In particular, distributed FTQC imposes stringent requirements on the fidelity of the Bell pairs, which must be encoded in quantum error-correcting codes to ensure faithful logical operations.
Since the physical Bell pairs generated over the network are generally noisy, it is necessary to execute protocols that turn noisy Bell pairs into high-fidelity ones.

In the standard setting for information theory and entanglement theory, \emph{entanglement distillation} refers to the process of converting many noisy Bell pairs into a smaller number of nearly maximally entangled states using noiseless local operations and classical communication (LOCC)~\cite{Bennett1996purification}. 
In this framework, LOCC are free, so the task is to characterize which noisy states are distillable, and at what rate, under this idealized model~\cite{Horodecki2009}. 
Canonical examples include recurrence protocols~\cite{Bennett1996purification,Deutsch1996} and one-way hashing protocols~\cite{Bennett1996mixedstate}, which established the foundations of entanglement distillation and highlighted the close relationship with quantum error-correcting codes~\cite{Bennett1996mixedstate,Dur2007}.

In contrast, in the context of distributed FTQC, the LOCC assumptions break down: local operations are noisy, and therefore, error correction must be performed explicitly to implement protocols for obtaining high-fidelity remote Bell pairs.
In practice, these costs can be comparable to, or even exceed, the costs associated with network usage, especially in hardware platforms enabling high-speed remote Bell pair generation methods, such as neutral atoms and trapped ions~\cite{Stephenson2020,Sunami2025,main2025,Li2024,Sinclair2025}.
A crucial metric for local operation cost is the spacetime volume, i.e., the number of physical qubits involved in a protocol multiplied by the duration of the computation, with a typical time unit being the number of syndrome extraction (SE) cycles, where a cycle corresponds to measuring all syndrome checks of a quantum error-correcting code once~\cite{litinski2019game,Fowler2012,Gidney2021howtofactorbit}.
This volume, in units of \textit{qubit-cycles}, comprises operations needed for local error correction and logical-level circuit execution on encoded qubits, typically involving hundreds to thousands of physical qubits and tens to hundreds of SE cycles, even for small-scale logical circuits for entanglement distillation.
Therefore, protocols must be judged not only by how few Bell pairs they consume, but also by the spacetime volume for local operations.

The reduction of spacetime volume has been the focus of the development of magic state distillation protocols in recent years~\cite{Litinski2019magicstate, litinski2019game}, leading to rapid improvements in protocols for preparing logical magic states, such as better layout design and multi-stage strategies~\cite{Itogawa2025,Gidney2024,Hirano2024,chen2025efficientmagicstatecultivation,Gidney2025resource}.
Critical differences in the operational constraints between logical magic state preparation and logical Bell pair preparation necessitate distinct metrics and strategies.
The physical magic states are nearly free: these can be obtained by local single-qubit gates with no latency at any location.
In contrast, the generation of remote physical Bell pairs is inherently different from local operations, achievable through only sequential generation at a finite throughput, which limits their usage.
While it is possible to buffer the required Bell pairs before starting a protocol, the associated space and time costs for the buffering cannot be ignored in practice.
This favors protocols with a careful balance between the Bell pair consumption, a standard metric for the LOCC framework, and the spacetime volume for the protocol, a metric used for magic state preparation; however, finding such a balance is challenging without a guiding principle.

There are currently two common approaches to preparing high-fidelity logical Bell pairs, each of which primarily focuses on only one of the two contrasting desired properties discussed above. 
First, the physical-to-logical approach utilizes the physical Bell pairs directly in the protocol, such as for syndrome extraction across the network in lattice-surgery-based protocols~\cite{Ramette2024,Shalby2025}, as well as through direct projections of many physical Bell pairs onto a code space via syndrome extraction~\cite{Glancy2006,Maeda2025,Ataides2025}.
These protocols require hundreds to thousands of physical Bell pairs to achieve the high fidelities required for large-scale FTQC\@.
Second, the injection-and-distillation approach begins by first injecting physical Bell pairs onto logical qubits and executing the distillation protocols with logical gates~\cite{Sunami2025}.
This includes LOCC protocols implementable by logical gates, such as recurrence protocols~\cite{Bennett1996purification,Deutsch1996} and concatenated stabilizer-code distillation~\cite{Pattison2025}.
While the physical-to-logical protocols are implementable with relatively small local circuit volume, the required number of Bell pairs is generally large, resulting in a substantial requirement for network performance.
On the other hand, while the injection-and-distillation protocols are efficient in Bell pair usage, the local circuit volume is significant due to the inherent overhead of fault-tolerant gates, thus potentially diminishing the benefits of modular scaling.

In this work, we develop an efficient protocol to turn noisy physical Bell pairs into high-fidelity logical Bell pairs encoded in a rotated surface code~\cite{Bombin2007}.
As a guiding metric for logical Bell pair generation, we propose a \textit{link-limited volume} (LLV) to quantify the overall cost of high-fidelity logical Bell pair preparation, in the presence of noisy local operations and the finite speed of physical remote Bell pair generation. 
This is defined in qubit-cycles~\cite{litinski2019game} and accounts for both the spacetime volume of local operations and the buffering cost imposed by the finite throughput of physical Bell pair generation.
By expressing the requirements for networks and local operations within a single quantity, LLV provides a common metric that enables efficient optimization for the realistic implementation of logical Bell pair preparation for distributed FTQC\@.

Guided by this metric, we propose \emph{entanglement boosting}, a physical-to-logical Bell pair preparation protocol for the rotated surface code.
This protocol combines (i) a code projection step that projects the physical Bell pairs onto the logical code space of a surface code with a small code distance, (ii) code expansion to the target surface code distance, and (iii) postselection based on soft-information decoding.
The techniques are partly inspired by the magic state cultivation approach for efficient logical magic state preparation~\cite{Gidney2024} while tailored for logical Bell pair preparation.
This allows for both efficient use of physical Bell pairs and a small circuit volume while maintaining scalable error suppression, thus achieving a large reduction in the LLV.
We also design an efficient pipelined approach for implementing entanglement distillation circuits using logical gates and parallel logical-qubit reconfiguration, complementing the boosting stage.
The combined approach improves the yield with additional local operations, achieving a lower LLV in regimes with limited network throughput and enabling arbitrary logical error suppression.

These results are key to the scalable realization of FTQC, where modular architecture is expected to play a central role. 
By formulating a unified metric that quantifies both network throughput and local circuit volume, our framework provides a principled basis for optimizing protocols for logical Bell pair preparation under realistic hardware constraints. 
The protocols proposed in this work demonstrate low-volume implementations with a flexible design adaptable to a wide range of interconnect speeds and fidelity. 
Beyond quantum computation, our theoretical framework is broadly applicable to distributed information processing settings based on remote entanglement generation with local fault-tolerant operations, such as device-independent quantum key distribution~\cite{zapatero2023advances} and the communication-based demonstration of energy-consumption advantage of quantum computation~\cite{PRXEnergy.4.023008}, offering a systematic recipe for scalable and efficient distributed quantum technologies.

This article is organized as follows. 
In Sec.~\ref{sec:preliminaries}, we cover preliminaries, including stabilizer codes and stabilizer entanglement distillation.
In Sec.~\ref{sec:cost-model}, we describe the setup for the distributed FTQC assumed in this work and introduce LLV\@. 
We present the entanglement boosting protocol in Sec.~\ref{sec:boosting}, together with circuit-level numerical simulation results.
We then discuss the implementation of pipelined entanglement distillation with parallel logical-qubit reconfiguration in Sec.~\ref{sec:distillation}.
In Sec.~\ref{sec:conclusion}, we conclude our results and provide an outlook.

\section{Preliminaries}\label{sec:preliminaries}

Here, we first summarize the basic notations for quantum error correction relevant to this work in Sec.~\ref{sec:stab_codes}, stabilizer entanglement distillation in Sec.~\ref{sec:ent-distillation}, and our assumptions on the noise model in Sec.~\ref{sec:noise_model}.

\subsection{Stabilizer codes} \label{sec:stab_codes}

The single-qubit Hilbert space $\mathbb{C}^2$ is spanned by the computational basis $\{\ket{0}, \ket{1}\}$, where $\ket{\pm} = (\ket{0}\pm\ket{1})/\sqrt{2}$. 
On this space, the Pauli operators are defined as $X := \ket{0}\bra{1} + \ket{1}\bra{0}$, $Y := \mathrm{i}\ket{1}\bra{0} - \mathrm{i}\ket{0}\bra{1}$, $Z := \ket{0}\bra{0} - \ket{1}\bra{1}$, $I := \ket{0}\bra{0} + \ket{1}\bra{1}$, where $\mathrm{i} = \sqrt{-1}$. 
The $n$-qubit Pauli group $\mathcal{P}_n$ consists of tensor products of single-qubit Pauli operators up to a global phase $\alpha \in \{\pm 1, \pm \mathrm{i}\}$.

A \emph{stabilizer} $\mathcal{S}$ is an Abelian subgroup of $\mathcal{P}_n$ that does not include $-I^{\otimes n}$. The corresponding \emph{stabilizer code} is the joint $+1$ eigenspace of $\mathcal{S}$:
\[
    \mathcal{C} = \{\ket{\psi} \in (\mathbb{C}^2)^{\otimes n} \mid s\ket{\psi} = \ket{\psi}, \; \forall s \in \mathcal{S}\}.
\]
If $\mathcal{S}$ is generated by $n-k$ independent stabilizer generators, then $\mathcal{C}$ encodes $k$ logical qubits, i.e., $\dim \mathcal{C} = 2^k$. 
Let $\mathcal{N}(\mathcal{S})\coloneqq \{P\in\mathcal{P}_n\mid P\mathcal{S}P^{-1}=\mathcal{S}\}$ be the normalizer of $\mathcal{S}$, where $P\mathcal{S}P^{-1}\coloneqq \{PsP^{-1}\mid s\in\mathcal{S}\}$.
Let $\mathcal{Z}\coloneqq \{\alpha I^{\otimes n}\mid\alpha\in\{\pm1, \pm \mathrm{i}\}\}$ be the center of $\mathcal{P}_n$, the set of elements that commute with all elements in $\mathcal{P}_n$.
The logical Pauli group is the quotient group $\mathcal{L}\coloneqq \mathcal{N}(\mathcal{S})/(\mathcal{S}\cdot \mathcal{Z})$, where $\mathcal{S}\cdot \mathcal{Z}=\{sz\mid s\in\mathcal{S}, z\in\mathcal{Z}\}$.
An operator $\mathcal{L}\in \mathcal{N}(\mathcal{S})$ is called a logical operator if we view it modulo $\mathcal{S}\cdot \mathcal{Z}$, i.e., via its coset $ [L]\in \mathcal{L}$.
It is called a \textit{nontrivial logical operator} if and only if its coset $[L]\in\mathcal{L}$ is nontrivial, i.e., $ [L]\neq [I]$.
The \emph{weight} $|P|$ of $P \in \mathcal{P}_n$ is the number of qubits on which $P$ acts nontrivially, i.e.~$|P|$ counts the tensor factors of $P = P_1 \otimes \cdots \otimes P_n$ that are $X$, $Y$, or $Z$, rather than the identity. 
The \emph{distance} $d$ of a stabilizer code $\mathcal{C}$ is the minimum weight of all the nontrivial logical operators. 
The logical Pauli operators $\overline{X}_i$ and $\overline{Z}_i$ ($i=1,\dots,k$)
are representatives of cosets in $L = \mathcal{N}(\mathcal{S})/(\mathcal{S}\cdot\mathcal{Z})$ that act nontrivially
on the $i$-th logical qubit. They commute with all $s\!\in\!S$ and satisfy
$[\overline{X}_i,\overline{Z}_j]=0$ ($i\neq j$) and
$\{\overline{X}_i,\overline{Z}_i\}=0$, where $[A,B]\coloneqq AB-BA$, and $\{A,B\}\coloneqq AB+BA$.

A stabilizer code encoding $k$ logical qubits into $n$ physical qubits with distance $d$ is called an $[[n,k,d]]$ code.
A particularly important family is the Calderbank-Shor-Steane (CSS) codes~\cite{Calderbank1996,Steane1996_CSS}, where stabilizer generators are tensor products of only $I$ and $X$ (X-type) or only $I$ and $Z$ (Z-type). 
Quantum error correction involves two processes: \textit{syndrome extraction} and \textit{decoding}.
Syndrome extraction measures stabilizer generators, referred to as \textit{stabilizer checks}, to obtain error information, which is called the \textit{error syndrome}.
Decoding is a classical computation process that, given an error syndrome, returns a recovery operator consistent with the error syndrome.

\begin{figure}[t]
	\centering
	\includegraphics[width=0.9\linewidth]{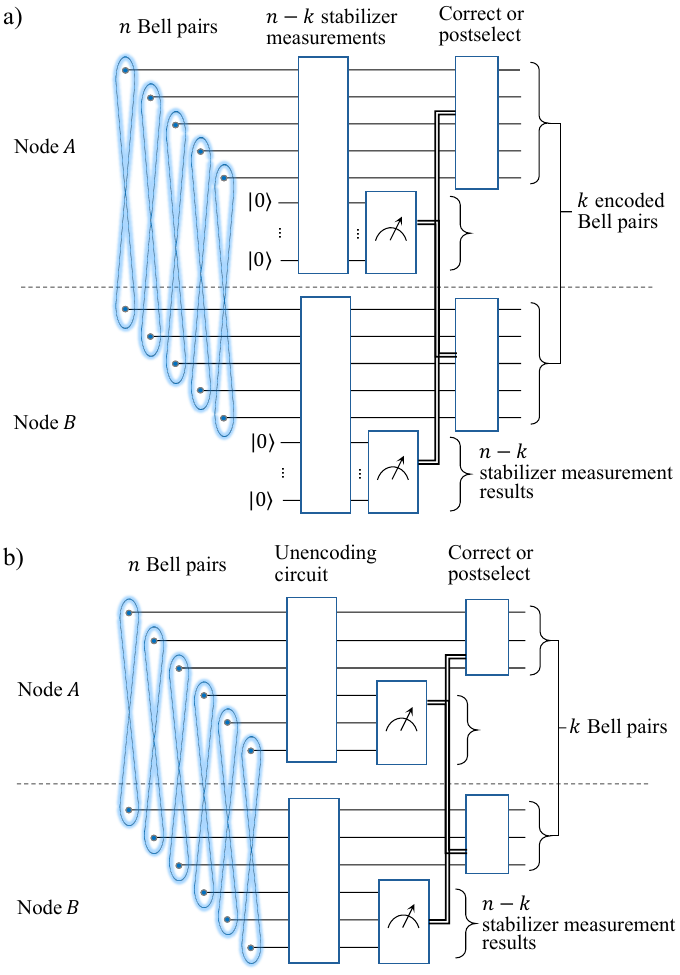}
	\caption{Two entanglement distillation protocols based on $[[n,k,d]]$ stabilizer code $\mathcal{C}$ used in this work.
		a) Auxiliary-qubit-assisted, projection-based preparation of logical Bell pairs from physical Bell pairs.
		Here, stabilizer checks of $n$ Bell pairs are measured by $n-k$ auxiliary qubits, allowing either postselection or error correction to achieve $k$ high-fidelity logical Bell pairs encoded in code $\mathcal{C}$.
		b) A decoding-based approach, which combines the protocol in a) and decoding of the encoded state, to result in $k$ output Bell pairs.
		The stabilizer measurements are performed by the measurement of $n-k$ qubits that do not constitute the output.
	}
	\label{fig:distillation_protocols}
\end{figure}

\subsection{Entanglement distillation with stabilizer codes}\label{sec:ent-distillation}

An $[[n,k,d]]$ stabilizer code $\mathcal{C}$ can be used to construct entanglement distillation protocols~\cite{matsumoto2003conversion, hostens2004equivalence,Glancy2006}.
Here, we outline two approaches used in this work, illustrated in Fig.~\ref{fig:distillation_protocols}.

The first approach (Fig.~\ref{fig:distillation_protocols}a) starts with $n$ Bell pairs and prepares $k$ logical Bell pairs encoded in stabilizer code $\mathcal{C}$ by measuring stabilizer checks of the code in nodes $A$ and $B$~\cite{Glancy2006}.
Specifically, stabilizer measurements are performed by $n-k$ auxiliary qubits in each node, and the measurement results are sent between $A$ and $B$, where $B$ combines the results by taking the parities of the corresponding stabilizer values.
The measurement results can be used to either postselect the output states via two-way communication (error detection), or to perform correcting operations identified by classical decoding (error correction) via one-way communication.
The stabilizer measurements and error detection (correction) on the joint checks leave the state stabilized by $\mathcal{S}' = \{s\otimes s, s\in \mathcal{S}\}$.
The effect of the above procedure on the logical qubits can be identified by the fact that the protocol begins with $n$ physical Bell pairs.
The $i$-th pair (for $i=1,...,n$) is stabilized by $X_{i,A}\otimes X_{i, B}$ and $Z_{i,A}\otimes Z_{i, B}$, where $X_{i, A(B)}$ and $Z_{i, A(B)}$ are the corresponding Pauli operators on the $i$-th physical qubit in node A(B).
For $X$ and $Z$ logical operators for $k$ logical qubits, $\mathcal{L}_X = \{\overline{X}_1,...\overline{X}_k\}, \mathcal{L}_Z = \{\overline{Z}_1,...\overline{Z}_k\}$, the resulting logical qubits are stabilized by $\overline{X}_{i,A}\otimes \overline{X}_{i,B}$ and $\overline{Z}_{i,A}\otimes \overline{Z}_{i,B}, $ for $i=1$ to $k$; therefore, the resulting states are logical Bell pairs~\cite{Ataides2025}.

The second approach does not leave the resulting state encoded in the chosen code. 
Let $U_\mathrm{enc}$ denote the unitary encoding map that prepares a codeword
\(
\ket{\bar{\psi}} = U_\mathrm{enc} (\ket{\psi} \otimes \ket{0}^{\otimes (n-k)})
\)
for \(k\)-qubit state \( \ket{\psi} \).
The inverse operation \( U_\mathrm{enc}^{-1} \), referred to as the \emph{unencoding unitary},
acts as $U_\mathrm{enc}^{-1} \ket{\bar{\psi}}
   = \ket{\psi} \otimes \ket{\mathrm{syndrome}},$
where the last \(n-k\) qubit states, denoted as syndrome qubits, can be measured in the computational basis to obtain the stabilizer values.
By appending an unencoding unitary to the protocol described above (Fig.~\ref{fig:distillation_protocols}a), we obtain $k$ unencoded Bell pairs as a result of successful execution.
This can be simplified to only an application of $U_\mathrm{enc}^{-1}$ to $n$ Bell pairs~\cite{Dur2007}, which results in $k$ output qubits along with $n-k$ qubits that can be measured to provide the stabilizer checks needed for error correction or detection (Fig.~\ref{fig:distillation_protocols}b).

\subsection{Noise model}\label{sec:noise_model}
In this work, we perform circuit-level simulations to numerically evaluate the logical Bell pair preparation protocols.
We assume that physical operations are associated with the following noise model with noise strength $p=10^{-3}$:
$\ket{0}$($\ket{+}$)-state qubit preparation is flipped to $\ket{1}$($\ket{-}$) with a probability of $p$, qubit measurement results are flipped with a probability of $p$, and single-qubit gates are followed by $X$, $Y$, or $Z$ by probabilities of $p/3$ each.
Two-qubit gates are followed by one of the two-qubit Pauli operators $ I\otimes X,\; I\otimes Y,\; I\otimes Z,\; ...,  Z\otimes Z$, except for the identity, with a probability of $p/15$ each. We consider no errors for qubit idling, as is appropriate for neutral atoms and trapped ions with coherence times many orders of magnitude longer than gate times~\cite{Bluvstein2022,Duan2010}.
Remote physical Bell pair generation is associated with an error rate $p_\mathrm{Bell}$.
Bell pairs $\ket{\Phi^+}_{AB} = \frac{1}{\sqrt{2}} \left( \ket{0}_A\ket{0}_B + \ket{1}_A\ket{1}_B \right)$ are shared between nodes $A$ and $B$, followed by qubit $B$ of the pair experiencing $X$, $Y$ or $Z$ error with probability $p_\text{Bell}/3$ each.

\section{Distributed FTQC and link-limited volume}\label{sec:cost-model}

\begin{figure}[t]
	\centering
	\includegraphics[width=3.4in]{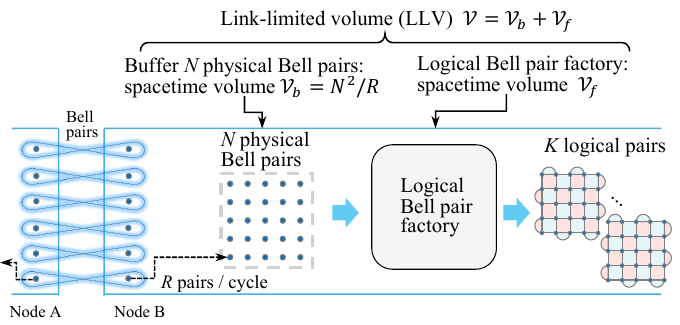}
	\caption{The link-limited volume (LLV). LLV consists of network-related volume $\mathcal{V}_b=N^2/R$ and Bell pair factory volume $\mathcal{V}_f$ in each node (Eq.~\eqref{eq:llv}), where $N$ is the number of physical Bell pairs needed for the factory and $R$ is the throughput of the physical Bell pairs, defined by the number of Bell pairs generated in the duration a single syndrome extraction cycle.
	}
	\label{fig:llv}
\end{figure}

We consider two computing nodes $A$ and $B$ linked by an interconnect.
The interconnect generates physical Bell pairs at a fixed speed (throughput), $R$ qubits in each syndrome extraction (SE) cycle of the rotated surface code (Fig.~\ref{fig:llv}).
In distributed computing, computing nodes are typically placed nearby, resulting in classical communications with high bandwidth and negligible latency compared to the required time for SE\@.
As such, we treat classical communication as free, be it one-way or two-way, throughout this article.

The \emph{spacetime volume} is an important metric for evaluating how costly a certain protocol is. 
In the context of FTQC, this is typically computed in units of qubit-cycles~\cite{litinski2019game}, where the space cost is counted by the number of physical qubits actively involved in the protocol, and the time cost is evaluated by the number of SE cycles.
For example, a transversal CNOT gate between two distance-$d_s$ rotated surface code patches requires $2d_s^2-1$ qubits per patch and is followed by $d_s$ cycles of SEs, with the leading-order term of the volume being $4d_s^3$, while the corresponding term for lattice-surgery CNOT is $12d_s^3$~\cite{Horsman2012,Gidney2024,Sahay2025}.

\begin{figure*}[t]
	\centering
	\includegraphics[width=0.99\linewidth]{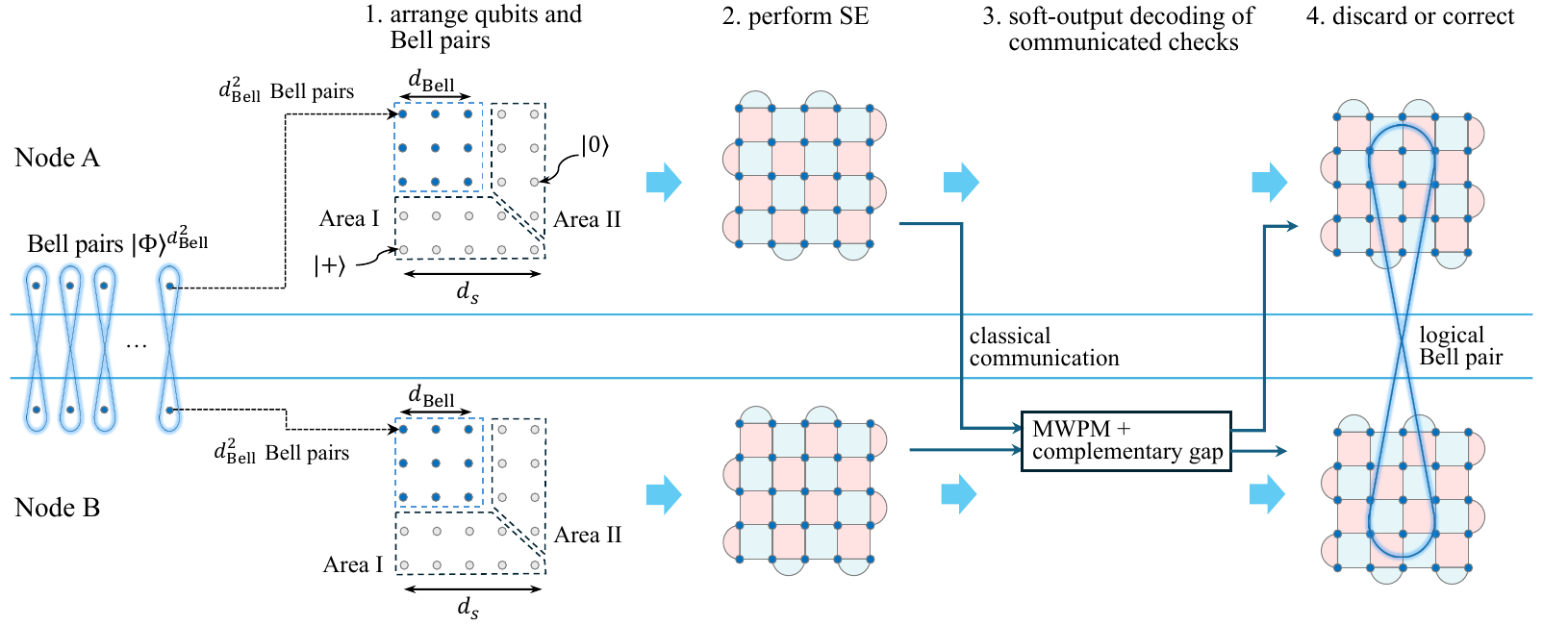}
	\caption{Entanglement boosting. 
		Entanglement boosting begins with the preparation of $d_\text{Bell}^2$ physical Bell pairs, which are to be arranged in a square grid, together with qubits in $\ket{0}$ and $\ket{+}$ around them to form a $d_s\times d_s$ square grid.
		This is followed by $d_s$ cycles of syndrome extraction (SE) and MWPM decoding.
		We additionally perform decodings for complementary logical outcomes to compute the complementary gap~\cite{Gidney2025}, which allows efficient postselection (see Appendix~\ref{sec:complementary-gap} for the details of complementary gap calculation).
	}
	\label{fig:boosting}
\end{figure*}

Our primary interest is the spacetime volume required to prepare logical Bell pairs, which we call LLV (Fig.~\ref{fig:llv}),
\begin{equation}\label{eq:llv}
    \mathcal{V} = \mathcal{V}_b + \mathcal{V}_f,
\end{equation}
with $\mathcal{V}_b$ and $\mathcal{V}_f$ specified in the following.
In~\eqref{eq:llv}, the first term $\mathcal{V}_b$ is the volume associated with buffering $N$ physical Bell pairs needed to perform an instance of the logical Bell pair preparation.
For physical Bell-pair generation throughput of $R$ pairs per cycle, the volume is $N^2/R$, since a buffer space of $N$ must be kept for $N/R$ cycles of SEs. 
The second term $\mathcal{V}_f$ in~\eqref{eq:llv} is the spacetime volume for local operations needed to prepare a high-fidelity logical Bell pair using the buffered physical Bell pairs.
This can be computed by counting the number of physical qubits involved in the logical Bell pair preparation protocol at each SE cycle and summing this value throughout the protocol.
LLV must be modified from the above for protocols that consume physical Bell pairs sequentially or in a pipelined manner.
For example, the remote lattice-surgery-based logical Bell pair preparation protocols~\cite{Ramette2024,Shalby2025} consume $O(d_s)$ Bell pairs in each SE cycle over $d_s$ cycles, instead of requiring all Bell pairs at the start of the protocol.
In such a case, if the physical Bell pair throughput $R$ is larger than the consumption speed, $\mathcal{V}_b$ becomes 0 since no buffering is required; 
if $R$ is smaller than the Bell pair consumption speed, then some of the Bell pairs must be buffered initially to ensure that SE cycles are performed without additional latency, thus finite $\mathcal{V}_b$ is required.
For protocols involving postselection with an acceptance rate of $q$, the LLV must be computed by the protocol volume multiplied by the expected attempt count $1/q$.
We further let $\mathcal{Y}$ denote the yield of the Bell pair factory operation as the number of output logical Bell pairs per input physical Bell pair, considering retries in the case of probabilistic protocols.
For example, a factory protocol requiring $N$ physical Bell pair inputs and $K$ logical Bell pair outputs, with a success probability of $q$, has a yield of $\mathcal{Y}=qK/N$.
Inverse yield $1/\mathcal{Y}=N/qK$ thus denotes the number of physical Bell pairs required to output one logical Bell pair.

LLV quantifies how the optimal balance between the physical Bell pair requirement $\mathcal{V}_b$ and the volume of local operations $\mathcal{V}_f$ changes with the Bell pair generation throughput $R$.
With a slow network with small $R$, where the shared entanglement is costly, protocols with small physical Bell pair consumption are favored; in contrast, with a fast interconnect with large $R$, LLV is affected more strongly by the volume of local operations, precisely reflecting the situation with less cost for creating Bell pairs.

The concrete values of the throughput $R$ vary by orders of magnitude depending on the remote entanglement generation protocols, qubit types, and the implementation details of the SE cycles.
For example, the state-of-the-art remote entanglement generation speed for trapped ions is on the order of $100~\mathrm{s}^{-1}$~\cite{Stephenson2020,Oreilly2024}, and an estimation of the duration of a surface-code SE cycle is on the order of $100~\upmu\mathrm{s}$ to $1~\mathrm{ms}$~\cite{Lekitsch2017,beverland2022}, giving $R$ that ranges from $10^{-2}$ to $10^{-1}$.
In general, however, $R$ can differ by many orders of magnitude across platforms.
The characteristic gate times already vary by over three orders of magnitude~\cite{beverland2022} among leading qubit technologies, and the achievable rate of remote entanglement generation is expected to span an even wider range: while optical losses and inefficient qubit-photon coupling may significantly reduce the speed, fast interconnects such as microwave channels~\cite{Magnard2020} and optical cavities~\cite{Li2024,Sunami2025,Hartung2025} may enhance the speed by orders of magnitude.
As a representative range, in this work, we consider $R$ between $10^{-3}$ and $10^{2}$.

\section{Entanglement boosting}\label{sec:boosting}

Entanglement boosting is a logical Bell pair generation protocol operated within a single surface code patch, combining code projection onto a small surface code patch, code expansion, and postselection based on soft-output decoding~\cite{Bombin2024,Gidney2024}.
Conceptually, entanglement boosting consists of two steps.
The first is preparing a logical Bell pair encoded in a $[[d_{\text{Bell}}^2,1,d_{\text{Bell}}]]$ rotated surface code of distance $d_{\text{Bell}}$ using the logical Bell pair preparation procedure of Fig~\ref{fig:distillation_protocols}a,
and the second is its expansion to a larger distance $d_s$ required to preserve the postselected states and for further operations, such as entanglement distillation, implemented by logical gates.
The two steps can, in fact, be implemented simultaneously, achieving both reduced local error effects and spacetime volume, as described below.

In the first step, we prepare $d_\text{Bell}^2$ physical Bell pairs between two parties $A$ and $B$.
The $i$th pair is stabilized by $X_i^{(A)}\otimes X_i^{(B)}$ and $Z_i^{(A)}\otimes Z_i^{(B)}$ where the superscript $A(B)$ denotes the operators acting on qubits in nodes $A (B)$.
Both parties arrange the respective endpoints of the Bell pairs into a square lattice and then locally perform syndrome extraction of the rotated surface code.
For the rotated surface code on a $d_\text{Bell}\times d_\text{Bell}$ square lattice, stabilizer checks are arranged in a checkerboard pattern~\cite{Bombin2007}, with plaquettes representing $X$($Z$)-type stabilizer checks $g_{X(Z)}$, shown as red and blue plaquettes in Fig.~\ref{fig:boosting}.
Each stabilizer check acts as the tensor product of Pauli $X$($Z$) operators on the four (or two at the boundary) qubits at the plaquette corners.
Syndrome extraction requires $d_\text{Bell}^2-1$ auxiliary qubits in each node.
Following the syndrome extraction, the parities of the corresponding measurement outcomes between the two parties are computed via classical communication, obtaining the values for joint stabilizer checks such as $g_{X}^{(j, A)} \otimes g_{X}^{(j, B)}$.
In the noiseless case, these values must all be even, whereas noise may flip them.
Upon successful execution of this protocol, the state is stabilized by $\overline{X}^{(A)}\otimes \overline{X}^{(B)}$ and $\overline{Z}^{(A)}\otimes \overline{Z}^{(B)}$, hence the output state is a $k=1$ logical Bell pair encoded in the rotated surface code~\cite{matsumoto2003conversion, hostens2004equivalence}.

\begin{figure*}[t]
	\centering
	\includegraphics[width=0.99\linewidth]{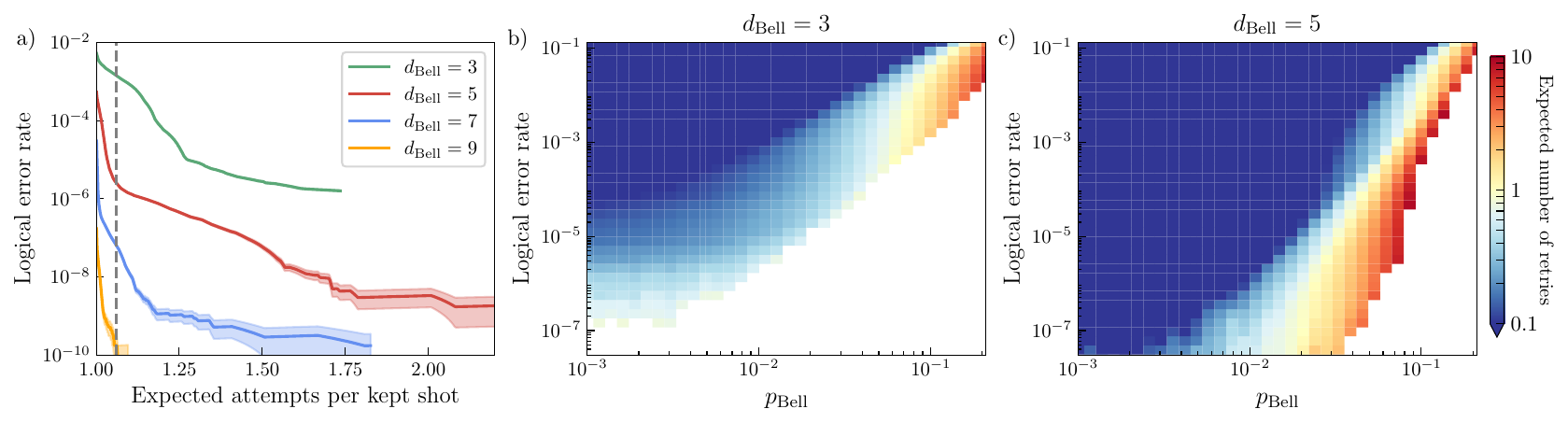}
	\caption{Numerical simulation results for the entanglement boosting protocol. 
		a) Numerical simulation results of the logical error rates of the logical Bell pairs generated by the entanglement boosting protocol, for varying postselection criteria, for $d_\mathrm{Bell}= 3, 5, 7$ and 9, with $d_s$ fixed at 19 and $p_\mathrm{Bell} = 0.01$.
        Varying threshold values for the complementary gap results in different acceptance rate $q_0$; here, the logical error rate is shown for varying expected attempt count per kept shots, $1/q_0$.
        The shaded region is the standard error of estimated probability from $10^{10}$ sampling results.
        At $d_\mathrm{Bell}=9$, logical error rate of $10^{-10}$ is reached at the expected attempts per kept shot of $1.06$ (vertical gray dashed line): a logical Bell pair with logical error rate of $10^{-10}$ is prepared with an average of 86 noisy physical Bell pairs having $p_\mathrm{Bell}=1\%$.
		b-c) Numerical simulation results for the expected number of retries (colors), $1/q_0 - 1$, to achieve target error rates (vertical axis) for varying input Bell pair error rates $p_\text{Bell}$ (horizontal axis) over two orders of magnitude, for $d_\text{Bell}=3$ and 5.
	}
	\label{fig:boosting-performance}
\end{figure*}

In the next step, the code is expanded to a larger code distance $d_s$, as in the protocol in Ref.~\cite{li2015magic}.
In this procedure, as shown in Fig.~\ref{fig:boosting}, additional physical qubits are prepared in $\ket{+}$ ($\ket{0}$) in areas I (II) around the initial surface code patch of distance $d_\text{Bell}$ separated by the diagonal line, and syndrome extraction for the expanded code is performed for $d_s$ cycles.

In entanglement boosting, the above two steps are performed simultaneously.
While the code expansion deforms the two-weight stabilizers along the bottom and right edges of the initial distance-$d_\mathrm{Bell}$ surface code to four-weight stabilizers, the additional qubits are arranged such that the new stabilizers yield the same outcomes as the original two-weight stabilizers;
for example, as shown in Fig.~\ref{fig:boosting}, each of the additional physical qubits below the initial patch is prepared in $\ket{+}$, and thus the two-weight $X$ stabilizer check in the original patch and the corresponding four-weight $X$ stabilizer check in the expanded patch yield the same outcome.
This allows the simultaneous operation of the two steps to reduce the effects of local errors and the overall volume.

Operationally, the combined procedure proceeds as follows.
Initially, both parties arrange $d_\text{Bell}^2$ physical Bell pairs and $d_s^2 - d_\text{Bell}^2$ physical qubits prepared in $\ket{+}$ and $\ket{0}$ (Fig.~\ref{fig:boosting}).
Next, $d_s$ cycles of syndrome extraction are performed for the distance-$d_s$ surface code, where the stabilizer checks are denoted by red and blue plaquettes in Fig.~\ref{fig:boosting}.
$A$ sends the outcomes of the first cycle of syndrome extraction to $B$ via classical communication, and $B$ obtains the error syndrome by computing the parities of the corresponding syndromes.

The error syndrome is then decoded using the minimum-weight perfect matching (MWPM) decoder. 
The complementary gap~\cite{Bombin2024, Gidney2025, Gidney2024} is also computed, which is the absolute difference between the minimum weights identified in the MWPM decoder conditioned on the original and complementary logical outcomes (see Appendix~\ref{sec:complementary-gap} for more details and illustrations of the complementary gap). 
If this value is below a chosen threshold, the boosting protocol is aborted.
Otherwise, the resulting logical qubit pair is kept, and error correction is performed.
The complementary gap captures error information from both the encoding and expansion steps, allowing postselection on the decoded error information and thereby achieving a substantial reduction in output error rates.
In our numerical simulation, we sampled the above protocol and obtained the complementary gap value and the existence of logical errors for each shot by evaluating the $\overline{X}^{(A)}\otimes \overline{X}^{(B)}$ and $\overline{Z}^{(A)}\otimes \overline{Z}^{(B)}$ logical operators.
For a given threshold complementary gap value, we first obtain the acceptance rate $q_0$ by dividing the number of accepted shots (the complementary gap value below the threshold) by the total number of samples;
then, we determine the logical error rate by counting the number of shots that have a gap value below the threshold and also contain a logical error, and dividing this count by the number of accepted shots (see Appendix~\ref{app:numerics} for the details of the numerical simulation).

\begin{figure*}[t]
	\centering
	\includegraphics[width=0.99\linewidth]{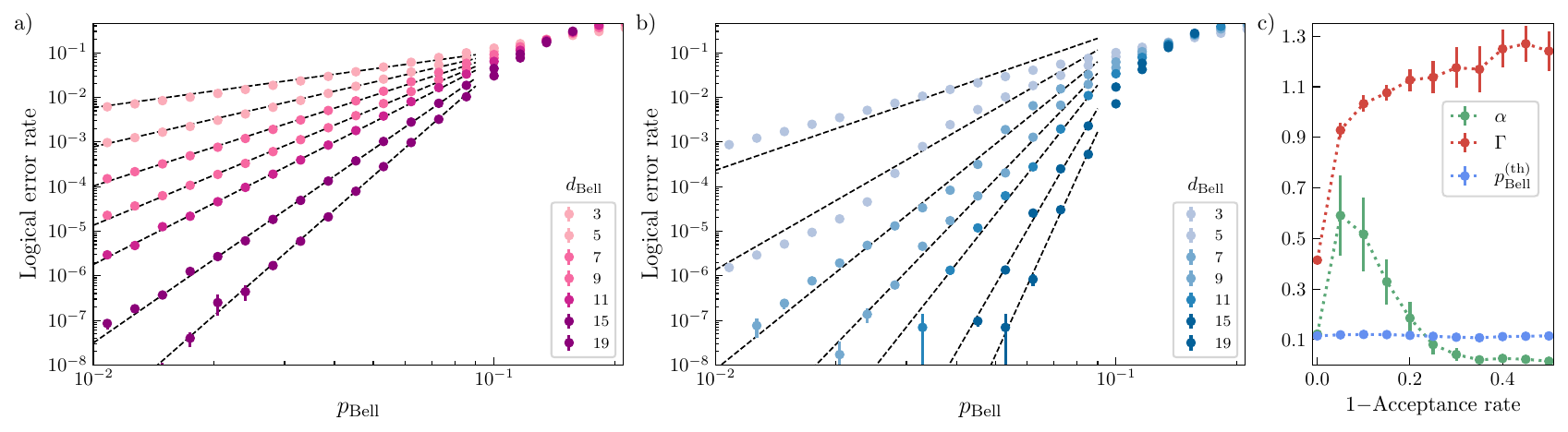}
	\caption{Scaling of the logical error rate of the Bell pairs produced with the entanglement boosting protocol.
    a-b) logical error rate as a function of the physical Bell pair error rate~$p_{\mathrm{Bell}}$ for different~$d_{\mathrm{Bell}}$, with acceptance rates $q_0$ of 100\% (a) and 90\% (b) based on the complementary gap. 
    Each point is obtained from circuit-level simulations of the entanglement boosting protocol, and the dashed line represents the fitting with Eq.~\eqref{eq:scaling}.
    Error bars represent the standard error of estimated probability from the $10^8$ sampling results.
    c) the dependence of the fitted values of parameters $\alpha$, $\Gamma$ and $p_\mathrm{Bell}^\mathrm{(th)}$ on the discard rates $1-q_0$ of entanglement boosting, where the error bars represent 95\% confidence intervals.
    While $p_\mathrm{Bell}^\text{(th)}$ remain nearly constant, the scaling prefactor $\Gamma$ increases as a function of the discard rates.
    }
	\label{fig:scaling}
\end{figure*}

There is a conceptual connection between entanglement boosting and recent work on magic state cultivation~\cite{Gidney2024,chen2025efficientmagicstatecultivation}. 
Both protocols make use of soft-output decoding, such as the complementary gap for code expansion of postselected logical qubits; the initial state is prepared by the logical double-checking ($H_{XY}$ measurement) for magic state cultivation, while the projection of Bell pairs to a logical Bell pair is used for entanglement boosting.
This difference arises from the fact that Bell pairs are stabilizer states that admit simpler methods for error detection via stabilizer codes~\cite{Glancy2006}, allowing the entanglement boosting to have a much simpler implementation, as described above.

However, from an operational viewpoint, fundamental differences in the physical implementation of the physical $T$ gates and remote Bell pair generation lead to different design considerations. 
In magic state cultivation, physical $T$ gates are operated locally with negligible latency at any location and time, providing flexibility in protocol design. 
By contrast, in distributed FTQC, the remote Bell pairs are generated only by the photonic interconnect, which has a limited generation speed; therefore, the protocol must balance Bell pair usage against the local spacetime volume. 
For this reason, entanglement boosting only partially adapts ideas from magic state cultivation and tailors them to the fundamentally different theoretical and operational characteristics of remote entanglement. 
By optimizing the protocol based on the LLV metric to meet the distinct requirements for logical Bell pair preparation, entanglement boosting fully leverages advances in magic state preparation to achieve efficient logical Bell pair preparation.

In Fig.~\ref{fig:boosting-performance}, we show the circuit-level simulation results for the boosting protocol, with the final distance of the rotated surface code being $d_s=19$ and the input Bell pair error rate being $p_\text{Bell} = 1\%$.
We then vary $d_\text{Bell}$ and the postselection criteria, which are set by the threshold values for the complementary gap to discard the trial. 
From the results of the numerical simulation, we identify the acceptance rate $q_0$ for a given threshold complementary gap value, and the logical error rate of kept shots is plotted against the \textit{expected attempts per kept shot}, $1/q_0$.

These plots highlight the tradeoff between the output logical error rate and the acceptance rate of the boosting protocol. 
In Fig.~\ref{fig:boosting-performance}a, tightening the postselection criterion (moving to the right along the horizontal axis) initially rapidly suppresses the logical error rate of the kept shots at the cost of requiring more attempts per successful output. 
In particular, for $d_\mathrm{Bell}=9$, $1/q_0 = 1.06$ is sufficient to reach a logical error rate below $10^{-10}$: 
therefore, on average, 86 physical Bell pairs at 1\% error rate are sufficient to obtain a logical Bell pair at logical error rates eight orders of magnitude lower than the initial Bell pairs, while maintaining all the operations within a surface code patch and with 2D local connectivity.
Figures~\ref{fig:boosting-performance}b and~\ref{fig:boosting-performance}c further show the expected number of attempts required to reach a given logical error rate for varying physical Bell pair error rates $p_{\mathrm{Bell}}$. 
The error suppression is observed for up to $p_\mathrm{Bell} \approx 10\%$, demonstrating a wide operating regime of this protocol.
Overall, these results demonstrate that entanglement boosting provides a tunable mechanism to trade physical Bell pair consumption and postselection criteria for output error rates, offering a high-performance and highly tunable method to produce logical Bell pairs while maintaining all operations within a single surface code patch.
In Appendix~\ref{app:idle-error}, we analyze the effect of idling errors by performing additional numerical simulations of entanglement boosting protocols, with $p_\mathrm{Bell}=1\%$ and $d_\mathrm{Bell}=3$ and 5, with the same assumptions as Fig.~\ref{fig:boosting-performance}a except for the presence of idling errors with noise strength $p=0.1\%$.
For these parameters, the logical error rates reached by the entanglement boosting protocol remain similar in the presence of idling errors, while the expected attempts per kept shot increase by up to 40\%.

To further analyze the scaling of the logical error rates as a function of $d_\mathrm{Bell}$ and acceptance rates $q_0$, we present the logical Bell pair error rate of the entanglement boosting protocol in Fig.~\ref{fig:scaling}a-b as a function of the physical Bell pair error rates~$p_{\mathrm{Bell}}$ and~$d_{\mathrm{Bell}}$, for acceptance rates of 100\% (Fig.~\ref{fig:scaling}a; error correction) and 90\% (Fig.~\ref{fig:scaling}b). 
Points with different colors correspond to different $d_\mathrm{Bell}$ values, illustrating how the logical error probability decreases with increasing $ d_{\mathrm{Bell}}$ and a reduced Bell pair error rate; noise strengths for the local physical operations are maintained at 0.1\% throughout the simulations. 
The dashed lines indicate the fitting with an approximate scaling function for the logical error rate $p_L$ as a function of $p_\mathrm{Bell}$, $d_\mathrm{Bell}$, and $p_\mathrm{Bell}^\mathrm{(th)}$,
\begin{equation}\label{eq:scaling}
    p_L = \alpha \left(\frac{p_\mathrm{Bell}}{p_\mathrm{Bell}^\mathrm{(th)}} \right)^{\Gamma d_\mathrm{Bell}},
\end{equation}
where the fit is performed once for each panel, considering all $d_\mathrm{Bell}$ shown in each panel.
From the fits, we obtain $\alpha$, $\Gamma$ and $p_\mathrm{Bell}^\mathrm{(th)}$ that depend on the discard rate of the protocol.
Figure~\ref{fig:scaling}c shows how the fitted parameters depend on the discard rates. 
As the discard rates increase, $\Gamma$ increases from 0.4 to 1.2, while the parameter $p_{\mathrm{Bell}}^\mathrm{(th)}$ remains constant, quantifying the enhanced error suppression thanks to the postselection based on soft-output decoding~\cite{chen2025scalableaccuracygainspostselection}.
We remark that, due to the physical operations with noise strengths at $p=0.1\%$ error probabilities, the observed scaling has a limited range of applicability, as discussed in Appendix~\ref{app:scaling}, resulting in the saturation of error suppression for $p_\mathrm{Bell}$ below 1\%.
Further, it is expected that the logical error rate of the boosting protocol will saturate at the logical error rates of the rotated surface codes with $d_s$, such as $10^{-12}$ for $d_s=19$.
The fitted scaling shown in Fig.~\ref{fig:scaling}c is used within these constraints throughout this work.
We also show the additional results, including the results for larger discard rates, in Appendix~\ref{app:scaling}.

\begin{figure}[t]
\centering
\includegraphics[width=0.99\linewidth]{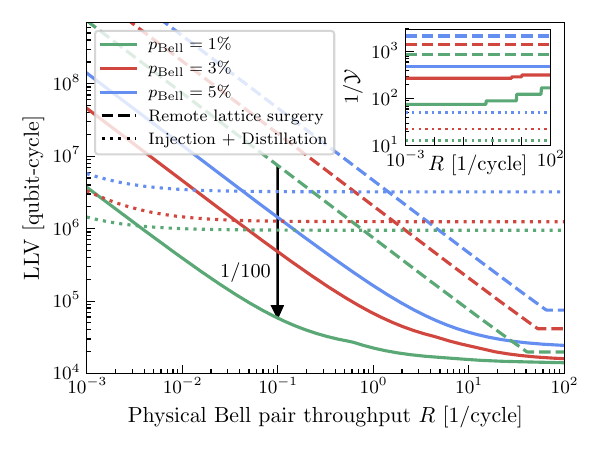}
	\caption{The link-limited volume (LLV) to prepare a logical Bell pair encoded in the rotated surface code at logical error rates of $10^{-10}$.
	We set the distance of the expanded surface code to be $d_s=19$.
	The dashed line is the corresponding LLV for the remote lattice surgery protocol for the rotated surface code (see Appendix~\ref{app:surgery} and Refs .~\cite {Ramette2024,Shalby2025}), and the dotted line is the concatenated injection-distillation protocol (see Appendix~\ref{app:concat-llv} and Ref.~\cite{Pattison2025}) with corresponding colors for each $p_\mathrm{Bell}$.
    At $R=0.1$, with $d_\mathrm{Bell}=9, d_s=19$ for entanglement boosting and $d_s=21$ for remote lattice surgery, the improvement of the LLV from the remote lattice surgery to the entanglement boosting protocol is 100, mainly from the order-of-magnitude improvement of the yield while maintaining nearly the same $V_f$.
	(inset) The number of physical Bell pairs consumed per output logical Bell pair (inverse yield, 1/$\mathcal{Y}$).
    The jumps in the plots are due to the changes in optimal $d_\mathrm{Bell}$ that minimizes the LLV for varying $R$.
}
\label{fig:boosting_llv}
\end{figure}

\begin{figure*}[t]
	\centering
	\includegraphics[width=0.8\linewidth]{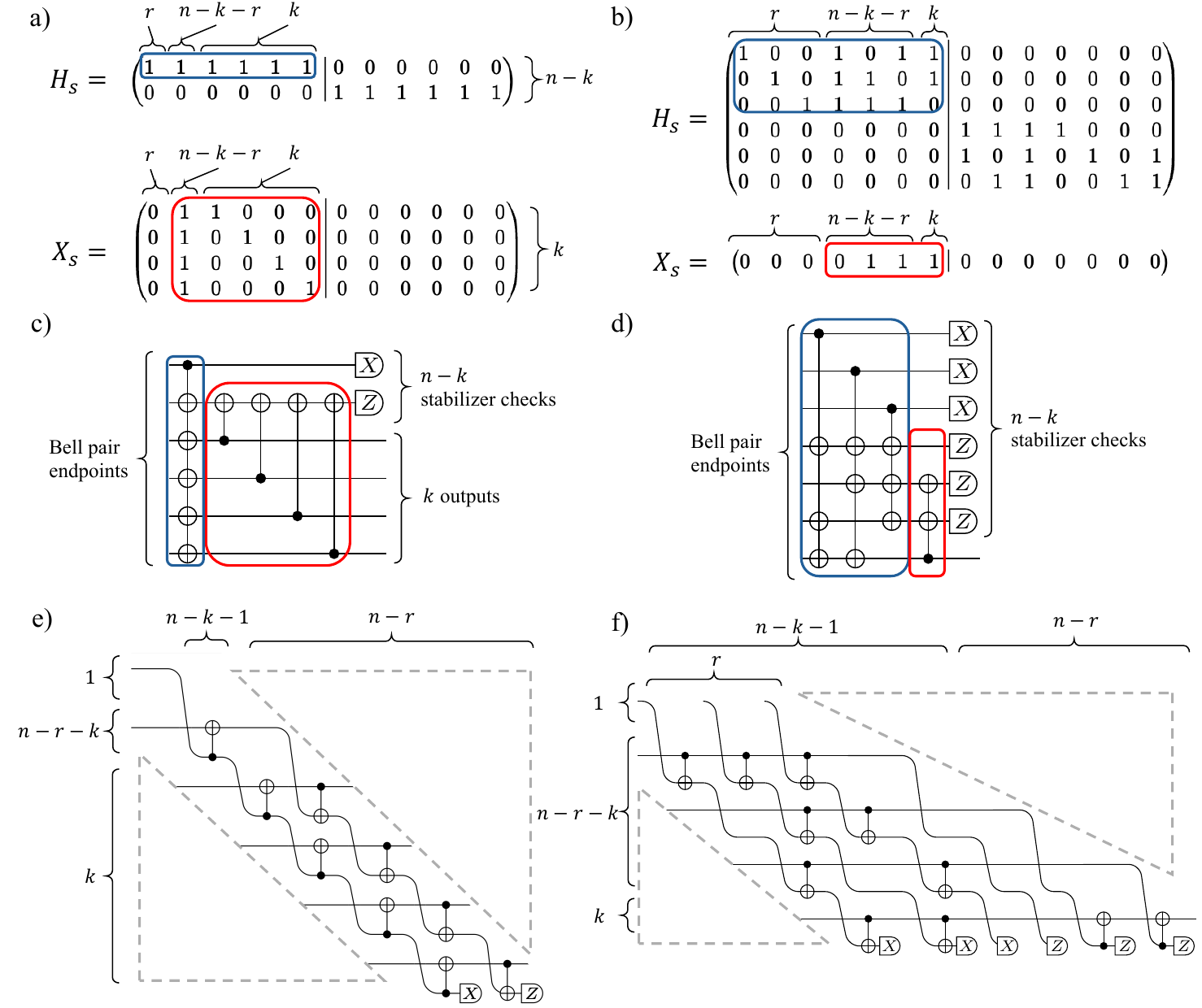}
	\caption{Pipelined entanglement distillation based on CSS codes.	
		(a,b) standard form of $(n-k)\times 2n$ binary matrices $H_s$ representing the stabilizer generators of $[[6,4,2]]$ quantum parity code and $[[7,1,3]]$ Steane code, along with binary matrix representation of logical $X$ gates.
		(c,d) entanglement distillation circuit (only for one of the nodes shown), based on the encoding circuit synthesis of Ref.~\cite{gottesman1997stabilizer}.
		(e,f) same circuits as (c,d), implemented with qubit reconfigurations (wire bending); see Fig.~\ref{fig:distillation}.
	}
	\label{fig:pipeline}
\end{figure*}

Finally, to assess the performance of entanglement boosting using a practically relevant circuit-volume metric (Sec.~\ref{sec:cost-model}), in Fig.~\ref{fig:boosting_llv}, we plot the LLV for preparing logical Bell pairs at a logical error rate of $10^{-10}$ encoded in the $d_s=19$ rotated surface code, starting from $p_\text{Bell}=1\%$, 3\%, and $5\%$, for varying physical Bell pair throughput $R$.
For this evaluation, we used the scaling of the logical error rates for varying postselection fractions, with discard rates $1-q_0$ of up to 50\% considered.
We compare the LLV with the remote lattice surgery protocol for logical Bell pair generation with rotated surface code~\cite{Ramette2024,Shalby2025}, which requires $O(d_s^2)$ Bell pairs over $d_s$ cycles of SE. 
We perform circuit-level simulation of the remote lattice surgery protocol with the same noise model as the one used for the simulation of the entanglement boosting protocol, as detailed in Appendix~\ref{app:surgery}, to evaluate the required $d_s$ to reach the target logical error rate of $10^{-10}$. 
Appendix~\ref{app:overhead-surgery} describes the LLV model for remote lattice surgery.
Further, we plot the LLV for the state-of-the-art injection-distillation protocol based on concatenated stabilizer entanglement distillation with an optimized code sequence~\cite{Pattison2025} (see Appendix~\ref{app:overhead} for the details of LLV evaluation for the injection-distillation protocol).
For the boosting protocol, for each $p_\text{Bell}$ and $R$, we choose the $d_\mathrm{Bell}$ and postselection criteria that minimize the LLV while keeping the final code distance at $d_s=19$. 
Thus, we observe jumps in the inverse yield for varying $R$ in the inset of Fig.~\ref{fig:boosting_llv};
for $p_\mathrm{Bell}=1\%$, the chosen $d_\mathrm{Bell}$ ranged between 9 and 13, while for $p_\mathrm{Bell}=3\%$, the range is between 13 and 17, with larger numbers favored for higher $R$; this is because the reduced volume for Bell pair buffering favors reduced retries, resulting in smaller volumes required for local operations.
For $p_\mathrm{Bell}=5\%$, $d_\mathrm{Bell}=17$ was favored throughout the range of $R$ considered.
For remote lattice surgery, the required surface-code distance to achieve the target Bell pair logical error rate is $d_s = 21, 27$ and 33, for physical Bell pair error rates of $p_\text{Bell}=1\%$, 3\%, and $5\%$, resulting in significant Bell pair consumption compared to the boosting protocol (see Appendix~\ref{app:surgery}).
For example, at $R=0.1$ and $p_\mathrm{Bell}=1\%$, the difference in LLV between entanglement boosting and remote lattice surgery is 100 (black arrow).
This arises from the order-of-magnitude improvement in yield ($1/\mathcal{Y}=86$ for entanglement boosting, while $1/\mathcal{Y}=861$ for remote lattice surgery), and the fact that the buffering volume $V_b$ scales quadratically with the number of Bell pairs required.
Entanglement boosting, therefore, achieves orders of magnitude improvements in the LLV compared to the remote lattice surgery protocol for a wide range of bell pair throughput, while both protocols operate within a single surface-code patch in each node and maintain 2D local connectivity.
In contrast, while the injection-distillation protocol~\cite{Pattison2025} achieves a higher yield, the  spacetime overhead for multi-stage distillation is large, resulting in a large LLV compared to entanglement boosting or remote lattice surgery for a wide range of the Bell pair throughput, $R>0.05$.

While we neglected the effect of idling errors in this work, as appropriate, e.g.,~for reconfigurable qubit platforms, we remark that the inclusion of idling errors in the entanglement boosting protocol (at 0.1\% per identity gate, see Sec.~\ref{app:idle-error}) increases the required attempts per kept shot by around 40\% (see Fig.~\ref{fig:idle-error}), which is likely relevant for qubit systems with short coherence times, such as superconducting qubits. 
Furthermore, additional circuit volumes are expected in the presence of finite decoding time; for example, Ref.~\cite{Gidney2024} considers nearly 10 SE cycles while waiting for decoding for superconducting qubits with a 1 microsecond cycle time and considering a 10 microsecond decoding time~\cite{Higgott2025}. 
Such decoding wait times will be negligible for qubit platforms with slower cycle times, such as neutral atoms and trapped ions.

\section{Pipelined entanglement distillation}\label{sec:distillation}

\begin{figure*}[t]
	\centering
	\includegraphics[width=0.85\linewidth]{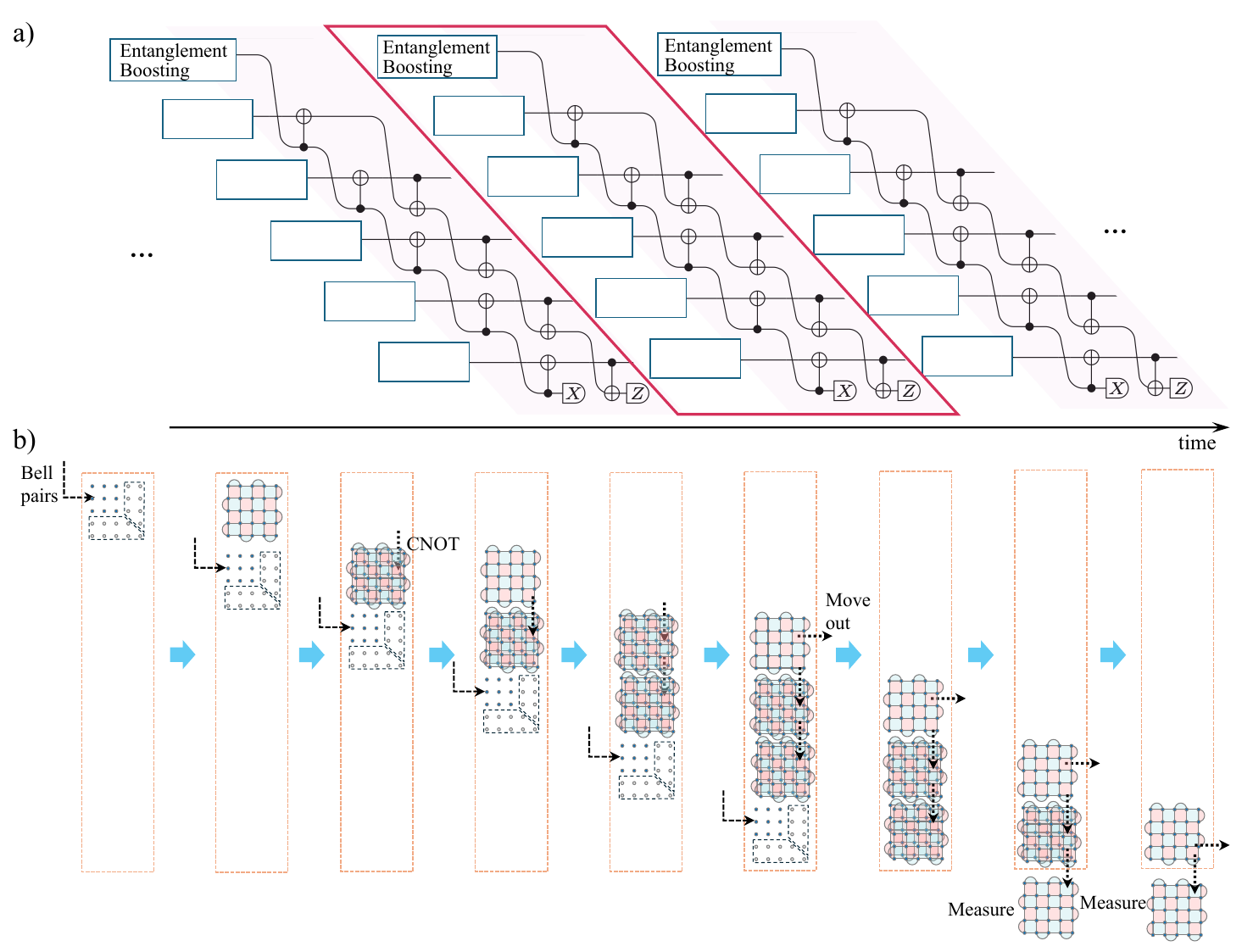}
	\caption{Reconfigurable-qubit implementation of pipelined entanglement distillation with $[[2m,2m-2,2]]$ quantum parity code.
		a) an example circuit for $[[6,4,2]]$ code, distilling 4 logical Bell pairs out of 6 logical Bell pairs.
		The curved wires in the circuit represent the logical qubit reconfigurations, which move the location of the logical qubits for transversal CNOT gates.
		For example, we assume the input to the circuit originates from the entanglement boosting protocol.
		b) More detailed illustration of qubit reconfiguration for one instance of the distillation in a) (marked by red rhombus).
		The first two input qubits are moved along the vertical direction inside the Bell pair factory (orange rectangle), interacting with other qubits. 
		Once the interactions are completed, they are moved out from the factory for logical $X$- and $Z$-basis measurements.
		Bell pair input and output can be directed along the horizontal axis, with sequential input of $n$ boosted Bell pairs during the execution of the distillation.
	}
	\label{fig:distillation}
\end{figure*}

The entanglement boosting stage can be complemented by logical-level entanglement distillation to further suppress logical error rates. 
In particular, we consider the entanglement distillation protocol based on $[[n,k,d]]$ stabilizer code $\mathcal{C}$ defined by a set of independent $n-k$ stabilizer generators $S = \{s_i\}_{i=1,...,n-k}$.
By choosing codes with a high encoding rate $k/n$, it is possible to improve the logical error rates of the Bell pairs with only a moderate reduction in yield $\mathcal{Y}$, at the cost of an increased circuit volume for local operations.
This is in contrast to the error suppression by the entanglement boosting only, where a quadratic increase in the number of Bell pairs is required to achieve stronger error suppression with increased $d_\mathrm{Bell}$.
Therefore, the combined approach is expected to provide improved LLV in the regime of small $R$ or very low target logical Bell pair error rates, complementing the entanglement boosting protocol for further scalability.

Here, we follow the protocol illustrated in Fig.~\ref{fig:distillation_protocols}b, with the $n$ Bell pair inputs encoded in rotated surface code, such as those from the boosting stage;  distillation is therefore implemented by logical gates on the rotated surface code.
Below, we describe how to construct a distillation circuit that can be implemented efficiently with reconfigurable qubit platforms by the parallel use of local one-way qubit shuttling.

In the stabilizer formalism, each stabilizer generator of $\mathcal{C}$ can be represented by a binary vector of length $2n$.  
For the first $n$ entries, the presence of $1$ at index $i$ indicates the $X$ operator on qubit $i$, and the remaining $n$ entries indicate the $Z$ operators.  
Collecting these binary vectors for all stabilizer generators gives an $(n-k)\times 2n$ matrix,
\begin{equation}
	H_q = [\,H_X \mid H_Z\,],
\end{equation}
The two $(n-k)\times n$ submatrices $H_X$ and $H_Z$ describe, respectively, the $X$ and $Z$ components of the set of stabilizer generators.
The stabilizer generators are not unique, and $H_q$ can be transformed by Gaussian elimination into a standard form following the procedure of Ref.~\cite{gottesman1997stabilizer}.
Here, with $r$ denoting the rank of $H_X$, the resulting matrix is
\begin{equation}
	H_s =
	\left[
	\begin{array}{ccc|ccc}
		I_1 & A_1 & A_2 & B & C_1 & C_2 \\
		0   & 0   & 0   & D & I_2 & E
	\end{array}
	\right],
	\label{eq:H_standard_form}
\end{equation}
where $I_1$ and $I_2$ are identity matrices of size $r\times r$ and $(n-k-r)\times (n-k-r)$, 
$B$ is $r\times r$, 
$A_1$ and $C_1$ are $r\times (n-k-r)$,  
$A_2$ and $C_2$ are $r\times k$, 
$D$ is an $(n-k-r)\times r$, 
$I_2$ is $(n-k-r)\times (n-k-r)$, 
and $E$ is $(n-k-r)\times k$.
The logical $X$ operator can be expressed in block-matrix form as
\begin{align}
	X_s &= [\,0\; E^{\mathsf T}\; I_3 \mid  V\; 0\; 0\,], \label{eq:X_logical}
\end{align}
where $I_3$ is a $k\times k$ identity matrix and $V = E^{\mathsf T}C_1^{\mathsf T}+C_2^{\mathsf T}$ modulo 2~\cite{gottesman1997stabilizer}, and ${\mathsf T}$ denotes matrix transpose.
We show the $H_s$ for $[[6,4,2]]$ quantum parity code~\cite{Gottesman1998} and $[[7,1,3]]$ Steane code~\cite{Steane1996codes} in Fig.~\ref{fig:pipeline}a-b, along with binary representations of $X$ logical operators.

\begin{figure*}[t]
	\centering
	\includegraphics[width=0.99\linewidth]{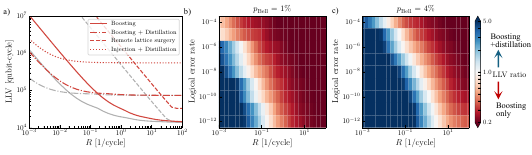}
	\caption{
    Comparison between the entanglement boosting and the combined boosting + pipelined distillation scheme, with an example of [[10, 8, 2]] quantum parity code used for the pipelined distillation. 
    a) Link-limited volume (LLV) as a function of the Bell pair generation throughput $R$, for target logical error rates of $10^{-12}$ (red) and $10^{-8}$ (gray), with $p_\mathrm{Bell}=1\%$. 
    Solid lines represent the entanglement boosting, the dash-dotted lines denote the boosting-and-pipelined distillation protocol, dashed lines indicate remote lattice surgery and dotted lines is the injection-distillation approach of Ref.~\cite{Pattison2025} (see Appendix~\ref{app:concat-llv}).
    With a higher $R$, boosting yields the smallest LLV, whereas for a lower $R$, combining with distillation becomes advantageous due to improved yield at the cost of larger local circuit volume. 
    b-c) Ratio of LLVs for the boosting and combined boosting-distillation protocols, as functions of $R$ for $p_{\mathrm{Bell}} = 1\%$ and $4\%$, respectively, identifying the crossover regime where the combined protocol outperforms boosting alone. 
    For $p_{\mathrm{Bell}} = 1\%$, the crossover lies between $R \approx 10^{-4}$ and $10^{-1}$ per cycle; for $p_{\mathrm{Bell}} = 4\%$, it extends up to $R \approx 1$ per cycle. 
    The distance-2 $[[2m,2m\!-\!2,2]]$ code used here serves as a representative toy model; higher-distance, high-rate codes, such as quantum Hamming or QLDPC codes, further enhance error suppression.
    }
	\label{fig:protocol-switching}
\end{figure*}

An encoding circuit can be synthesized from the standard form following Ref.~\cite{gottesman1997stabilizer}, and its reverse (unencoding circuit) can be used for entanglement distillation, which we describe below for CSS codes. 
First, $n$ Bell pairs are prepared in nodes $A$ and $B$, and both parties first apply gates following the first $r$ rows of $H_s$: for the $i$th row, if the $j$th column entry is 1, place a CNOT between qubits $i$ and $j$, with qubit $i$ being the control. 
For CSS codes, entries $n+1, ..., 2n$ of the $i$th row are $0$; hence, this completes the first step (blue rectangles in Fig.~\ref{fig:pipeline}a-d).
Second, following Eq.~\eqref{eq:X_logical}, CNOT gates are placed following the matrix $E^{\mathsf T}$: if $j$th entry of $i$th row is 1, then place CNOT between qubits $n-k+i$ and $j$, where qubit $j$ is the target (red rectangles in Fig.~\ref{fig:pipeline}a-d).
In general, for non-CSS codes, not only CNOT gates but CZ and controlled-$Y$ gates appear in the circuit~\cite{gottesman1997stabilizer}.
Following the application of CNOT gates, the first $r$ qubits are measured in the $X$ basis, while $n-r-k$ qubits are measured in the $Z$ basis.
These measurement results correspond to the stabilizer checks of the code $\mathcal{C}$, allowing the postselection or error correction based on the measurement outcomes.
We note that the same circuit is executed in the node $A$ and $B$ for entanglement distillation (Fig.~\ref{fig:distillation_protocols}b).
For the case of error detection, if any of the stabilizer generators are measured to be in an odd parity between nodes $A$ and $B$, the output state is discarded; if the error rate of the Bell pair is $p_\mathrm{Bell}$, and the code distance of $\mathcal{C}$ is $d$, then the error rate of the post-selected output state is $O\left(p_\mathrm{Bell}^d\right)$.

For reconfigurable qubits with efficient parallel qubit shuttling capability, such as neutral atoms and trapped ions, we propose to implement the distillation circuits illustrated in Fig.~\ref{fig:pipeline}c-d using an equivalent \emph{pipelined} implementation illustrated in Fig.~\ref{fig:pipeline}e-f. 
Here, instead of preparing $n$ qubits at the start and applying the gates following the above procedure, $n-r$ qubits are first prepared, and $r$ remaining qubits are sequentially moved across the $n-r$ qubits in a pipelined manner.
This is illustrated in the modified circuits of Fig.~\ref{fig:pipeline}e-f as the bending of the wires, corresponding to qubit reconfigurations.
Since reconfigurations occur in only one direction at each step, this is compatible with fully parallel qubit shuttling.
Reconfiguration and transversal CNOT gates for the first $r$ qubits implement the gates in the blue rectangle in Figs.~\ref{fig:pipeline}a-d, 
and $n-r-k$ qubits are then shuttled in the same direction, realizing the gates in the red rectangle, completing the required gate network.
This construction enables the concentration of the intrinsic idle volume of the circuit, as illustrated by the dashed triangles in Figs.~\ref{fig:pipeline}e and f, which can be utilized efficiently for another instance of the same or a different distillation circuit.

In Fig.~\ref{fig:distillation}, we further clarify the implementation of the pipelined distillation using $[[2m,2m-2,2]]$ quantum parity code~\cite{Gottesman1998}, which has stabilizer generators $X_1 X_2... X_{2m}$ and $Z_1 Z_2 ... Z_{2m}$.
We use $m=3$ to illustrate a small-scale example in Fig.~\ref{fig:distillation}, while larger $m$ results in a better encoding rate $k/n = (2m-2)/2m$.
Figure~\ref{fig:distillation}a shows the time-multiplexed operation of multiple instances of the distillation circuits, where the idle volumes indicated in Figs.~\ref{fig:pipeline}e-f are used for another instance of the distillation circuit.
Figure~\ref{fig:distillation}b is a more concrete qubit reconfiguration procedure, taking into account the input and output of the logical Bell pair factory (orange dotted rectangle).
Qubit shuttling along the vertical direction and transversal gates implement the logical circuit for entanglement distillation, while the Bell pair inputs (e.g., entanglement boosting) and outputs can be moved along the horizontal direction, allowing fully pipelined operations based on qubit reconfiguration.
In general, the circuit volume of this protocol is approximated by
\begin{equation}\label{eq:total-volume}
    \begin{aligned}
    &\mathcal{V} \approx k\mathcal{V}_k + r\mathcal{V}_r + (n-r-k) \mathcal{V}_{(n-r-k)}, \\ 
    &\mathcal{V}_k = (n-k) (2d_s^3-d_s),   \\
    &\mathcal{V}_{r} = (n-r) (2d_s^3-d_s) \\
    &\mathcal{V}_{(n-r-k)} = (n-1) (2d_s^3-d_s), 
    \end{aligned}
\end{equation}
for the distance-$d_s$ rotated surface code (see Appendix~\ref{app:pipeline-volume} for a more detailed description of the above model).
The fully parallelized reconfiguration of the logical patches in each circuit depth, as illustrated in Fig.~\ref{fig:distillation}b, maintains a small reconfiguration time cost between circuit layer executions.

Figure~\ref{fig:protocol-switching} quantitatively compares the LLV of the entanglement boosting and the combined approach of entanglement boosting and the pipelined distillation with the $[[10,8,2]]$ quantum parity code. 
Here, we utilized a larger code than shown in Fig.~\ref{fig:distillation} to achieve a better encoding rate, which improves the yield at the cost of slightly weaker error suppression in entanglement distillation (see Appendix~\ref{app:parity-sim}).
Figure~\ref{fig:protocol-switching}a shows the LLV as a function of the Bell pair generation throughput~$R$, for output logical error rates of $10^{-12}$ (red) and $10^{-8}$ (gray), where the logical error rates of the entanglement boosting are obtained from the scaling reported in Fig.~\ref{fig:boosting-performance}.
We also show the LLV for the remote lattice surgery protocol~\cite{Ramette2024,Shalby2025} and a state-of-the-art injection-distillation protocol based on concatenated distillation~\cite{Pattison2025} (see Appendix~\ref{app:overhead} for the details of LLV evaluation for these protocols), both of which have an order-of-magnitude larger LLV than the boosting or boosting+distillation protocols for a wide range of $R$.
With the pipelined distillation operation using the high-rate code, the inverse yield $1/\mathcal{Y}$ is smaller than that of the boosting-only protocol, while the local operation volume is larger.
These distinct features lead to the crossover of the LLV as a function of the Bell pair generation throughput~$R$: for high $R$, the boosting-only approach (solid line) is efficient, while for lower $R$, it is cheaper to increase the local circuit volume and increase the yield by using a combination of boosting and pipelined distillation (dash-dotted lines).
The crossover point is also dependent on the output logical error rate, with a lower target error rate resulting in larger crossover throughput $R$, since the required number of physical Bell pairs is larger.
The remote lattice surgery protocol has a significantly larger LLV due to its large physical Bell pair consumption (dashed lines).

In Fig.~\ref{fig:protocol-switching}b-c, we show the ratio of LLVs for the boosting-only protocol and the combined boosting-and-distillation protocol with $[[10,8,2]]$ code used for the pipelined distillation in order to identify the crossover point as a function of varying output logical error rates.
For $p_{\mathrm{Bell}} = 1\%$ and a target logical error rate of $10^{-12}$, the crossover is at $R \approx 10^{-1}$ per cycle, while for $p_{\mathrm{Bell}} = 4\%$, the crossover increases to $R \approx 1$ per cycle, suggesting that the combined protocol is more efficient for a wider range of parameters where the physical Bell pair error rates are higher.
With a fast quantum interconnect of $R \gtrsim 1$ per cycle, the boosting protocol is favored for the range of target error rates shown.

It should be emphasized that the distance-2 code used here serves only as a simple example to demonstrate the pipelined distillation and its general behavior. 
In realistic architectures, higher-distance, high-rate codes, such as the quantum Hamming codes~\cite{Steane1996,Yamasaki2024}, high-rate quantum low-density parity-check (QLDPC) codes~\cite{Tillich2014,Leverrier2015}, and quantum Bose-Chaudhuri-Hocquenghem (BCH) codes~\cite{Steane1996,Grassl1997}, would potentially provide even more scalable implementations with stronger error suppression.
The concatenation of pipelined distillation protocols also supports a scalable approach, with automated code sequence optimization recently demonstrated for reducing inverse yield and memory footprint~\cite{Pattison2025}.
While the full optimization of the combination of entanglement boosting with pipelined entanglement distillation across a range of code choices, protocol variants, and more realistic error models is left for future investigation, we show in Fig.~\ref{fig:hamming_code_combined} a comparison of the combined boosting-and-distillation protocol with pipelined entanglement distillation implemented using quantum parity codes and quantum Hamming codes.

\section{Conclusion}\label{sec:conclusion}

In this work, we have proposed entanglement boosting to efficiently transform noisy physical Bell pairs into logical Bell pairs encoded in rotated surface codes.
With all operations kept within a rotated surface code patch, this approach achieves a substantial reduction of the logical error rates by using variable $d_\mathrm{Bell}^2$ physical Bell pair inputs, thus achieving low spacetime volume to prepare logical Bell pairs.
This protocol can be complemented by a pipelined implementation of stabilizer entanglement distillation executed with logical gates of the surface code, designed for a reconfigurable qubit platform with parallel qubit shuttling.
The existing protocols for logical Bell pair preparation are constrained by two opposing characteristics: achieving small local circuit volume demands substantial physical Bell-pair consumption, while injection-and-distillation protocols incur large circuit volume.
By contrast, the combined approach presented here resolves these limitations by providing a flexible interpolation between the two regimes.

As an outlook, further performance improvement is expected through the use of soft information regarding physical Bell pair errors, readily accessible in experiments via the photon detection times~\cite{Li2024,kikura2025taming} and erasure information~\cite{Wu2022erasure,kikura2025passive}.
In the presence of finite error biases on the Bell pairs, for example, with phase errors dominating over bit-flip errors~\cite{Li2024,kikura2025taming}, better performance is expected from the use of Clifford-deformed surface codes, such as the XZZX surface  code~\cite{bonilla2021xzzx} and the XY surface code~\cite{Tuckett2018}.
Practically, the finite coherence time of physical qubits may result in error accumulation during buffering, with larger $p_\text{Bell}$ for the Bell pairs created earlier than those created near the end of the buffering;
this reinforces the benefit of a fast interconnect with high $R$ and a protocol that achieves both high yield and low circuit volume to maintain low overhead for logical Bell pair generation.
Furthermore, such an error model presents an interesting opportunity for optimizing the protocol to tolerate a non-uniform error model, for example, by optimizing the placement of Bell pairs with heterogeneous error properties~\cite{Tiurev2023correctingnon, khosravani2026heterogeneous}.
Correlated decoding methods for the transversal-gate FTQC protocols~\cite{zhou2025nature, sunami2025transversal}, with $O(1)$ syndrome extraction cycles between logical gates, further enhance the efficiency of pipelined entanglement distillation.
Finally, the probabilistic nature of the postselected logical Bell pair preparation protocol proposed here calls for efficient runtime resource state management, similarly to the magic state resource management~\cite{hirano2024magicpool}.

\section*{Acknowledgements}
We acknowledge discussions with Nicholas Fazio and Mert G{\"o}kduman, and we thank Akihisa Goban and Shiro Tamiya for discussions and their detailed feedback on the manuscript.

\section*{Code availability}
Code used for the numerical simulation in this work is available at \url{https://github.com/nano-qt/entanglement-boosting}.

\section*{Declaration of Competing Interests}
S. Sunami and H. Yamasaki are employees, and Y. Hirano and T. Hinokuma are interns of Nanofiber Quantum Technologies, Inc.

\appendix

\setcounter{table}{0}
\setcounter{figure}{0}
\renewcommand{\thetable}{A\arabic{table}}
\renewcommand{\thefigure}{A\arabic{figure}}
\renewcommand{\theHtable}{Supplement.\thetable}
\renewcommand{\theHfigure}{Supplement.\thefigure}

\newpage
\section*{Appendices}

Appendices are organized as follows. 
In Appendix~\ref{app:simulation}, we summarize our notation and provide a detailed account of the numerical simulations presented in the main text.
In Appendix~\ref{app:parity-sim}, we present the details of the performance scaling of the entanglement distillation with quantum parity codes.
Appendix~\ref{app:overhead} describes the detailed models for the LLV we used for the results in the main text.

\section{Numerical simulations}\label{app:simulation}

Our terminology is summarized in Table~\ref{tab:notation}.

\begin{table}[b]
	\centering
	\begin{tabular}{c|c}
		\hline
		Description & symbol  \\
		\hline
		Bell pair generation throughput & $R$ \\
		Bell pair error rate & $p_\text{Bell}$ \\
		\hline 
		noise strength for physical operations (see Table~\ref{tab:local-noise-model}) & $p$ \\
		acceptance rate & $q_0$ \\
		code distance for Bell pair projection & $d_\text{Bell}$ \\
		code distance of full surface code after expansion & $d_s$ \\
		spacetime volume of single trial & $\mathcal{V}_{0}$ \\
		\hline
	\end{tabular}
	\caption{Notation for the entanglement boosting protocol used in this paper.}
	\label{tab:notation}
\end{table}

\begin{table}[t]
\centering
\begin{tabular}{l|l}
 \hline
 physical operations   & noise  \\
 \hline
 reset ($X$)          & $Z$ error with probability $p$ \\
 reset ($Z$)          & $X$ error with probability $p$ \\
 measurement ($X$)    & $Z$ error with probability $p$ \\
 measurement ($Z$)    & $X$ error with probability $p$ \\
 single-qubit gates & $X$, $Y$ or $Z$ error with with probability $p/3$ each \\
 two-qubit gates    & two-qubit Pauli errors except $I\otimes I$,\\ 
 & with probability $p/15$ each\\
 \hline
\end{tabular}
\caption{Noise model of local operations for circuit-level simulation of the entanglement boosting stage. Noise model for the physical Bell pairs is given in Sec.~\ref{sec:noise_model}.}
\label{tab:local-noise-model}
\end{table}

\subsection{Circuit-level simulation of entanglement boosting }\label{app:numerics}

\begin{figure}[b]
  \centering
  \includegraphics[width=0.99\linewidth]{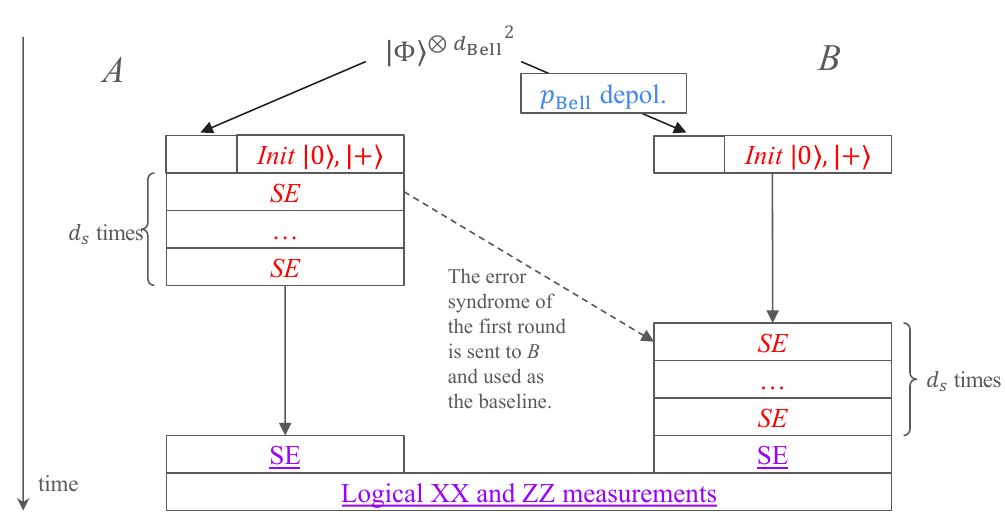}
  \caption{
    Numerical simulation of the boosting stage described in Sec.~\ref{sec:boosting}.
    Steps in red are subject to the noise model shown in Table~\ref{tab:local-noise-model}, while the steps in purple (with underlines) are noiseless.
    We initially prepare $d_\text{Bell}^2$ Bell pairs and subject one of the arms of the pairs to a single-qubit depolarizing channel.
    The remaining $d^2-d_\text{Bell}^2$ physical qubits are prepared with noise strength 0.1\% in each party.
    We simulate the SE steps of two parties in series for implementation purposes, which is equivalent to simultaneous SE operations followed by classical communication to exchange the syndrome information in an actual operation.
  }
  \label{fig:projection-stim-process}
\end{figure}

We perform a Monte Carlo sampling simulation to evaluate the performance of the entanglement boosting protocol in Sec.~\ref{sec:boosting} using Stim~\cite{gidney2021stim} and PyMatching~\cite{Higgott2025}.
We adopt the following noise model for the simulation: for the input Bell pairs, we first prepare ideal Bell pairs and subject one of the arms of the pairs to a single-qubit depolarizing channel with parameter $p_\text{Bell}$.
Other local operations experience noise as described in Table~\ref{tab:local-noise-model} with a noise strength of 0.1\%, while we do not consider idle errors; this is justified by the fact that idling error rates are small in atomic qubits such as neutral atoms~\cite{Bluvstein2022}.

Figure \ref{fig:projection-stim-process} depicts the implementation of the protocol.
We have two surface code patches of distance $d_s$, one for $A$ and one for $B$.
Each party prepares its surface code patch using the endpoints of $d_\text{Bell}^2$ physical Bell pairs, and the remaining qubits are separable, prepared following the layout in Fig.~\ref{fig:distillation}.

Subsequently, $A$ and $B$ each perform $d_s$ cycles of syndrome extraction using local physical operations.
While $A$ performs syndrome extraction independently of $B$, in our simulation, $B$ uses $A$'s first-cycle error syndrome as the baseline error syndrome of its code.
This corresponds to computing error syndrome parities, as described in Sec.~\ref{sec:boosting}, which is equivalent to communicating the readout results after the syndrome extraction.

After $d_s$ cycles of syndrome extraction for each party, we perform one cycle of \textit{noise-free} syndrome extraction, followed by noise-free logical $\overline{XX}$ and $\overline{ZZ}$ measurements.
The syndrome values are used to compute the complementary gap~\cite{Gidney2025}, described in more detail in Appendix~\ref{sec:complementary-gap}, while the final measurements provide the reference logical error to be compared with the decoding result.

Following the decoding, a sample is labeled as discarded if its complementary gap is below a certain threshold.
A sample is ``valid'' if it is not discarded and the outcomes of the $\overline{XX}$ and $\overline{ZZ}$ measurements match the expected values.
A sample is ``wrong'' if it is not discarded and the outcomes of the $\overline{XX}$ and $\overline{ZZ}$ measurements deviate from the expected values.
The logical error rate is $\frac{\#\text{wrong}}{\#\text{valid} + \#\text{wrong}}$, and the acceptance rate is $\frac{\#\text{valid} + \#\text{wrong}}{\#\text{discarded} + \#\text{valid} + \#\text{wrong}}$, which are reported in the main text.

\subsection{Complementary gap}\label{sec:complementary-gap}
\begin{figure}[t]
    \centering\includegraphics[width=0.99\linewidth]{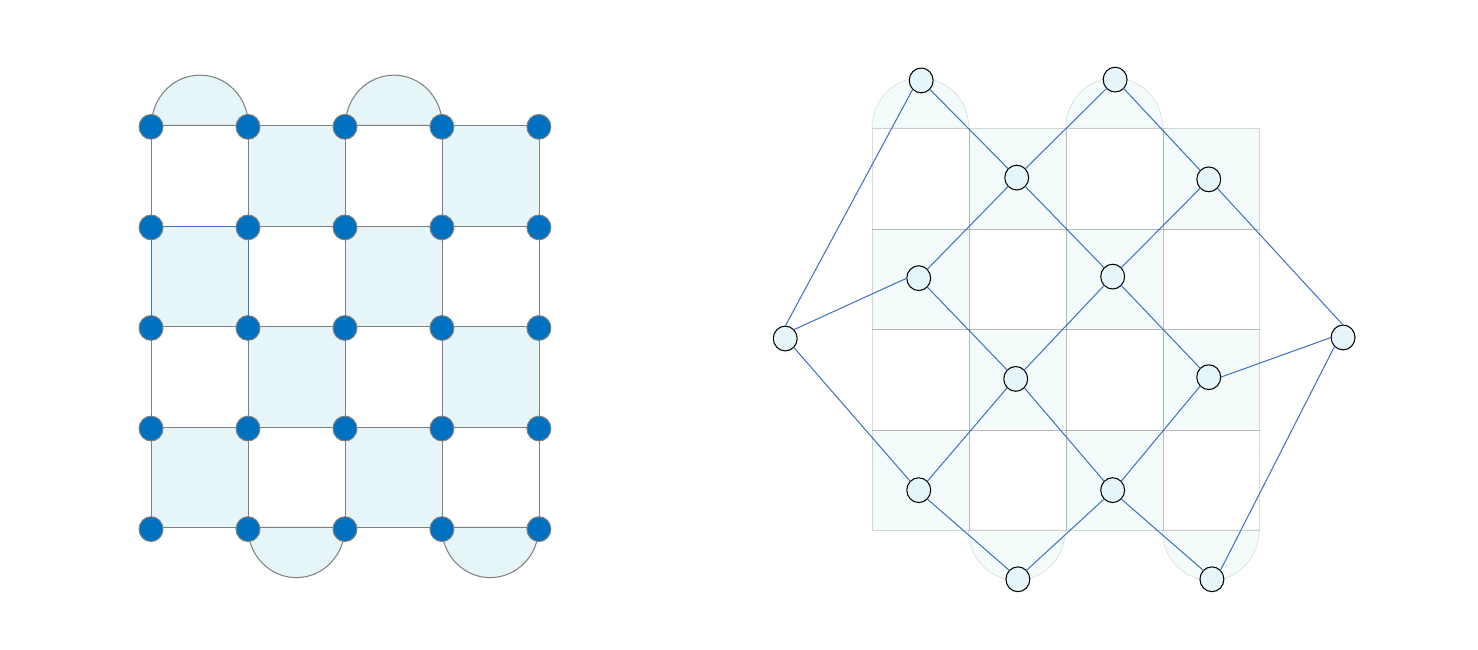}
    \caption{
    Left: $X$ stabilizer checks on the distance-5 rotated surface code.
    Each filled circle represents $d_s^2$ qubits comprising the code, and each light-blue square plaquette represents an $X$ stabilizer check.
    Right: the corresponding error graph.
    Each vertex (open circles), except for the leftmost and rightmost, corresponds to an $X$ stabilizer check, and each edge corresponds to a data qubit shared by two stabilizer checks.
    The leftmost (rightmost) vertex is a virtual vertex referred to as the left (right) boundary node.
    }
    \label{fig:surface-error-graph}
\end{figure}

The complementary gap is a value that represents the decoder's confidence in its decoding results.
As a simple example, we consider the distance-5 rotated surface code (Fig.~\ref{fig:surface-error-graph}).
For simplicity, in this subsection, we focus on $X$ stabilizer checks that detect $Z$ errors.
We additionally assume that syndrome extraction is noise-free.

\begin{figure}[t]
    \centering
    \includegraphics[width=0.99\linewidth]{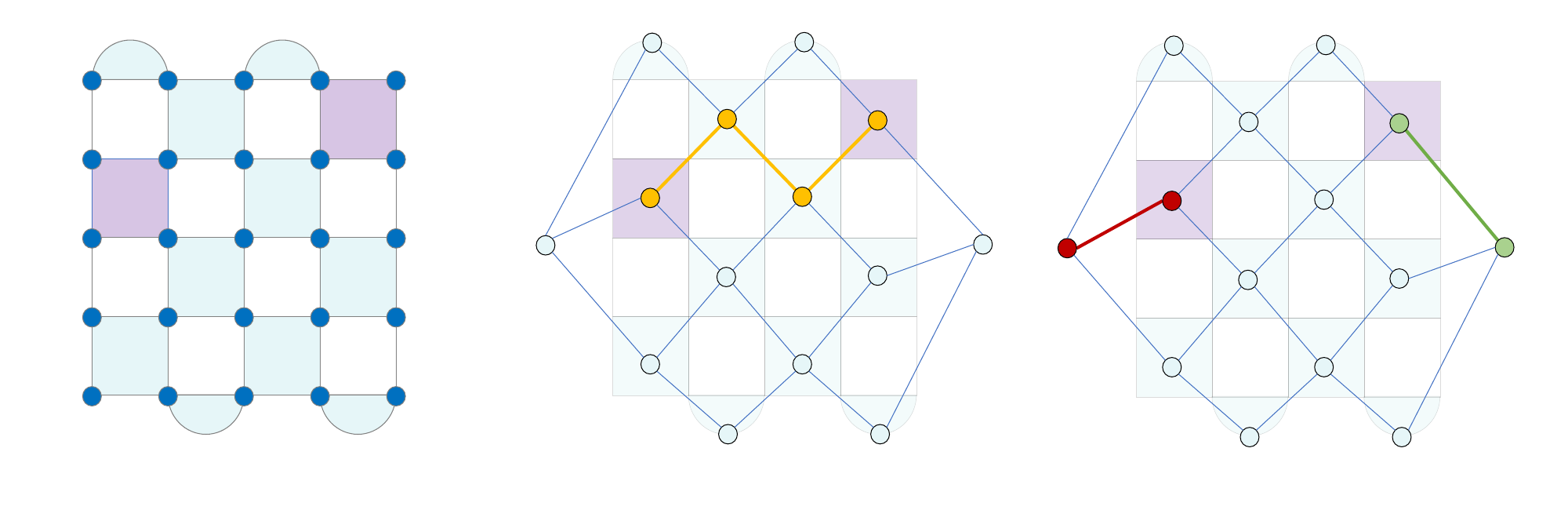}
    \caption{
    Left: $X$ stabilizer checks on the distance-5 rotated surface code. $Z$ errors are detected by the purple checks.
    Middle: a solution of the MWPM problem conditioned with the left boundary check off.
    Right: a solution of the MWPM problem conditioned with the left boundary check on.
    Matching paths are shown in different colors.
    }
    \label{fig:complementary-gap}
\end{figure}

Figure \ref{fig:surface-error-graph} (left) depicts the $X$ stabilizer checks on the distance-5 rotated surface code.
The minimum-weight perfect matching (MWPM) decoder~\cite{Fowler2012} performs decoding by solving a matching problem on a graph known as the \textit{error graph}.
The right panel of Fig.~\ref{fig:surface-error-graph} shows the corresponding error graph.
Each vertex corresponds to an $X$ stabilizer check, and each edge connecting two vertices represents the data qubit shared by those stabilizer checks.
There are two virtual vertices, i.e, the left and right boundary nodes, to handle data qubits checked by only one stabilizer check.
Each edge is associated with a weight, $w_i = -\log p_i$, where $p_i$ denotes the $Z$-error probability of qubit $i$.
The MWPM decoder finds the minimum-weight matching on this error graph.

The complementary gap, defined as the absolute difference between the minimum weights conditioned on the complementary logical outcomes, is illustrated in Fig.~\ref{fig:complementary-gap}.
The complementary gap is computed by running the MWPM decoder with the left boundary node forced on and off, and taking the absolute difference between the two resulting weights.
If the complementary gap is small, the decoder's confidence in its decision is low.
In the left panel, purple checks detect $Z$ errors.
Two matchings conditioned on the complementary logical outcomes are illustrated in the middle and right panels.
Both of these are valid interpretations of the error syndrome, differing by a logical $Z$ chain.
In Fig.~\ref{fig:boosting}, we use this value for postselection; that is, if the complementary gap is smaller than the chosen threshold, the distillation attempt is rejected.

\subsection{Effect of idling errors}\label{app:idle-error}
\begin{figure}[t]
    \centering
    \includegraphics[width=0.99\linewidth]{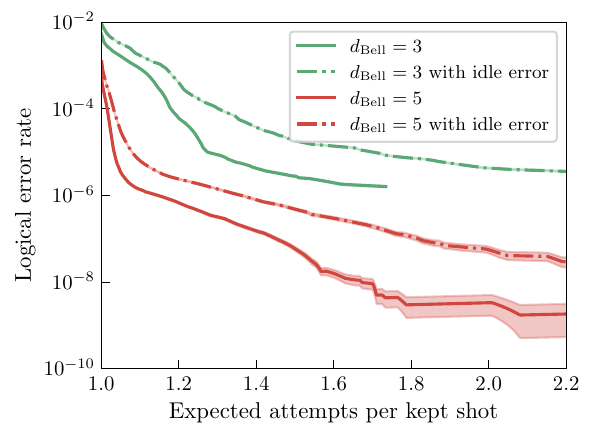}
    \caption{Effect of idling errors on the logical error rates of logical Bell pairs generated by entanglement boosting.
    Here, we set $p_\mathrm{Bell}=1\%$ and $d_s=19$, and physical operation errors shown in Table~\ref{tab:local-noise-model}.
    }
    \label{fig:idle-error}
\end{figure}

We have neglected the effect of idling errors throughout this work, which is justifiable for reconfigurable qubits such as neutral atoms and trapped ions with coherence times of orders of magnitude longer than typical gate times (for example, for neutral atoms, the coherence times are on the order of seconds while the Rydberg gate times are on the order of 100 ns).
In Fig.~\ref{fig:idle-error}, we plot the logical error rates of Bell pairs from the entanglement boosting protocol with and without idling error at $p=$0.1\%. 
Here, we use $p_\mathrm{Bell}=1$\% and $d_s=19$, the same configuration as Fig.~\ref{fig:boosting-performance}a.
In the presence of idle errors, the required number of attempts increases by up to 40\% to reach the same logical error rates, for the range of parameters shown in this plot.

\subsection{Additional circuit-level simulation data and the scaling of logical error rate}\label{app:scaling}
\begin{figure*}[t]
	\centering
	\includegraphics[width=0.99\linewidth]{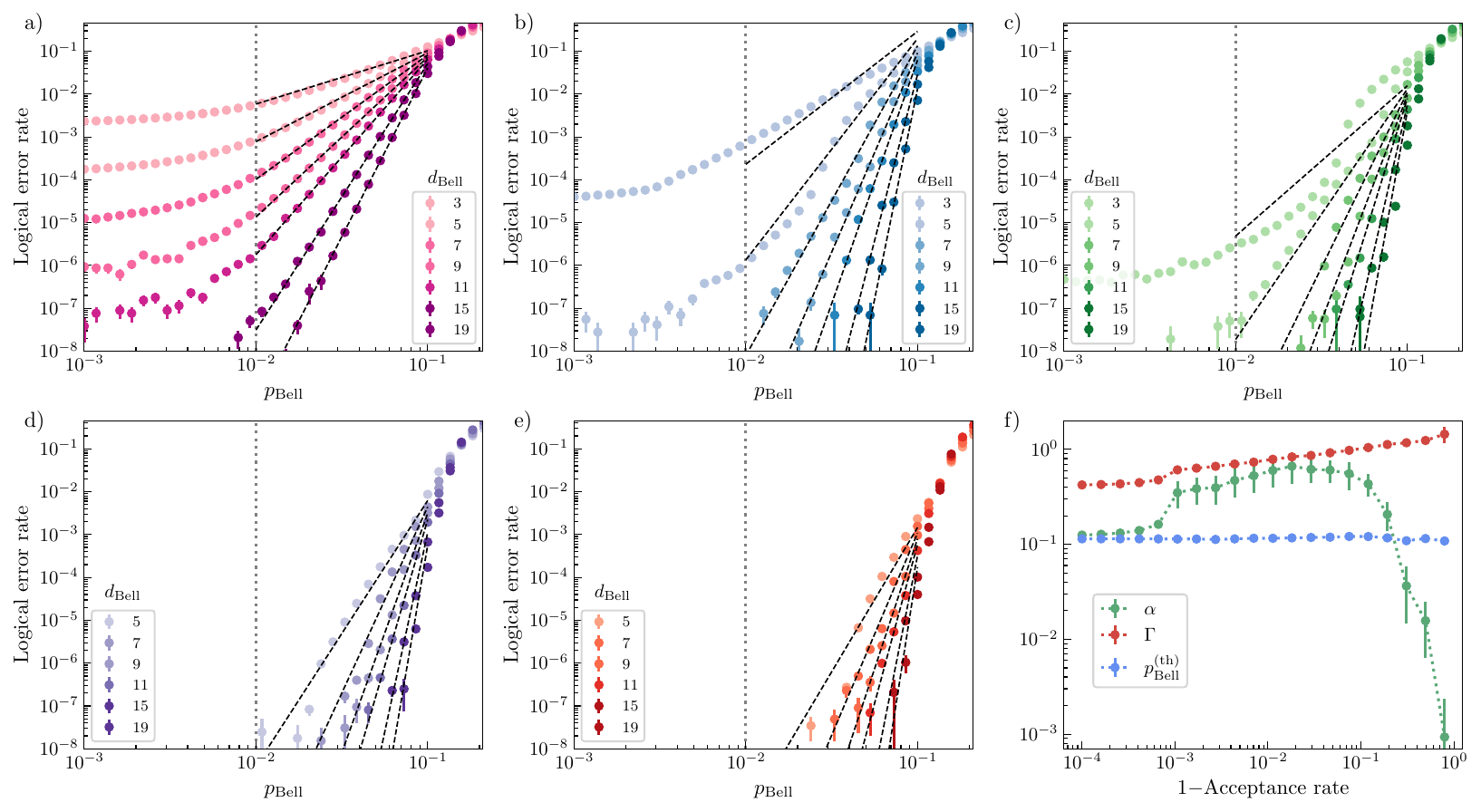}
	\caption{Additional plots showing the scaling of the logical error rate for logical Bell pairs produced with the entanglement boosting protocol.
    a-e) logical error rate as a function of the physical Bell-pair error rate~$p_{\mathrm{Bell}}$ for different~$d_{\mathrm{Bell}}$, with discard probabilities of 0\% (a), 10\% (b), 35\% (c), 50\% (d) and 70 \% (e), based on the complementary gap. 
    Each point is obtained from circuit-level simulations of the entanglement boosting protocol with soft-output decoding and postselection based on varying complementary-gap thresholds that result in different discard probabilities. 
    The dashed line represents the fitting with Eq.~\eqref{eq:scaling} in the range $10^{-2} < p_\mathrm{Bell} < 9\times10^{-2}$.
    In panels d and e, the statistics of the $d_\mathrm{Bell}=3$ simulation results are insufficient to obtain cases with complementary-gap-based discard rates reaching 50\% or above; thus, no points are shown for $d_\mathrm{Bell}=3$.
    f) the dependence of the fitted values of parameters $\alpha$, $\Gamma$ and $p_\mathrm{Bell}^\mathrm{(th)}$ on the discard probabilities of entanglement boosting.}
	\label{fig:scaling-supp}
\end{figure*}

Here, we show additional data from the circuit-level simulation of the entanglement boosting protocol. 
In Fig.~\ref{fig:scaling-supp}, we show the logical error rates for varying $d_\mathrm{Bell}$, $p_\mathrm{Bell}$, and discard rates 0, 10, 35, 50, and 70\%.
With small $d_\mathrm{Bell}$ and $p_\mathrm{Bell}$, we observe no gap values that can be used to discard large fractions, such as $d_\mathrm{Bell}=3$ with discard rates of 50\% and 70\%, based on the $10^{8}$ sampling that we performed.
Therefore, we have not shown data with $d_\mathrm{Bell}=3$ in Fig.~\ref{fig:scaling-supp}d-e.
A notable feature of Fig.~\ref{fig:scaling-supp}a-c is the saturation of the error rates for $p_\mathrm{Bell} < 1\%$ (vertical dashed line); as such, we perform the fits with Eq.~\eqref{eq:scaling} only for $1\% < p_\mathrm{Bell} < 8\%$.

\subsection{Remote lattice surgery protocol for rotated surface code}\label{app:surgery}

We perform numerical simulations of the remote lattice surgery by adopting the protocol of Ref.~\cite{Ramette2024} for a rotated surface code, following the syndrome extraction schedule proposed in Refs.~\cite{Shalby2025} and using the same error model as that used for the simulation of entanglement boosting, summarized in Table~\ref{tab:local-noise-model}.
This allows a fair LLV comparison between the two protocols, as shown in Fig.~\ref{fig:boosting_llv}.

\begin{figure}[t]
    \centering
    \includegraphics[width=0.99\linewidth]{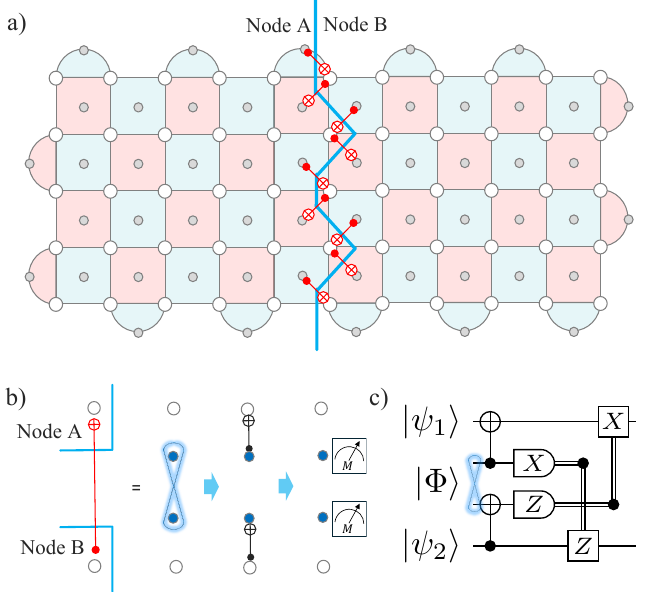}
    \caption{Remote lattice surgery with rotated surface code.
    	a) The zig-zag interface for remote lattice surgery of a merged patch~\cite{Shalby2025}, where the blue line represents the interface between the two nodes and red CNOTs represent the teleported CNOT gates implemented by Bell pairs.
    	b-c) Gate teleportation by physical Bell pairs. Remote Bell pair (blue circles) interacts with qubits in each node (white circles) via CNOT gates, before the Bell pair is measured for feedforward Pauli gates, as shown in c).
    }
    \label{fig:surgery_layout}
\end{figure}

\begin{figure}[b]
	\centering
	\includegraphics[width=0.85\linewidth]{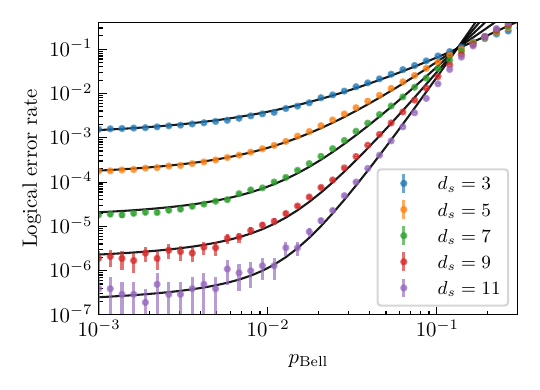}
	\caption{Circuit-level simulation results of the logical error rate of logical Bell pairs generated by remote lattice surgery protocol, as a function of the physical Bell pair error rates $p_\mathrm{Bell}$, with noise model of Table~\ref{tab:local-noise-model}.
	Solid curves are the fits to the data below the threshold with Eq.~\eqref{eq:fit-func}.
    Error bars represent the standard error of estimated probability from the $10^7$ sampling results.
	}
	\label{fig:surgery_numerics}
\end{figure}

More concretely, we simulate logical Bell pair generation via the splitting of a merged patch~\cite{Horsman2012}, where the inter-patch remote physical CNOT gates are implemented by gate teleportation, and the resource state is the input physical Bell pairs with error rates $p_\text{Bell}$~\cite{Ramette2024,Shalby2025}.
Local operations follow the error model of Table~\ref{tab:local-noise-model}, and the boundary condition follows the `zig-zag interface' of Ref.~\cite{Shalby2025}, illustrated in Fig.~\ref{fig:surgery_layout}, which avoids the hook errors arising in other configurations~\cite{Shalby2025}.
For a simulation with a distance-$d_s$ rotated surface code, we prepare a merged code patch with an $X$ boundary of distance $d_s$ and a $Z$ boundary of distance $2d_s+1$, prepared in $\ket{0}^{\otimes(2n+d_s)}$.
Next, we perform $d_s$ cycles of syndrome extraction on the merged patch, followed by the split operation, which measures the linking region in the $Z$ basis.
Finally, we perform a noiseless syndrome extraction cycle on split patches, followed by noiseless $\overline{ZZ}$ and $\overline{XX}$ measurements, obtaining the logical error rate.
The simulation is implemented with stim~\cite{gidney2021stim}, and decoding is performed by PyMatching~\cite{Higgott2025}.
The resulting error scaling is plotted in Fig.~\ref{fig:surgery_numerics}, showing the remote Bell pair error threshold of 15.3\% for the case of $p=0.1\%$ noise strength for local physical operations.
To analyze the required distance to achieve a certain logical error rate, we obtain the thresholds for both local and remote error rates, following the analysis in Ref.~\cite{Ramette2024}, and use the scaling, 
\begin{widetext}
\begin{align}\label{eq:fit-func}
    p_{\text{out}} = \kappa (d_s+1)^{\eta} \left(
    \left(\frac{p_{\text{Bell}}}{p_{\text{th}}^{\text{Bell}}}\right)^{(d_s+1)/2} + 
    \left(\frac{p}{p_{\text{th}}^{\text{local}}}\right)^{(d_s+1)/2} +
    \sum_{\gamma_S = 1}^{d_s}
    \left[
        \frac{p_{\text{Bell}}}{p_{\text{th}}^{\text{Bell}}}
        \left( 1 + \alpha_c p 
        \frac{p_{\text{th}}^{\text{Bell}}}{1 - \sqrt{p/p_{\text{th}}^{\text{local}}}} \right)^2 \
    \right]^{\gamma_S/2} \left[ \frac{p}{p_{\text{th}}^{\text{local}}} \right]^{\frac{d_s+1 - \gamma_S}{2}} \right)
\end{align}
\end{widetext}
where $\kappa (d_s+1)^{\eta}$ is the approximation of $\text{poly}(L)$ shown in Ref.~\cite{Ramette2024}, and for odd $d_s$, the threshold values are $p_\text{th}^\text{local} = 0.0102$ and $p_\text{th}^\text{Bell} = 0.153$ and $p_\text{th}^\text{Bell} = 0.198$, obtained from separate simulations.
The fitted values are $\kappa = 5.44 \times 10^{-2}$, $\eta = 5.34 \times 10^{-1}$ and $\alpha_c = 3.15 \times 10^{2}$ obtained from fitting shown in Fig.~\ref{fig:surgery_numerics}.
We remark that remote syndrome extraction based on Bell measurement~\cite{haug2025latticesurgerybellmeasurements} can reduce the required number of physical Bell pairs by a factor of 2, at the cost of reduced effective distance.

\section{Entanglement distillation with quantum parity codes and quantum Hamming codes}\label{app:parity-sim}

Here, we perform simplified evaluations of the scaling of entanglement distillation with quantum parity codes $\mathcal{C}_m$, which we use for the combined boosting+distillation protocol in Fig.~\ref{fig:protocol-switching}.
For distillation based on the $[[2m,2m-2,2]]$ quantum parity code, we first prepare $2m$ ideal Bell pairs and subject one of the arms of the pairs individually to single-qubit Pauli errors with probability $p_\mathrm{in}$, where we apply Pauli $X$, $Y$ or $Z$ with probability $p_\mathrm{in}/3$ each, and assign the $2m$ endpoints to $A$ and the $2m$ remaining endpoints of the pairs to $B$.
Both $A$ and $B$ run their respective input states through an ideal distillation circuit locally and perform measurements in the $X$ and $Z$ bases. 
Conditioned on the measurement patterns being the same for $A$ and $B$, we obtain $2m-2$ output pairs, and the probability of such measurement patterns provides the success probability of the distillation protocol.
To obtain the output error rates, we track the propagation of Bell pair Pauli errors through the distillation circuit using the tableau simulator of stim~\cite{gidney2021stim} and identify the leading-order error probabilities of the postselected output Bell pairs.
The error rates of the postselected output states depend on the code size $2m$, and we fit the output error rate as a function of $m$ and $p_\text{in}$ with a polynomial model, which yields 
\begin{equation}\label{eq:parity-code-distillation}
    p_\text{out} = 0.69 m^{1.36}p_\text{in}^2.
\end{equation}
This is used for the evaluation of the combined boosting+distillation protocol, shown in Fig.~\ref{fig:protocol-switching}.
Furthermore, as a representative example of a distance-3 code, we similarly analyze the entanglement distillation with $[[2^\ell-1, 2^\ell-2\ell-1, 3]]$ quantum Hamming codes. 
We fit the output error rate as a function of $\ell$ and $p_\mathrm{in}$ with an exponential model, which yields
\begin{equation}\label{eq:parity-code-distillation}
    p_\text{out} = 0.84 e^{1.07\ell} p_\text{in}^3.
\end{equation}

\section{Volume calculations}\label{app:overhead}

In this section, we describe the details of LLV used for Fig.~\ref{fig:boosting_llv}, and Eq.~\eqref{eq:total-volume} used for Fig.~\ref{fig:protocol-switching}.

\subsection{Remote lattice surgery}\label{app:overhead-surgery}
For the lattice-surgery-based protocol described in Appendix~\ref{app:surgery}, $2d_s-1$ Bell pairs are consumed in each SE cycle, repeated for $d_s$ cycles, with a total of $d_s(2d_s-1)$ physical Bell pairs consumed.
If the physical Bell pair generation throughput cannot keep up with the consumption speed, i.e.,~$R < (2d_s-1)$, then it is necessary to accumulate Bell pairs before initiating this protocol; for this, we only need to prepare $N = \max \left[ d_s(2d_s-1) - R d_s, 0 \right]$, accounting for the number of Bell pairs generated during the SE cycles. 
Therefore, total LLV is
\begin{equation}
    \mathcal{V}_\mathrm{surgery} = \frac{N^2}{R}  + d_s\times \left[ (2d_s^2-1)+\frac{2d_s-1}{2}\right],
\end{equation}
where the first term represents the spacetime volume for Bell pair accumulation, the second represents the consumption of the accumulated pairs over $d_s$ cycles, and the third is the volume for local operations; the second term in the square brackets is the additional boundary qubits for lattice surgery, with a factor of 1/2 representing the boundary qubit cost $2d_s-1$ split over the two parties involved.

\subsection{Entanglement boosting}
As discussed in Sec.~\ref{sec:boosting}, the boosting stage proceeds by first preparing $d_\text{Bell}^2$ physical Bell pairs and $d_s^2-d_\text{Bell}^2$ qubits, followed by $d_s$ cycles of SEs.
To initiate this protocol, we first wait for $d_\text{Bell}^2/R$ cycles, during which $d_\text{Bell}^2$ physical Bell pairs are accumulated (Fig.~\ref{fig:llv}). 
The single-trial LLV is hence
\begin{equation}\label{eq:overhead-boosting}
    \mathcal{V}_0 = \frac{d_\text{Bell}^4}{R} + d_s \times (2d_s^2-1).
\end{equation}
For the acceptance rate of $q_0$, the resulting LLV is $\mathcal{V}_\mathrm{boosting} = \mathcal{V}_0 / q_0$.

\subsection{Pipelined entanglement distillation}\label{app:pipeline-volume}

The general expression in Eq.~\eqref{eq:total-volume} for the pipelined entanglement distillation with qubit reconfigurations is derived as follows.
First, $\mathcal{V}_k$ corresponds to the $k$ output patches that remain stationary throughout the circuit in Fig.~\ref{fig:pipeline}e–f. 
Each of these patches interacts with up to $(n-k)$ incoming qubits through transversal CNOT layers before being moved out from the factory, and the associated spacetime volume is obtained by multiplying this layer count by the volume of $d_s$ cycles of syndrome extraction (SE) for a single patch.
Next, $\mathcal{V}_r$ accounts for the $r$ patches that are sequentially reconfigured across the other $(n-r)$ patches, as illustrated in Fig.~\ref{fig:pipeline}e–f. 
Each moving patch performs up to $(n-r)$ transversal CNOT layers during its traversal, again followed by $d_s$ SE cycles per layer, yielding the contribution $\mathcal{V}_r$.
Finally, the remaining $(n-r-k)$ patches interact with both the traversing $r$ patches and the $k$ output patches. 
During the first stage, they undergo up to $r$ transversal CNOT gates with the moving patches, followed by additional layers with the rest of $(n-r-k)$ patches and output patches, up to $(n-r-k-1)$ and $k$ CNOT layers, respectively, as depicted in Fig.~\ref{fig:pipeline}c–f. 
Summing these contributions gives the total spacetime volume in Eq.~\eqref{eq:total-volume}.

\begin{figure}[t]
    \centering
    \includegraphics[width=0.9\linewidth]{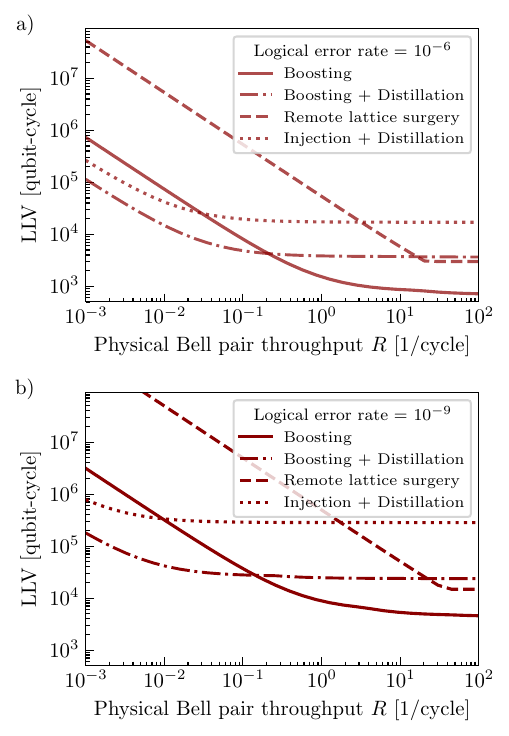}
    \caption{
    Additional plots for the LLV as a function of physical Bell pair throughput $R$, for target logical error rates of $10^{-6}$ with $d_{\text{s}}=7$ (a) and $10^{-9}$ with $d_{\text{s}}=13$ (b), with the same error model as for Fig.~\ref{fig:protocol-switching}a, i.e. $p_\mathrm{Bell}=1\%$.
    Solid lines represent the entanglement boosting, the dash-dotted lines denote the boosting-and-pipelined distillation protocol with $[[10,8,2]]$ quantum parity code, dashed lines indicate remote lattice surgery, and doted lines are the injection-distillation approach using optimized sequence of $[2,1,2]_X, [2,1,2]_Y$, and $[[8,6,2]]$ for (a), and $[2,1,2]_X, [2,1,2]_Y,$ and $[[8,3,3]]$ for (b). These optimized sequences are obtained using the open-source code accompanying Ref.~\cite{Pattison2025} with buffer space of 10 logical qubits (see Appendix.~\ref{app:additional-data}). 
    }
\label{fig:llv_additional}
\end{figure}

\begin{figure}[t]
	\centering
	\includegraphics[width=0.99\linewidth]{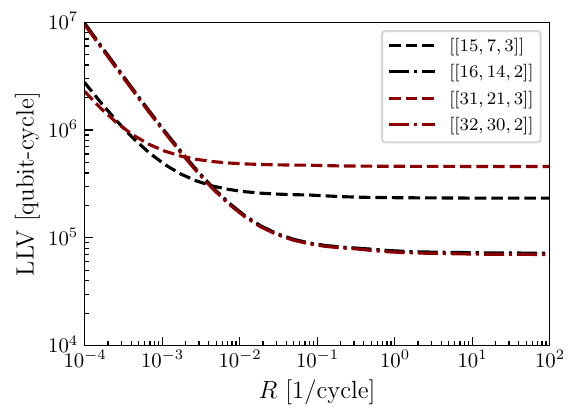}
	\caption{Additional evaluation of the LLV for the combined boosting + pipelined distillation protocol, with four different quantum error-correcting codes used for the pipeliend distillation, with $p_\mathrm{Bell}=0.01$ and target logical error rate of $10^{-12}$. 
    Here, quantum parity codes $[[2m, 2m-2, 2]]$ with $m=8$ (black dash-dotted line) and $m=16$ (red dash-dotted line), and quantum Hamming codes $[[2^\ell-1,2^\ell-2\ell-1,3]]$ with $\ell=4$ (black dashed line) and $\ell=5$ (red dashed line) are used for the pipeliend entanglement distillation.
	}
    \label{fig:hamming_code_combined}
\end{figure}

\subsection{Concatenated entanglement distillation}\label{app:concat-llv}

In Fig.~\ref{fig:boosting_llv} and Fig.~\ref{fig:protocol-switching}a, we show the LLV for concatenated entanglement distillation in Ref.~\cite{Pattison2025}, to achieve the target output logical error rates with buffer space (the space allocated for logical entanglement distillation) of 10 rotated surface code patches. 
This is the same space required for the boosting+distillation with the $[[10,8,2]]$ code shown in Fig.~\ref{fig:protocol-switching}a.
For this evaluation, we assume the use of $d_s=19$ rotated surface code, which is the same as the boosting and boosting+distillation protocols.
For an input Bell pair error rate of 1\% and a target logical error rate of $10^{-12}$, the optimized code sequence identified in Ref.~\cite{Pattison2025} is $[3,1,3]_X, [2,1,2]_Y, [2,1,2]_X,$ and $[[6,4,2]]$, where $[n,1,n]$ denotes the classical repetition code used for entanglement purification~\cite{Bennett1996purification} along a specific basis (subscripts).
For input Bell pair error rates of 1\%, 3\%, and 5\%, and a target logical error rate of $10^{-10}$, plotted in Fig.~\ref{fig:boosting_llv}, we used the open-source code accompanying Ref.~\cite{Pattison2025} to obtain the optimized code sequence, which are $[2,1,2]_Y, [2,1,2]_X$ and $ [[8,3,3]]$ for $p_\mathrm{Bell}=1\%$, $[3,1,3]_X, [2,1,2]_Y , [2,1,2]_X$ and $ [[6,4,2]]$ for $p_\mathrm{Bell}=3\%$, and $[2,1,2]_Y, [2,1,2]_X , [2,1,2]_Y$ and $ [[5,1,3]]$ for $p_\mathrm{Bell}=5\%$, where $[[8,3,3]]$ and $[[5,1,3]]$ are the non-CSS stabilizer codes of Refs.~\cite{Gottesman1996class, laflamme1996perfectquantumerrorcorrection}. 
To evaluate the LLV, we first note that the number of logical CNOT layers required to perform entanglement purification with the $[2,1,2]$ code is 1, and 2 for $[3,1,3]$, and that $Y$-basis purification requires another gate layer for a logical $S$ gate.
For the highest level of concatenation with the $[[2m,2m-2,2]]$ code, we use the circuit volume model for the pipelined implementation (Eq.~\eqref{eq:total-volume} and Sec.~\ref{app:pipeline-volume}), 
and for the $[[8,3,3]]$ and $[[5,1,3]]$ codes, we adopt the optimized encoding circuits presented in Refs.~\cite{10.1093/nsr/nwab011} and \cite{PhysRevA.88.052307}, respectively, which can directly be used to construct the entanglement distillation circuit; in these circuits, the number of logical gate layers is $9$ and $5$, respectively, which we use to obtain circuit volumes of $8\times9\times d_s (2d_s^2-1) \approx 144 d_s^3$ for the $[[8,3,3]]$ code and $5\times5\times d_s (2d_s^2-1) \approx 50 d_s^3$ for $[[5,1,3]]$ code.
We note that the unencoding circuits of non-CSS codes generally do not admit the pipelined implementation presented in Sec.~\ref{sec:distillation}, and therefore we adopted an alternative approach with optimized circuit volume.

We obtain the success rates of state injection and each distillation step from the open-source code accompanying Ref.~\cite{Pattison2025}, and calculate the required number of distillation trials at each concatenation level to achieve a single successful output.
Adding the volumes described above, along with an additional term for the buffering cost of Bell pairs ($\mathcal{V}_b$ in Fig.~\ref{fig:llv}), we obtain the LLV of the injection-distillation approach shown in Fig.~\ref{fig:boosting_llv}.
We remark that, for simplicity, the above evaluation for the concatenated protocol in Ref.~\cite{Pattison2025} neglects the additional spacetime volume associated with inter-level buffering required while accumulating the outputs of lower level distillations for use in the next concatenation level, which may be significant for operations within a limited space; such costs are ignored by assuming that all required distillation instances operate in parallel.

\subsection{Additional LLV data}\label{app:additional-data}

Here, we provide additional results on the LLV evaluations.
In Fig.~\ref{fig:llv_additional}, we show the LLV of the four protocols discussed in this section (remote lattice surgery, entanglement boosting, pipelined entanglement distillation, and concatenated entanglement distillation) at different target logical error rates and $d_s$ than those shown in Fig.~\ref{fig:boosting_llv} and Fig.~\ref{fig:protocol-switching}.
Fig.~\ref{fig:llv_additional}a shows the results for a target logical error rate of $10^{-6}$, where $d_s$ for the entanglement boosting is set to $d_s=7$.
For the injection-distillation protocol, the optimized code sequence is $[2,1,2]_X, [2,1,2]_Y$ and $[[8,6,2]]$, and we evaluate the LLV with $d_s=7$.
Fig.~\ref{fig:llv_additional}b is for a target logical error rate of $10^{-9}$, with $d_s=13$ for the boosting protocol.
For the injection-distillation protocol, the optimized code sequence is $[2,1,2]_X, [2,1,2]_Y$ and $[[8,3,3]]$, and we evaluate the LLV with $d_s=13$.
For both configurations, the qualitative feature of the plot is similar to that shown in Fig.~\ref{fig:protocol-switching}a: entanglement boosting achieves the lowest LLV for large $R$, while the boosting+distillation protocol demonstrates the lowest LLV for low $R$.

In Fig.~\ref{fig:hamming_code_combined}, we compare the LLV of the combined boosting + distillation with different quantum error-correcting codes employed for the entanglement distillation, namely, quantum parity codes $[[2m, 2m-2, 2]]$ with $m=8$ (black dash-dotted line) and $m=16$ (red dash-dotted line), and quantum Hamming codes $[[2^\ell-1,2^\ell-2\ell-1,3]]$ with $\ell=4$ (black dashed line) and $\ell=5$ (red dashed line).
The stronger error suppression attained by the entanglement distillation with quantum Hamming code enables a higher yield compared to the case of quantum parity codes, at the cost of a larger circuit volume to implement the entanglement distillation.
As a result, the LLV shows a crossover at around $R=3\times10^{-3}$, below which the boosting+distillation protocol with quantum Hamming code demonstrates a lower LLV.


\begin{thebibliography}{75}%
\makeatletter
\providecommand \@ifxundefined [1]{%
 \@ifx{#1\undefined}
}%
\providecommand \@ifnum [1]{%
 \ifnum #1\expandafter \@firstoftwo
 \else \expandafter \@secondoftwo
 \fi
}%
\providecommand \@ifx [1]{%
 \ifx #1\expandafter \@firstoftwo
 \else \expandafter \@secondoftwo
 \fi
}%
\providecommand \natexlab [1]{#1}%
\providecommand \enquote  [1]{``#1''}%
\providecommand \bibnamefont  [1]{#1}%
\providecommand \bibfnamefont [1]{#1}%
\providecommand \citenamefont [1]{#1}%
\providecommand \href@noop [0]{\@secondoftwo}%
\providecommand \href [0]{\begingroup \@sanitize@url \@href}%
\providecommand \@href[1]{\@@startlink{#1}\@@href}%
\providecommand \@@href[1]{\endgroup#1\@@endlink}%
\providecommand \@sanitize@url [0]{\catcode `\\12\catcode `\$12\catcode `\&12\catcode `\#12\catcode `\^12\catcode `\_12\catcode `\%12\relax}%
\providecommand \@@startlink[1]{}%
\providecommand \@@endlink[0]{}%
\providecommand \url  [0]{\begingroup\@sanitize@url \@url }%
\providecommand \@url [1]{\endgroup\@href {#1}{\urlprefix }}%
\providecommand \urlprefix  [0]{URL }%
\providecommand \Eprint [0]{\href }%
\providecommand \doibase [0]{https://doi.org/}%
\providecommand \selectlanguage [0]{\@gobble}%
\providecommand \bibinfo  [0]{\@secondoftwo}%
\providecommand \bibfield  [0]{\@secondoftwo}%
\providecommand \translation [1]{[#1]}%
\providecommand \BibitemOpen [0]{}%
\providecommand \bibitemStop [0]{}%
\providecommand \bibitemNoStop [0]{.\EOS\space}%
\providecommand \EOS [0]{\spacefactor3000\relax}%
\providecommand \BibitemShut  [1]{\csname bibitem#1\endcsname}%
\let\auto@bib@innerbib\@empty
\bibitem [{\citenamefont {Panayi}\ \emph {et~al.}(2014)\citenamefont {Panayi}, \citenamefont {Razavi}, \citenamefont {Ma},\ and\ \citenamefont {Lütkenhaus}}]{Panayi2014}%
  \BibitemOpen
  \bibfield  {author} {\bibinfo {author} {\bibfnamefont {C.}~\bibnamefont {Panayi}}, \bibinfo {author} {\bibfnamefont {M.}~\bibnamefont {Razavi}}, \bibinfo {author} {\bibfnamefont {X.}~\bibnamefont {Ma}},\ and\ \bibinfo {author} {\bibfnamefont {N.}~\bibnamefont {Lütkenhaus}},\ }\bibfield  {title} {\bibinfo {title} {Memory-assisted measurement-device-independent quantum key distribution},\ }\href {https://doi.org/10.1088/1367-2630/16/4/043005} {\bibfield  {journal} {\bibinfo  {journal} {New Journal of Physics}\ }\textbf {\bibinfo {volume} {16}},\ \bibinfo {pages} {043005} (\bibinfo {year} {2014})}\BibitemShut {NoStop}%
\bibitem [{\citenamefont {Azuma}\ \emph {et~al.}(2023)\citenamefont {Azuma}, \citenamefont {Economou}, \citenamefont {Elkouss}, \citenamefont {Hilaire}, \citenamefont {Jiang}, \citenamefont {Lo},\ and\ \citenamefont {Tzitrin}}]{Azuma2023}%
  \BibitemOpen
  \bibfield  {author} {\bibinfo {author} {\bibfnamefont {K.}~\bibnamefont {Azuma}}, \bibinfo {author} {\bibfnamefont {S.~E.}\ \bibnamefont {Economou}}, \bibinfo {author} {\bibfnamefont {D.}~\bibnamefont {Elkouss}}, \bibinfo {author} {\bibfnamefont {P.}~\bibnamefont {Hilaire}}, \bibinfo {author} {\bibfnamefont {L.}~\bibnamefont {Jiang}}, \bibinfo {author} {\bibfnamefont {H.-K.}\ \bibnamefont {Lo}},\ and\ \bibinfo {author} {\bibfnamefont {I.}~\bibnamefont {Tzitrin}},\ }\bibfield  {title} {\bibinfo {title} {Quantum repeaters: From quantum networks to the quantum internet},\ }\href {https://doi.org/10.1103/RevModPhys.95.045006} {\bibfield  {journal} {\bibinfo  {journal} {Rev. Mod. Phys.}\ }\textbf {\bibinfo {volume} {95}},\ \bibinfo {pages} {045006} (\bibinfo {year} {2023})}\BibitemShut {NoStop}%
\bibitem [{\citenamefont {Fitzsimons}(2017)}]{Fitzsimons2017}%
  \BibitemOpen
  \bibfield  {author} {\bibinfo {author} {\bibfnamefont {J.~F.}\ \bibnamefont {Fitzsimons}},\ }\bibfield  {title} {\bibinfo {title} {Private quantum computation: an introduction to blind quantum computing and related protocols},\ }\href {https://doi.org/10.1038/S41534-017-0025-3} {\bibfield  {journal} {\bibinfo  {journal} {npj Quantum Information}\ }\textbf {\bibinfo {volume} {3}},\ \bibinfo {pages} {23} (\bibinfo {year} {2017})}\BibitemShut {NoStop}%
\bibitem [{\citenamefont {Gottesman}\ \emph {et~al.}(2012)\citenamefont {Gottesman}, \citenamefont {Jennewein},\ and\ \citenamefont {Croke}}]{Gottesman2012}%
  \BibitemOpen
  \bibfield  {author} {\bibinfo {author} {\bibfnamefont {D.}~\bibnamefont {Gottesman}}, \bibinfo {author} {\bibfnamefont {T.}~\bibnamefont {Jennewein}},\ and\ \bibinfo {author} {\bibfnamefont {S.}~\bibnamefont {Croke}},\ }\bibfield  {title} {\bibinfo {title} {Longer-baseline telescopes using quantum repeaters},\ }\href {https://doi.org/10.1103/PhysRevLett.109.070503} {\bibfield  {journal} {\bibinfo  {journal} {Phys. Rev. Lett.}\ }\textbf {\bibinfo {volume} {109}},\ \bibinfo {pages} {070503} (\bibinfo {year} {2012})}\BibitemShut {NoStop}%
\bibitem [{\citenamefont {Monroe}\ \emph {et~al.}(2014)\citenamefont {Monroe}, \citenamefont {Raussendorf}, \citenamefont {Ruthven}, \citenamefont {Brown}, \citenamefont {Maunz}, \citenamefont {Duan},\ and\ \citenamefont {Kim}}]{Monroe2014}%
  \BibitemOpen
  \bibfield  {author} {\bibinfo {author} {\bibfnamefont {C.}~\bibnamefont {Monroe}}, \bibinfo {author} {\bibfnamefont {R.}~\bibnamefont {Raussendorf}}, \bibinfo {author} {\bibfnamefont {A.}~\bibnamefont {Ruthven}}, \bibinfo {author} {\bibfnamefont {K.~R.}\ \bibnamefont {Brown}}, \bibinfo {author} {\bibfnamefont {P.}~\bibnamefont {Maunz}}, \bibinfo {author} {\bibfnamefont {L.-M.}\ \bibnamefont {Duan}},\ and\ \bibinfo {author} {\bibfnamefont {J.}~\bibnamefont {Kim}},\ }\bibfield  {title} {\bibinfo {title} {Large-scale modular quantum-computer architecture with atomic memory and photonic interconnects},\ }\href {https://doi.org/10.1103/PhysRevA.89.022317} {\bibfield  {journal} {\bibinfo  {journal} {Phys. Rev. A}\ }\textbf {\bibinfo {volume} {89}},\ \bibinfo {pages} {022317} (\bibinfo {year} {2014})}\BibitemShut {NoStop}%
\bibitem [{\citenamefont {Sunami}\ \emph {et~al.}(2025{\natexlab{a}})\citenamefont {Sunami}, \citenamefont {Tamiya}, \citenamefont {Inoue}, \citenamefont {Yamasaki},\ and\ \citenamefont {Goban}}]{Sunami2025}%
  \BibitemOpen
  \bibfield  {author} {\bibinfo {author} {\bibfnamefont {S.}~\bibnamefont {Sunami}}, \bibinfo {author} {\bibfnamefont {S.}~\bibnamefont {Tamiya}}, \bibinfo {author} {\bibfnamefont {R.}~\bibnamefont {Inoue}}, \bibinfo {author} {\bibfnamefont {H.}~\bibnamefont {Yamasaki}},\ and\ \bibinfo {author} {\bibfnamefont {A.}~\bibnamefont {Goban}},\ }\bibfield  {title} {\bibinfo {title} {Scalable networking of neutral-atom qubits: Nanofiber-based approach for multiprocessor fault-tolerant quantum computers},\ }\href {https://doi.org/10.1103/PRXQuantum.6.010101} {\bibfield  {journal} {\bibinfo  {journal} {PRX Quantum}\ }\textbf {\bibinfo {volume} {6}},\ \bibinfo {pages} {010101} (\bibinfo {year} {2025}{\natexlab{a}})}\BibitemShut {NoStop}%
\bibitem [{\citenamefont {Bennett}\ \emph {et~al.}(1996{\natexlab{a}})\citenamefont {Bennett}, \citenamefont {Brassard}, \citenamefont {Popescu}, \citenamefont {Schumacher}, \citenamefont {Smolin},\ and\ \citenamefont {Wootters}}]{Bennett1996purification}%
  \BibitemOpen
  \bibfield  {author} {\bibinfo {author} {\bibfnamefont {C.~H.}\ \bibnamefont {Bennett}}, \bibinfo {author} {\bibfnamefont {G.}~\bibnamefont {Brassard}}, \bibinfo {author} {\bibfnamefont {S.}~\bibnamefont {Popescu}}, \bibinfo {author} {\bibfnamefont {B.}~\bibnamefont {Schumacher}}, \bibinfo {author} {\bibfnamefont {J.~A.}\ \bibnamefont {Smolin}},\ and\ \bibinfo {author} {\bibfnamefont {W.~K.}\ \bibnamefont {Wootters}},\ }\bibfield  {title} {\bibinfo {title} {Purification of noisy entanglement and faithful teleportation via noisy channels},\ }\href {https://doi.org/10.1103/PhysRevLett.76.722} {\bibfield  {journal} {\bibinfo  {journal} {Phys. Rev. Lett.}\ }\textbf {\bibinfo {volume} {76}},\ \bibinfo {pages} {722} (\bibinfo {year} {1996}{\natexlab{a}})}\BibitemShut {NoStop}%
\bibitem [{\citenamefont {Horodecki}\ \emph {et~al.}(2009)\citenamefont {Horodecki}, \citenamefont {Horodecki}, \citenamefont {Horodecki},\ and\ \citenamefont {Horodecki}}]{Horodecki2009}%
  \BibitemOpen
  \bibfield  {author} {\bibinfo {author} {\bibfnamefont {R.}~\bibnamefont {Horodecki}}, \bibinfo {author} {\bibfnamefont {P.}~\bibnamefont {Horodecki}}, \bibinfo {author} {\bibfnamefont {M.}~\bibnamefont {Horodecki}},\ and\ \bibinfo {author} {\bibfnamefont {K.}~\bibnamefont {Horodecki}},\ }\bibfield  {title} {\bibinfo {title} {Quantum entanglement},\ }\href {https://doi.org/10.1103/RevModPhys.81.865} {\bibfield  {journal} {\bibinfo  {journal} {Rev. Mod. Phys.}\ }\textbf {\bibinfo {volume} {81}},\ \bibinfo {pages} {865} (\bibinfo {year} {2009})}\BibitemShut {NoStop}%
\bibitem [{\citenamefont {Deutsch}\ \emph {et~al.}(1996)\citenamefont {Deutsch}, \citenamefont {Ekert}, \citenamefont {Jozsa}, \citenamefont {Macchiavello}, \citenamefont {Popescu},\ and\ \citenamefont {Sanpera}}]{Deutsch1996}%
  \BibitemOpen
  \bibfield  {author} {\bibinfo {author} {\bibfnamefont {D.}~\bibnamefont {Deutsch}}, \bibinfo {author} {\bibfnamefont {A.}~\bibnamefont {Ekert}}, \bibinfo {author} {\bibfnamefont {R.}~\bibnamefont {Jozsa}}, \bibinfo {author} {\bibfnamefont {C.}~\bibnamefont {Macchiavello}}, \bibinfo {author} {\bibfnamefont {S.}~\bibnamefont {Popescu}},\ and\ \bibinfo {author} {\bibfnamefont {A.}~\bibnamefont {Sanpera}},\ }\bibfield  {title} {\bibinfo {title} {Quantum privacy amplification and the security of quantum cryptography over noisy channels},\ }\href {https://doi.org/10.1103/PhysRevLett.77.2818} {\bibfield  {journal} {\bibinfo  {journal} {Phys. Rev. Lett.}\ }\textbf {\bibinfo {volume} {77}},\ \bibinfo {pages} {2818} (\bibinfo {year} {1996})}\BibitemShut {NoStop}%
\bibitem [{\citenamefont {Bennett}\ \emph {et~al.}(1996{\natexlab{b}})\citenamefont {Bennett}, \citenamefont {DiVincenzo}, \citenamefont {Smolin},\ and\ \citenamefont {Wootters}}]{Bennett1996mixedstate}%
  \BibitemOpen
  \bibfield  {author} {\bibinfo {author} {\bibfnamefont {C.~H.}\ \bibnamefont {Bennett}}, \bibinfo {author} {\bibfnamefont {D.~P.}\ \bibnamefont {DiVincenzo}}, \bibinfo {author} {\bibfnamefont {J.~A.}\ \bibnamefont {Smolin}},\ and\ \bibinfo {author} {\bibfnamefont {W.~K.}\ \bibnamefont {Wootters}},\ }\bibfield  {title} {\bibinfo {title} {Mixed-state entanglement and quantum error correction},\ }\href {https://doi.org/10.1103/PhysRevA.54.3824} {\bibfield  {journal} {\bibinfo  {journal} {Phys. Rev. A}\ }\textbf {\bibinfo {volume} {54}},\ \bibinfo {pages} {3824} (\bibinfo {year} {1996}{\natexlab{b}})}\BibitemShut {NoStop}%
\bibitem [{\citenamefont {Dür}\ and\ \citenamefont {Briegel}(2007)}]{Dur2007}%
  \BibitemOpen
  \bibfield  {author} {\bibinfo {author} {\bibfnamefont {W.}~\bibnamefont {Dür}}\ and\ \bibinfo {author} {\bibfnamefont {H.~J.}\ \bibnamefont {Briegel}},\ }\bibfield  {title} {\bibinfo {title} {Entanglement purification and quantum error correction},\ }\href {https://doi.org/10.1088/0034-4885/70/8/R03} {\bibfield  {journal} {\bibinfo  {journal} {Reports on Progress in Physics}\ }\textbf {\bibinfo {volume} {70}},\ \bibinfo {pages} {1381} (\bibinfo {year} {2007})}\BibitemShut {NoStop}%
\bibitem [{\citenamefont {Stephenson}\ \emph {et~al.}(2020)\citenamefont {Stephenson}, \citenamefont {Nadlinger}, \citenamefont {Nichol}, \citenamefont {An}, \citenamefont {Drmota}, \citenamefont {Ballance}, \citenamefont {Thirumalai}, \citenamefont {Goodwin}, \citenamefont {Lucas},\ and\ \citenamefont {Ballance}}]{Stephenson2020}%
  \BibitemOpen
  \bibfield  {author} {\bibinfo {author} {\bibfnamefont {L.~J.}\ \bibnamefont {Stephenson}}, \bibinfo {author} {\bibfnamefont {D.~P.}\ \bibnamefont {Nadlinger}}, \bibinfo {author} {\bibfnamefont {B.~C.}\ \bibnamefont {Nichol}}, \bibinfo {author} {\bibfnamefont {S.}~\bibnamefont {An}}, \bibinfo {author} {\bibfnamefont {P.}~\bibnamefont {Drmota}}, \bibinfo {author} {\bibfnamefont {T.~G.}\ \bibnamefont {Ballance}}, \bibinfo {author} {\bibfnamefont {K.}~\bibnamefont {Thirumalai}}, \bibinfo {author} {\bibfnamefont {J.~F.}\ \bibnamefont {Goodwin}}, \bibinfo {author} {\bibfnamefont {D.~M.}\ \bibnamefont {Lucas}},\ and\ \bibinfo {author} {\bibfnamefont {C.~J.}\ \bibnamefont {Ballance}},\ }\bibfield  {title} {\bibinfo {title} {High-rate, high-fidelity entanglement of qubits across an elementary quantum network},\ }\href {https://doi.org/10.1103/PhysRevLett.124.110501} {\bibfield  {journal} {\bibinfo  {journal} {Phys. Rev. Lett.}\ }\textbf {\bibinfo {volume} {124}},\ \bibinfo {pages} {110501} (\bibinfo {year}
  {2020})}\BibitemShut {NoStop}%
\bibitem [{\citenamefont {Main}\ \emph {et~al.}(2025)\citenamefont {Main}, \citenamefont {Drmota}, \citenamefont {Nadlinger}, \citenamefont {Ainley}, \citenamefont {Agrawal}, \citenamefont {Nichol}, \citenamefont {Srinivas}, \citenamefont {Araneda},\ and\ \citenamefont {Lucas}}]{main2025}%
  \BibitemOpen
  \bibfield  {author} {\bibinfo {author} {\bibfnamefont {D.}~\bibnamefont {Main}}, \bibinfo {author} {\bibfnamefont {P.}~\bibnamefont {Drmota}}, \bibinfo {author} {\bibfnamefont {D.~P.}\ \bibnamefont {Nadlinger}}, \bibinfo {author} {\bibfnamefont {E.~M.}\ \bibnamefont {Ainley}}, \bibinfo {author} {\bibfnamefont {A.}~\bibnamefont {Agrawal}}, \bibinfo {author} {\bibfnamefont {B.~C.}\ \bibnamefont {Nichol}}, \bibinfo {author} {\bibfnamefont {R.}~\bibnamefont {Srinivas}}, \bibinfo {author} {\bibfnamefont {G.}~\bibnamefont {Araneda}},\ and\ \bibinfo {author} {\bibfnamefont {D.~M.}\ \bibnamefont {Lucas}},\ }\bibfield  {title} {\bibinfo {title} {Distributed quantum computing across an optical network link},\ }\href {https://doi.org/10.1038/s41586-024-08404-x} {\bibfield  {journal} {\bibinfo  {journal} {Nature}\ }\textbf {\bibinfo {volume} {638}},\ \bibinfo {pages} {383} (\bibinfo {year} {2025})}\BibitemShut {NoStop}%
\bibitem [{\citenamefont {Li}\ and\ \citenamefont {Thompson}(2024)}]{Li2024}%
  \BibitemOpen
  \bibfield  {author} {\bibinfo {author} {\bibfnamefont {Y.}~\bibnamefont {Li}}\ and\ \bibinfo {author} {\bibfnamefont {J.~D.}\ \bibnamefont {Thompson}},\ }\bibfield  {title} {\bibinfo {title} {High-rate and high-fidelity modular interconnects between neutral atom quantum processors},\ }\href {https://doi.org/10.1103/PRXQuantum.5.020363} {\bibfield  {journal} {\bibinfo  {journal} {PRX Quantum}\ }\textbf {\bibinfo {volume} {5}},\ \bibinfo {pages} {020363} (\bibinfo {year} {2024})}\BibitemShut {NoStop}%
\bibitem [{\citenamefont {Sinclair}\ \emph {et~al.}(2025)\citenamefont {Sinclair}, \citenamefont {Ramette}, \citenamefont {Grinkemeyer}, \citenamefont {Bluvstein}, \citenamefont {Lukin},\ and\ \citenamefont {Vuleti\ifmmode~\acute{c}\else \'{c}\fi{}}}]{Sinclair2025}%
  \BibitemOpen
  \bibfield  {author} {\bibinfo {author} {\bibfnamefont {J.}~\bibnamefont {Sinclair}}, \bibinfo {author} {\bibfnamefont {J.}~\bibnamefont {Ramette}}, \bibinfo {author} {\bibfnamefont {B.}~\bibnamefont {Grinkemeyer}}, \bibinfo {author} {\bibfnamefont {D.}~\bibnamefont {Bluvstein}}, \bibinfo {author} {\bibfnamefont {M.~D.}\ \bibnamefont {Lukin}},\ and\ \bibinfo {author} {\bibfnamefont {V.}~\bibnamefont {Vuleti\ifmmode~\acute{c}\else \'{c}\fi{}}},\ }\bibfield  {title} {\bibinfo {title} {Fault-tolerant optical interconnects for neutral-atom arrays},\ }\href {https://doi.org/10.1103/PhysRevResearch.7.013313} {\bibfield  {journal} {\bibinfo  {journal} {Phys. Rev. Res.}\ }\textbf {\bibinfo {volume} {7}},\ \bibinfo {pages} {013313} (\bibinfo {year} {2025})}\BibitemShut {NoStop}%
\bibitem [{\citenamefont {Litinski}(2019{\natexlab{a}})}]{litinski2019game}%
  \BibitemOpen
  \bibfield  {author} {\bibinfo {author} {\bibfnamefont {D.}~\bibnamefont {Litinski}},\ }\bibfield  {title} {\bibinfo {title} {A {Game} of {Surface} {Codes}: {Large}-{Scale} {Quantum} {Computing} with {Lattice} {Surgery}},\ }\href {https://doi.org/10.22331/q-2019-03-05-128} {\bibfield  {journal} {\bibinfo  {journal} {Quantum}\ }\textbf {\bibinfo {volume} {3}},\ \bibinfo {pages} {128} (\bibinfo {year} {2019}{\natexlab{a}})}\BibitemShut {NoStop}%
\bibitem [{\citenamefont {Fowler}\ \emph {et~al.}(2012)\citenamefont {Fowler}, \citenamefont {Mariantoni}, \citenamefont {Martinis},\ and\ \citenamefont {Cleland}}]{Fowler2012}%
  \BibitemOpen
  \bibfield  {author} {\bibinfo {author} {\bibfnamefont {A.~G.}\ \bibnamefont {Fowler}}, \bibinfo {author} {\bibfnamefont {M.}~\bibnamefont {Mariantoni}}, \bibinfo {author} {\bibfnamefont {J.~M.}\ \bibnamefont {Martinis}},\ and\ \bibinfo {author} {\bibfnamefont {A.~N.}\ \bibnamefont {Cleland}},\ }\bibfield  {title} {\bibinfo {title} {Surface codes: Towards practical large-scale quantum computation},\ }\href {https://doi.org/10.1103/PhysRevA.86.032324} {\bibfield  {journal} {\bibinfo  {journal} {Phys. Rev. A}\ }\textbf {\bibinfo {volume} {86}},\ \bibinfo {pages} {032324} (\bibinfo {year} {2012})}\BibitemShut {NoStop}%
\bibitem [{\citenamefont {Gidney}\ and\ \citenamefont {Eker{\aa{}}}(2021)}]{Gidney2021howtofactorbit}%
  \BibitemOpen
  \bibfield  {author} {\bibinfo {author} {\bibfnamefont {C.}~\bibnamefont {Gidney}}\ and\ \bibinfo {author} {\bibfnamefont {M.}~\bibnamefont {Eker{\aa{}}}},\ }\bibfield  {title} {\bibinfo {title} {How to factor 2048 bit {RSA} integers in 8 hours using 20 million noisy qubits},\ }\href {https://doi.org/10.22331/q-2021-04-15-433} {\bibfield  {journal} {\bibinfo  {journal} {{Quantum}}\ }\textbf {\bibinfo {volume} {5}},\ \bibinfo {pages} {433} (\bibinfo {year} {2021})}\BibitemShut {NoStop}%
\bibitem [{\citenamefont {Litinski}(2019{\natexlab{b}})}]{Litinski2019magicstate}%
  \BibitemOpen
  \bibfield  {author} {\bibinfo {author} {\bibfnamefont {D.}~\bibnamefont {Litinski}},\ }\bibfield  {title} {\bibinfo {title} {Magic {S}tate {D}istillation: {N}ot as {C}ostly as {Y}ou {T}hink},\ }\href {https://doi.org/10.22331/q-2019-12-02-205} {\bibfield  {journal} {\bibinfo  {journal} {{Quantum}}\ }\textbf {\bibinfo {volume} {3}},\ \bibinfo {pages} {205} (\bibinfo {year} {2019}{\natexlab{b}})}\BibitemShut {NoStop}%
\bibitem [{\citenamefont {Itogawa}\ \emph {et~al.}(2025)\citenamefont {Itogawa}, \citenamefont {Takada}, \citenamefont {Hirano},\ and\ \citenamefont {Fujii}}]{Itogawa2025}%
  \BibitemOpen
  \bibfield  {author} {\bibinfo {author} {\bibfnamefont {T.}~\bibnamefont {Itogawa}}, \bibinfo {author} {\bibfnamefont {Y.}~\bibnamefont {Takada}}, \bibinfo {author} {\bibfnamefont {Y.}~\bibnamefont {Hirano}},\ and\ \bibinfo {author} {\bibfnamefont {K.}~\bibnamefont {Fujii}},\ }\bibfield  {title} {\bibinfo {title} {Efficient magic state distillation by zero-level distillation},\ }\href {https://doi.org/10.1103/thxx-njr6} {\bibfield  {journal} {\bibinfo  {journal} {PRX Quantum}\ }\textbf {\bibinfo {volume} {6}},\ \bibinfo {pages} {020356} (\bibinfo {year} {2025})}\BibitemShut {NoStop}%
\bibitem [{\citenamefont {Gidney}\ \emph {et~al.}(2024)\citenamefont {Gidney}, \citenamefont {Shutty},\ and\ \citenamefont {Jones}}]{Gidney2024}%
  \BibitemOpen
  \bibfield  {author} {\bibinfo {author} {\bibfnamefont {C.}~\bibnamefont {Gidney}}, \bibinfo {author} {\bibfnamefont {N.}~\bibnamefont {Shutty}},\ and\ \bibinfo {author} {\bibfnamefont {C.}~\bibnamefont {Jones}},\ }\href {https://arxiv.org/abs/2409.17595} {\bibinfo {title} {Magic state cultivation: growing t states as cheap as cnot gates}} (\bibinfo {year} {2024}),\ \Eprint {https://arxiv.org/abs/2409.17595} {arXiv:2409.17595 [quant-ph]} \BibitemShut {NoStop}%
\bibitem [{\citenamefont {Hirano}\ \emph {et~al.}(2024{\natexlab{a}})\citenamefont {Hirano}, \citenamefont {Itogawa},\ and\ \citenamefont {Fujii}}]{Hirano2024}%
  \BibitemOpen
  \bibfield  {author} {\bibinfo {author} {\bibfnamefont {Y.}~\bibnamefont {Hirano}}, \bibinfo {author} {\bibfnamefont {T.}~\bibnamefont {Itogawa}},\ and\ \bibinfo {author} {\bibfnamefont {K.}~\bibnamefont {Fujii}},\ }\bibfield  {title} {\bibinfo {title} {Leveraging zero-level distillation to generate high-fidelity magic states},\ }in\ \href {https://doi.org/10.1109/QCE60285.2024.00104} {\emph {\bibinfo {booktitle} {2024 IEEE International Conference on Quantum Computing and Engineering (QCE)}}},\ Vol.~\bibinfo {volume} {01}\ (\bibinfo {year} {2024})\ pp.\ \bibinfo {pages} {843--853}\BibitemShut {NoStop}%
\bibitem [{\citenamefont {Chen}\ \emph {et~al.}(2025{\natexlab{a}})\citenamefont {Chen}, \citenamefont {Chen}, \citenamefont {Lu},\ and\ \citenamefont {Pan}}]{chen2025efficientmagicstatecultivation}%
  \BibitemOpen
  \bibfield  {author} {\bibinfo {author} {\bibfnamefont {Z.-H.}\ \bibnamefont {Chen}}, \bibinfo {author} {\bibfnamefont {M.-C.}\ \bibnamefont {Chen}}, \bibinfo {author} {\bibfnamefont {C.-Y.}\ \bibnamefont {Lu}},\ and\ \bibinfo {author} {\bibfnamefont {J.-W.}\ \bibnamefont {Pan}},\ }\href {https://arxiv.org/abs/2503.18657} {\bibinfo {title} {Efficient magic state cultivation on $\mathbb{RP}^2$}} (\bibinfo {year} {2025}{\natexlab{a}}),\ \Eprint {https://arxiv.org/abs/2503.18657} {arXiv:2503.18657 [quant-ph]} \BibitemShut {NoStop}%
\bibitem [{\citenamefont {Gidney}(2025)}]{Gidney2025resource}%
  \BibitemOpen
  \bibfield  {author} {\bibinfo {author} {\bibfnamefont {C.}~\bibnamefont {Gidney}},\ }\href {https://arxiv.org/abs/2505.15917} {\bibinfo {title} {How to factor 2048 bit rsa integers with less than a million noisy qubits}} (\bibinfo {year} {2025}),\ \Eprint {https://arxiv.org/abs/2505.15917} {arXiv:2505.15917 [quant-ph]} \BibitemShut {NoStop}%
\bibitem [{\citenamefont {Ramette}\ \emph {et~al.}(2024)\citenamefont {Ramette}, \citenamefont {Sinclair}, \citenamefont {Breuckmann},\ and\ \citenamefont {Vulet\'ic}}]{Ramette2024}%
  \BibitemOpen
  \bibfield  {author} {\bibinfo {author} {\bibfnamefont {J.}~\bibnamefont {Ramette}}, \bibinfo {author} {\bibfnamefont {J.}~\bibnamefont {Sinclair}}, \bibinfo {author} {\bibfnamefont {N.~P.}\ \bibnamefont {Breuckmann}},\ and\ \bibinfo {author} {\bibfnamefont {V.}~\bibnamefont {Vulet\'ic}},\ }\bibfield  {title} {\bibinfo {title} {Fault-tolerant connection of error-corrected qubits with noisy links},\ }\href {https://doi.org/10.1038/s41534-024-00855-4} {\bibfield  {journal} {\bibinfo  {journal} {npj Quantum Information}\ }\textbf {\bibinfo {volume} {10}},\ \bibinfo {pages} {58} (\bibinfo {year} {2024})}\BibitemShut {NoStop}%
\bibitem [{\citenamefont {Shalby}\ \emph {et~al.}(2025)\citenamefont {Shalby}, \citenamefont {Wang}, \citenamefont {Sedov},\ and\ \citenamefont {Pryadko}}]{Shalby2025}%
  \BibitemOpen
  \bibfield  {author} {\bibinfo {author} {\bibfnamefont {M.~A.}\ \bibnamefont {Shalby}}, \bibinfo {author} {\bibfnamefont {R.}~\bibnamefont {Wang}}, \bibinfo {author} {\bibfnamefont {D.}~\bibnamefont {Sedov}},\ and\ \bibinfo {author} {\bibfnamefont {L.~P.}\ \bibnamefont {Pryadko}},\ }\bibfield  {title} {\bibinfo {title} {Optimized noise-resilient surface code teleportation interfaces},\ }\href {https://doi.org/10.1103/xqrn-wdw1} {\bibfield  {journal} {\bibinfo  {journal} {Phys. Rev. A}\ }\textbf {\bibinfo {volume} {112}},\ \bibinfo {pages} {L020403} (\bibinfo {year} {2025})}\BibitemShut {NoStop}%
\bibitem [{\citenamefont {Glancy}\ \emph {et~al.}(2006)\citenamefont {Glancy}, \citenamefont {Knill},\ and\ \citenamefont {Vasconcelos}}]{Glancy2006}%
  \BibitemOpen
  \bibfield  {author} {\bibinfo {author} {\bibfnamefont {S.}~\bibnamefont {Glancy}}, \bibinfo {author} {\bibfnamefont {E.}~\bibnamefont {Knill}},\ and\ \bibinfo {author} {\bibfnamefont {H.~M.}\ \bibnamefont {Vasconcelos}},\ }\bibfield  {title} {\bibinfo {title} {Entanglement purification of any stabilizer state},\ }\href {https://doi.org/10.1103/PhysRevA.74.032319} {\bibfield  {journal} {\bibinfo  {journal} {Phys. Rev. A}\ }\textbf {\bibinfo {volume} {74}},\ \bibinfo {pages} {032319} (\bibinfo {year} {2006})}\BibitemShut {NoStop}%
\bibitem [{\citenamefont {Maeda}\ \emph {et~al.}(2025)\citenamefont {Maeda}, \citenamefont {Suzuki}, \citenamefont {Kobayashi}, \citenamefont {Yamamoto}, \citenamefont {Tokunaga},\ and\ \citenamefont {Fujii}}]{Maeda2025}%
  \BibitemOpen
  \bibfield  {author} {\bibinfo {author} {\bibfnamefont {Y.}~\bibnamefont {Maeda}}, \bibinfo {author} {\bibfnamefont {Y.}~\bibnamefont {Suzuki}}, \bibinfo {author} {\bibfnamefont {T.}~\bibnamefont {Kobayashi}}, \bibinfo {author} {\bibfnamefont {T.}~\bibnamefont {Yamamoto}}, \bibinfo {author} {\bibfnamefont {Y.}~\bibnamefont {Tokunaga}},\ and\ \bibinfo {author} {\bibfnamefont {K.}~\bibnamefont {Fujii}},\ }\href {https://arxiv.org/abs/2503.14894} {\bibinfo {title} {Logical entanglement distribution between distant 2d array qubits}} (\bibinfo {year} {2025}),\ \Eprint {https://arxiv.org/abs/2503.14894} {arXiv:2503.14894 [quant-ph]} \BibitemShut {NoStop}%
\bibitem [{\citenamefont {Bonilla~Ataides}\ \emph {et~al.}(2025)\citenamefont {Bonilla~Ataides}, \citenamefont {Zhou}, \citenamefont {Xu}, \citenamefont {Baranes}, \citenamefont {Li}, \citenamefont {Lukin},\ and\ \citenamefont {Jiang}}]{Ataides2025}%
  \BibitemOpen
  \bibfield  {author} {\bibinfo {author} {\bibfnamefont {J.~P.}\ \bibnamefont {Bonilla~Ataides}}, \bibinfo {author} {\bibfnamefont {H.}~\bibnamefont {Zhou}}, \bibinfo {author} {\bibfnamefont {Q.}~\bibnamefont {Xu}}, \bibinfo {author} {\bibfnamefont {G.}~\bibnamefont {Baranes}}, \bibinfo {author} {\bibfnamefont {B.}~\bibnamefont {Li}}, \bibinfo {author} {\bibfnamefont {M.~D.}\ \bibnamefont {Lukin}},\ and\ \bibinfo {author} {\bibfnamefont {L.}~\bibnamefont {Jiang}},\ }\bibfield  {title} {\bibinfo {title} {Constant-overhead fault-tolerant bell-pair distillation using high-rate codes},\ }\href {https://doi.org/10.1103/s39k-r2kq} {\bibfield  {journal} {\bibinfo  {journal} {Phys. Rev. Lett.}\ }\textbf {\bibinfo {volume} {135}},\ \bibinfo {pages} {130804} (\bibinfo {year} {2025})}\BibitemShut {NoStop}%
\bibitem [{\citenamefont {Pattison}\ \emph {et~al.}(2025)\citenamefont {Pattison}, \citenamefont {Baranes}, \citenamefont {Bonilla~Ataides}, \citenamefont {Lukin},\ and\ \citenamefont {Zhou}}]{Pattison2025}%
  \BibitemOpen
  \bibfield  {author} {\bibinfo {author} {\bibfnamefont {C.}~\bibnamefont {Pattison}}, \bibinfo {author} {\bibfnamefont {G.}~\bibnamefont {Baranes}}, \bibinfo {author} {\bibfnamefont {J.~P.}\ \bibnamefont {Bonilla~Ataides}}, \bibinfo {author} {\bibfnamefont {M.~D.}\ \bibnamefont {Lukin}},\ and\ \bibinfo {author} {\bibfnamefont {H.}~\bibnamefont {Zhou}},\ }\bibfield  {title} {\bibinfo {title} {Constant-rate entanglement distillation for fast quantum interconnects},\ }in\ \href {https://doi.org/10.1145/3695053.3731069} {\emph {\bibinfo {booktitle} {Proceedings of the 52nd Annual International Symposium on Computer Architecture}}},\ \bibinfo {series and number} {ISCA '25}\ (\bibinfo  {publisher} {Association for Computing Machinery},\ \bibinfo {address} {New York, NY, USA},\ \bibinfo {year} {2025})\ p.\ \bibinfo {pages} {257–270}\BibitemShut {NoStop}%
\bibitem [{\citenamefont {Bombin}\ and\ \citenamefont {Martin-Delgado}(2007)}]{Bombin2007}%
  \BibitemOpen
  \bibfield  {author} {\bibinfo {author} {\bibfnamefont {H.}~\bibnamefont {Bombin}}\ and\ \bibinfo {author} {\bibfnamefont {M.~A.}\ \bibnamefont {Martin-Delgado}},\ }\bibfield  {title} {\bibinfo {title} {Optimal resources for topological two-dimensional stabilizer codes: Comparative study},\ }\href {https://doi.org/10.1103/PhysRevA.76.012305} {\bibfield  {journal} {\bibinfo  {journal} {Phys. Rev. A}\ }\textbf {\bibinfo {volume} {76}},\ \bibinfo {pages} {012305} (\bibinfo {year} {2007})}\BibitemShut {NoStop}%
\bibitem [{\citenamefont {Zapatero}\ \emph {et~al.}(2023)\citenamefont {Zapatero}, \citenamefont {van Leent}, \citenamefont {Arnon-Friedman}, \citenamefont {Liu}, \citenamefont {Zhang}, \citenamefont {Weinfurter},\ and\ \citenamefont {Curty}}]{zapatero2023advances}%
  \BibitemOpen
  \bibfield  {author} {\bibinfo {author} {\bibfnamefont {V.}~\bibnamefont {Zapatero}}, \bibinfo {author} {\bibfnamefont {T.}~\bibnamefont {van Leent}}, \bibinfo {author} {\bibfnamefont {R.}~\bibnamefont {Arnon-Friedman}}, \bibinfo {author} {\bibfnamefont {W.-Z.}\ \bibnamefont {Liu}}, \bibinfo {author} {\bibfnamefont {Q.}~\bibnamefont {Zhang}}, \bibinfo {author} {\bibfnamefont {H.}~\bibnamefont {Weinfurter}},\ and\ \bibinfo {author} {\bibfnamefont {M.}~\bibnamefont {Curty}},\ }\bibfield  {title} {\bibinfo {title} {Advances in device-independent quantum key distribution},\ }\href {https://www.nature.com/articles/s41534-023-00684-x} {\bibfield  {journal} {\bibinfo  {journal} {npj quantum information}\ }\textbf {\bibinfo {volume} {9}},\ \bibinfo {pages} {10} (\bibinfo {year} {2023})}\BibitemShut {NoStop}%
\bibitem [{\citenamefont {Meier}\ and\ \citenamefont {Yamasaki}(2025)}]{PRXEnergy.4.023008}%
  \BibitemOpen
  \bibfield  {author} {\bibinfo {author} {\bibfnamefont {F.}~\bibnamefont {Meier}}\ and\ \bibinfo {author} {\bibfnamefont {H.}~\bibnamefont {Yamasaki}},\ }\bibfield  {title} {\bibinfo {title} {Energy-consumption advantage of quantum computation},\ }\href {https://doi.org/10.1103/PRXEnergy.4.023008} {\bibfield  {journal} {\bibinfo  {journal} {PRX Energy}\ }\textbf {\bibinfo {volume} {4}},\ \bibinfo {pages} {023008} (\bibinfo {year} {2025})}\BibitemShut {NoStop}%
\bibitem [{\citenamefont {Calderbank}\ and\ \citenamefont {Shor}(1996)}]{Calderbank1996}%
  \BibitemOpen
  \bibfield  {author} {\bibinfo {author} {\bibfnamefont {A.~R.}\ \bibnamefont {Calderbank}}\ and\ \bibinfo {author} {\bibfnamefont {P.~W.}\ \bibnamefont {Shor}},\ }\bibfield  {title} {\bibinfo {title} {Good quantum error-correcting codes exist},\ }\href {https://doi.org/10.1103/PhysRevA.54.1098} {\bibfield  {journal} {\bibinfo  {journal} {Phys. Rev. A}\ }\textbf {\bibinfo {volume} {54}},\ \bibinfo {pages} {1098} (\bibinfo {year} {1996})}\BibitemShut {NoStop}%
\bibitem [{\citenamefont {Steane}(1996{\natexlab{a}})}]{Steane1996_CSS}%
  \BibitemOpen
  \bibfield  {author} {\bibinfo {author} {\bibfnamefont {A.}~\bibnamefont {Steane}},\ }\bibfield  {title} {\bibinfo {title} {Multiple-particle interference and quantum error correction},\ }\href {https://doi.org/10.1098/rspa.1996.0136} {\bibfield  {journal} {\bibinfo  {journal} {Proceedings of the Royal Society of London. Series A: Mathematical, Physical and Engineering Sciences}\ }\textbf {\bibinfo {volume} {452}},\ \bibinfo {pages} {2551} (\bibinfo {year} {1996}{\natexlab{a}})}\BibitemShut {NoStop}%
\bibitem [{\citenamefont {Matsumoto}(2003)}]{matsumoto2003conversion}%
  \BibitemOpen
  \bibfield  {author} {\bibinfo {author} {\bibfnamefont {R.}~\bibnamefont {Matsumoto}},\ }\bibfield  {title} {\bibinfo {title} {Conversion of a general quantum stabilizer code to an entanglement distillation protocol},\ }\href {https://doi.org/10.1088/0305-4470/36/29/316} {\bibfield  {journal} {\bibinfo  {journal} {Journal of Physics A: Mathematical and General}\ }\textbf {\bibinfo {volume} {36}},\ \bibinfo {pages} {8113} (\bibinfo {year} {2003})}\BibitemShut {NoStop}%
\bibitem [{\citenamefont {Hostens}\ \emph {et~al.}(2004)\citenamefont {Hostens}, \citenamefont {Dehaene},\ and\ \citenamefont {Moor}}]{hostens2004equivalence}%
  \BibitemOpen
  \bibfield  {author} {\bibinfo {author} {\bibfnamefont {E.}~\bibnamefont {Hostens}}, \bibinfo {author} {\bibfnamefont {J.}~\bibnamefont {Dehaene}},\ and\ \bibinfo {author} {\bibfnamefont {B.~D.}\ \bibnamefont {Moor}},\ }\href {https://arxiv.org/abs/quant-ph/0406017} {\bibinfo {title} {The equivalence of two approaches to the design of entanglement distillation protocols}} (\bibinfo {year} {2004}),\ \Eprint {https://arxiv.org/abs/quant-ph/0406017} {arXiv:quant-ph/0406017 [quant-ph]} \BibitemShut {NoStop}%
\bibitem [{\citenamefont {Bluvstein}\ \emph {et~al.}(2022)\citenamefont {Bluvstein}, \citenamefont {Levine}, \citenamefont {Semeghini}, \citenamefont {Wang}, \citenamefont {Ebadi}, \citenamefont {Kalinowski}, \citenamefont {Keesling}, \citenamefont {Maskara}, \citenamefont {Pichler}, \citenamefont {Greiner}, \citenamefont {Vuletić},\ and\ \citenamefont {Lukin}}]{Bluvstein2022}%
  \BibitemOpen
  \bibfield  {author} {\bibinfo {author} {\bibfnamefont {D.}~\bibnamefont {Bluvstein}}, \bibinfo {author} {\bibfnamefont {H.}~\bibnamefont {Levine}}, \bibinfo {author} {\bibfnamefont {G.}~\bibnamefont {Semeghini}}, \bibinfo {author} {\bibfnamefont {T.~T.}\ \bibnamefont {Wang}}, \bibinfo {author} {\bibfnamefont {S.}~\bibnamefont {Ebadi}}, \bibinfo {author} {\bibfnamefont {M.}~\bibnamefont {Kalinowski}}, \bibinfo {author} {\bibfnamefont {A.}~\bibnamefont {Keesling}}, \bibinfo {author} {\bibfnamefont {N.}~\bibnamefont {Maskara}}, \bibinfo {author} {\bibfnamefont {H.}~\bibnamefont {Pichler}}, \bibinfo {author} {\bibfnamefont {M.}~\bibnamefont {Greiner}}, \bibinfo {author} {\bibfnamefont {V.}~\bibnamefont {Vuletić}},\ and\ \bibinfo {author} {\bibfnamefont {M.~D.}\ \bibnamefont {Lukin}},\ }\bibfield  {title} {\bibinfo {title} {A quantum processor based on coherent transport of entangled atom arrays},\ }\href {https://doi.org/10.1038/s41586-022-04592-6} {\bibfield  {journal} {\bibinfo  {journal} {Nature}\ }\textbf
  {\bibinfo {volume} {604}},\ \bibinfo {pages} {451} (\bibinfo {year} {2022})}\BibitemShut {NoStop}%
\bibitem [{\citenamefont {Duan}\ and\ \citenamefont {Monroe}(2010)}]{Duan2010}%
  \BibitemOpen
  \bibfield  {author} {\bibinfo {author} {\bibfnamefont {L.-M.}\ \bibnamefont {Duan}}\ and\ \bibinfo {author} {\bibfnamefont {C.}~\bibnamefont {Monroe}},\ }\bibfield  {title} {\bibinfo {title} {Colloquium: Quantum networks with trapped ions},\ }\href {https://doi.org/10.1103/RevModPhys.82.1209} {\bibfield  {journal} {\bibinfo  {journal} {Rev. Mod. Phys.}\ }\textbf {\bibinfo {volume} {82}},\ \bibinfo {pages} {1209} (\bibinfo {year} {2010})}\BibitemShut {NoStop}%
\bibitem [{\citenamefont {Horsman}\ \emph {et~al.}(2012)\citenamefont {Horsman}, \citenamefont {Fowler}, \citenamefont {Devitt},\ and\ \citenamefont {Meter}}]{Horsman2012}%
  \BibitemOpen
  \bibfield  {author} {\bibinfo {author} {\bibfnamefont {D.}~\bibnamefont {Horsman}}, \bibinfo {author} {\bibfnamefont {A.~G.}\ \bibnamefont {Fowler}}, \bibinfo {author} {\bibfnamefont {S.}~\bibnamefont {Devitt}},\ and\ \bibinfo {author} {\bibfnamefont {R.~V.}\ \bibnamefont {Meter}},\ }\bibfield  {title} {\bibinfo {title} {Surface code quantum computing by lattice surgery},\ }\href {https://doi.org/10.1088/1367-2630/14/12/123011} {\bibfield  {journal} {\bibinfo  {journal} {New Journal of Physics}\ }\textbf {\bibinfo {volume} {14}},\ \bibinfo {pages} {123011} (\bibinfo {year} {2012})}\BibitemShut {NoStop}%
\bibitem [{\citenamefont {Sahay}\ \emph {et~al.}(2025)\citenamefont {Sahay}, \citenamefont {Lin}, \citenamefont {Huang}, \citenamefont {Brown},\ and\ \citenamefont {Puri}}]{Sahay2025}%
  \BibitemOpen
  \bibfield  {author} {\bibinfo {author} {\bibfnamefont {K.}~\bibnamefont {Sahay}}, \bibinfo {author} {\bibfnamefont {Y.}~\bibnamefont {Lin}}, \bibinfo {author} {\bibfnamefont {S.}~\bibnamefont {Huang}}, \bibinfo {author} {\bibfnamefont {K.~R.}\ \bibnamefont {Brown}},\ and\ \bibinfo {author} {\bibfnamefont {S.}~\bibnamefont {Puri}},\ }\bibfield  {title} {\bibinfo {title} {Error correction of transversal cnot gates for scalable surface-code computation},\ }\href {https://doi.org/10.1103/PRXQuantum.6.020326} {\bibfield  {journal} {\bibinfo  {journal} {PRX Quantum}\ }\textbf {\bibinfo {volume} {6}},\ \bibinfo {pages} {020326} (\bibinfo {year} {2025})}\BibitemShut {NoStop}%
\bibitem [{\citenamefont {Gidney}\ \emph {et~al.}(2025)\citenamefont {Gidney}, \citenamefont {Newman}, \citenamefont {Brooks},\ and\ \citenamefont {Jones}}]{Gidney2025}%
  \BibitemOpen
  \bibfield  {author} {\bibinfo {author} {\bibfnamefont {C.}~\bibnamefont {Gidney}}, \bibinfo {author} {\bibfnamefont {M.}~\bibnamefont {Newman}}, \bibinfo {author} {\bibfnamefont {P.}~\bibnamefont {Brooks}},\ and\ \bibinfo {author} {\bibfnamefont {C.}~\bibnamefont {Jones}},\ }\bibfield  {title} {\bibinfo {title} {Yoked surface codes},\ }\href {https://doi.org/10.1038/s41467-025-59714-1} {\bibfield  {journal} {\bibinfo  {journal} {Nature Communications}\ }\textbf {\bibinfo {volume} {16}},\ \bibinfo {pages} {4498} (\bibinfo {year} {2025})}\BibitemShut {NoStop}%
\bibitem [{\citenamefont {O'Reilly}\ \emph {et~al.}(2024)\citenamefont {O'Reilly}, \citenamefont {Toh}, \citenamefont {Goetting}, \citenamefont {Saha}, \citenamefont {Shalaev}, \citenamefont {Carter}, \citenamefont {Risinger}, \citenamefont {Kalakuntla}, \citenamefont {Li}, \citenamefont {Verma},\ and\ \citenamefont {Monroe}}]{Oreilly2024}%
  \BibitemOpen
  \bibfield  {author} {\bibinfo {author} {\bibfnamefont {J.}~\bibnamefont {O'Reilly}}, \bibinfo {author} {\bibfnamefont {G.}~\bibnamefont {Toh}}, \bibinfo {author} {\bibfnamefont {I.}~\bibnamefont {Goetting}}, \bibinfo {author} {\bibfnamefont {S.}~\bibnamefont {Saha}}, \bibinfo {author} {\bibfnamefont {M.}~\bibnamefont {Shalaev}}, \bibinfo {author} {\bibfnamefont {A.~L.}\ \bibnamefont {Carter}}, \bibinfo {author} {\bibfnamefont {A.}~\bibnamefont {Risinger}}, \bibinfo {author} {\bibfnamefont {A.}~\bibnamefont {Kalakuntla}}, \bibinfo {author} {\bibfnamefont {T.}~\bibnamefont {Li}}, \bibinfo {author} {\bibfnamefont {A.}~\bibnamefont {Verma}},\ and\ \bibinfo {author} {\bibfnamefont {C.}~\bibnamefont {Monroe}},\ }\bibfield  {title} {\bibinfo {title} {Fast photon-mediated entanglement of continuously cooled trapped ions for quantum networking},\ }\href {https://doi.org/10.1103/PhysRevLett.133.090802} {\bibfield  {journal} {\bibinfo  {journal} {Phys. Rev. Lett.}\ }\textbf {\bibinfo {volume} {133}},\ \bibinfo {pages}
  {090802} (\bibinfo {year} {2024})}\BibitemShut {NoStop}%
\bibitem [{\citenamefont {Lekitsch}\ \emph {et~al.}(2017)\citenamefont {Lekitsch}, \citenamefont {Weidt}, \citenamefont {Fowler}, \citenamefont {Mølmer}, \citenamefont {Devitt}, \citenamefont {Wunderlich},\ and\ \citenamefont {Hensinger}}]{Lekitsch2017}%
  \BibitemOpen
  \bibfield  {author} {\bibinfo {author} {\bibfnamefont {B.}~\bibnamefont {Lekitsch}}, \bibinfo {author} {\bibfnamefont {S.}~\bibnamefont {Weidt}}, \bibinfo {author} {\bibfnamefont {A.~G.}\ \bibnamefont {Fowler}}, \bibinfo {author} {\bibfnamefont {K.}~\bibnamefont {Mølmer}}, \bibinfo {author} {\bibfnamefont {S.~J.}\ \bibnamefont {Devitt}}, \bibinfo {author} {\bibfnamefont {C.}~\bibnamefont {Wunderlich}},\ and\ \bibinfo {author} {\bibfnamefont {W.~K.}\ \bibnamefont {Hensinger}},\ }\bibfield  {title} {\bibinfo {title} {Blueprint for a microwave trapped ion quantum computer},\ }\href {https://doi.org/10.1126/sciadv.1601540} {\bibfield  {journal} {\bibinfo  {journal} {Science Advances}\ }\textbf {\bibinfo {volume} {3}},\ \bibinfo {pages} {e1601540} (\bibinfo {year} {2017})}\BibitemShut {NoStop}%
\bibitem [{\citenamefont {Beverland}\ \emph {et~al.}(2022)\citenamefont {Beverland}, \citenamefont {Murali}, \citenamefont {Troyer}, \citenamefont {Svore}, \citenamefont {Hoefler}, \citenamefont {Kliuchnikov}, \citenamefont {Low}, \citenamefont {Soeken}, \citenamefont {Sundaram},\ and\ \citenamefont {Vaschillo}}]{beverland2022}%
  \BibitemOpen
  \bibfield  {author} {\bibinfo {author} {\bibfnamefont {M.~E.}\ \bibnamefont {Beverland}}, \bibinfo {author} {\bibfnamefont {P.}~\bibnamefont {Murali}}, \bibinfo {author} {\bibfnamefont {M.}~\bibnamefont {Troyer}}, \bibinfo {author} {\bibfnamefont {K.~M.}\ \bibnamefont {Svore}}, \bibinfo {author} {\bibfnamefont {T.}~\bibnamefont {Hoefler}}, \bibinfo {author} {\bibfnamefont {V.}~\bibnamefont {Kliuchnikov}}, \bibinfo {author} {\bibfnamefont {G.~H.}\ \bibnamefont {Low}}, \bibinfo {author} {\bibfnamefont {M.}~\bibnamefont {Soeken}}, \bibinfo {author} {\bibfnamefont {A.}~\bibnamefont {Sundaram}},\ and\ \bibinfo {author} {\bibfnamefont {A.}~\bibnamefont {Vaschillo}},\ }\href {https://arxiv.org/abs/2211.07629} {\bibinfo {title} {Assessing requirements to scale to practical quantum advantage}} (\bibinfo {year} {2022}),\ \Eprint {https://arxiv.org/abs/2211.07629} {arXiv:2211.07629 [quant-ph]} \BibitemShut {NoStop}%
\bibitem [{\citenamefont {Magnard}\ \emph {et~al.}(2020)\citenamefont {Magnard}, \citenamefont {Storz}, \citenamefont {Kurpiers}, \citenamefont {Sch\"ar}, \citenamefont {Marxer}, \citenamefont {L\"utolf}, \citenamefont {Walter}, \citenamefont {Besse}, \citenamefont {Gabureac}, \citenamefont {Reuer}, \citenamefont {Akin}, \citenamefont {Royer}, \citenamefont {Blais},\ and\ \citenamefont {Wallraff}}]{Magnard2020}%
  \BibitemOpen
  \bibfield  {author} {\bibinfo {author} {\bibfnamefont {P.}~\bibnamefont {Magnard}}, \bibinfo {author} {\bibfnamefont {S.}~\bibnamefont {Storz}}, \bibinfo {author} {\bibfnamefont {P.}~\bibnamefont {Kurpiers}}, \bibinfo {author} {\bibfnamefont {J.}~\bibnamefont {Sch\"ar}}, \bibinfo {author} {\bibfnamefont {F.}~\bibnamefont {Marxer}}, \bibinfo {author} {\bibfnamefont {J.}~\bibnamefont {L\"utolf}}, \bibinfo {author} {\bibfnamefont {T.}~\bibnamefont {Walter}}, \bibinfo {author} {\bibfnamefont {J.-C.}\ \bibnamefont {Besse}}, \bibinfo {author} {\bibfnamefont {M.}~\bibnamefont {Gabureac}}, \bibinfo {author} {\bibfnamefont {K.}~\bibnamefont {Reuer}}, \bibinfo {author} {\bibfnamefont {A.}~\bibnamefont {Akin}}, \bibinfo {author} {\bibfnamefont {B.}~\bibnamefont {Royer}}, \bibinfo {author} {\bibfnamefont {A.}~\bibnamefont {Blais}},\ and\ \bibinfo {author} {\bibfnamefont {A.}~\bibnamefont {Wallraff}},\ }\bibfield  {title} {\bibinfo {title} {Microwave quantum link between superconducting circuits housed in spatially
  separated cryogenic systems},\ }\href {https://doi.org/10.1103/PhysRevLett.125.260502} {\bibfield  {journal} {\bibinfo  {journal} {Phys. Rev. Lett.}\ }\textbf {\bibinfo {volume} {125}},\ \bibinfo {pages} {260502} (\bibinfo {year} {2020})}\BibitemShut {NoStop}%
\bibitem [{\citenamefont {Hartung}\ \emph {et~al.}(2024)\citenamefont {Hartung}, \citenamefont {Seubert}, \citenamefont {Welte}, \citenamefont {Distante},\ and\ \citenamefont {Rempe}}]{Hartung2025}%
  \BibitemOpen
  \bibfield  {author} {\bibinfo {author} {\bibfnamefont {L.}~\bibnamefont {Hartung}}, \bibinfo {author} {\bibfnamefont {M.}~\bibnamefont {Seubert}}, \bibinfo {author} {\bibfnamefont {S.}~\bibnamefont {Welte}}, \bibinfo {author} {\bibfnamefont {E.}~\bibnamefont {Distante}},\ and\ \bibinfo {author} {\bibfnamefont {G.}~\bibnamefont {Rempe}},\ }\bibfield  {title} {\bibinfo {title} {A quantum-network register assembled with optical tweezers in an optical cavity},\ }\href {https://doi.org/10.1126/science.ado6471} {\bibfield  {journal} {\bibinfo  {journal} {Science}\ }\textbf {\bibinfo {volume} {385}},\ \bibinfo {pages} {179} (\bibinfo {year} {2024})}\BibitemShut {NoStop}%
\bibitem [{\citenamefont {Bomb\'{\i}n}\ \emph {et~al.}(2024)\citenamefont {Bomb\'{\i}n}, \citenamefont {Pant}, \citenamefont {Roberts},\ and\ \citenamefont {Seetharam}}]{Bombin2024}%
  \BibitemOpen
  \bibfield  {author} {\bibinfo {author} {\bibfnamefont {H.}~\bibnamefont {Bomb\'{\i}n}}, \bibinfo {author} {\bibfnamefont {M.}~\bibnamefont {Pant}}, \bibinfo {author} {\bibfnamefont {S.}~\bibnamefont {Roberts}},\ and\ \bibinfo {author} {\bibfnamefont {K.~I.}\ \bibnamefont {Seetharam}},\ }\bibfield  {title} {\bibinfo {title} {Fault-tolerant postselection for low-overhead magic state preparation},\ }\href {https://doi.org/10.1103/PRXQuantum.5.010302} {\bibfield  {journal} {\bibinfo  {journal} {PRX Quantum}\ }\textbf {\bibinfo {volume} {5}},\ \bibinfo {pages} {010302} (\bibinfo {year} {2024})}\BibitemShut {NoStop}%
\bibitem [{\citenamefont {Li}(2015)}]{li2015magic}%
  \BibitemOpen
  \bibfield  {author} {\bibinfo {author} {\bibfnamefont {Y.}~\bibnamefont {Li}},\ }\bibfield  {title} {\bibinfo {title} {A magic state's fidelity can be superior to the operations that created it},\ }\href {https://doi.org/10.1088/1367-2630/17/2/023037} {\bibfield  {journal} {\bibinfo  {journal} {New Journal of Physics}\ }\textbf {\bibinfo {volume} {17}},\ \bibinfo {pages} {023037} (\bibinfo {year} {2015})}\BibitemShut {NoStop}%
\bibitem [{\citenamefont {Chen}\ \emph {et~al.}(2025{\natexlab{b}})\citenamefont {Chen}, \citenamefont {Xu}, \citenamefont {Sommers}, \citenamefont {Huse}, \citenamefont {Thompson},\ and\ \citenamefont {Gopalakrishnan}}]{chen2025scalableaccuracygainspostselection}%
  \BibitemOpen
  \bibfield  {author} {\bibinfo {author} {\bibfnamefont {H.}~\bibnamefont {Chen}}, \bibinfo {author} {\bibfnamefont {D.}~\bibnamefont {Xu}}, \bibinfo {author} {\bibfnamefont {G.~M.}\ \bibnamefont {Sommers}}, \bibinfo {author} {\bibfnamefont {D.~A.}\ \bibnamefont {Huse}}, \bibinfo {author} {\bibfnamefont {J.~D.}\ \bibnamefont {Thompson}},\ and\ \bibinfo {author} {\bibfnamefont {S.}~\bibnamefont {Gopalakrishnan}},\ }\href {https://arxiv.org/abs/2510.05222} {\bibinfo {title} {Scalable accuracy gains from postselection in quantum error correcting codes}} (\bibinfo {year} {2025}{\natexlab{b}}),\ \Eprint {https://arxiv.org/abs/2510.05222} {arXiv:2510.05222 [cond-mat.stat-mech]} \BibitemShut {NoStop}%
\bibitem [{\citenamefont {Gottesman}(1997)}]{gottesman1997stabilizer}%
  \BibitemOpen
  \bibfield  {author} {\bibinfo {author} {\bibfnamefont {D.}~\bibnamefont {Gottesman}},\ }\href {https://arxiv.org/abs/quant-ph/9705052} {\bibinfo {title} {Stabilizer codes and quantum error correction}} (\bibinfo {year} {1997}),\ \Eprint {https://arxiv.org/abs/quant-ph/9705052} {arXiv:quant-ph/9705052 [quant-ph]} \BibitemShut {NoStop}%
\bibitem [{\citenamefont {Higgott}\ and\ \citenamefont {Gidney}(2025)}]{Higgott2025}%
  \BibitemOpen
  \bibfield  {author} {\bibinfo {author} {\bibfnamefont {O.}~\bibnamefont {Higgott}}\ and\ \bibinfo {author} {\bibfnamefont {C.}~\bibnamefont {Gidney}},\ }\bibfield  {title} {\bibinfo {title} {Sparse {B}lossom: correcting a million errors per core second with minimum-weight matching},\ }\href {https://doi.org/10.22331/q-2025-01-20-1600} {\bibfield  {journal} {\bibinfo  {journal} {{Quantum}}\ }\textbf {\bibinfo {volume} {9}},\ \bibinfo {pages} {1600} (\bibinfo {year} {2025})}\BibitemShut {NoStop}%
\bibitem [{\citenamefont {Gottesman}(1998)}]{Gottesman1998}%
  \BibitemOpen
  \bibfield  {author} {\bibinfo {author} {\bibfnamefont {D.}~\bibnamefont {Gottesman}},\ }\bibfield  {title} {\bibinfo {title} {Theory of fault-tolerant quantum computation},\ }\href {https://doi.org/10.1103/PhysRevA.57.127} {\bibfield  {journal} {\bibinfo  {journal} {Phys. Rev. A}\ }\textbf {\bibinfo {volume} {57}},\ \bibinfo {pages} {127} (\bibinfo {year} {1998})}\BibitemShut {NoStop}%
\bibitem [{\citenamefont {Steane}(1996{\natexlab{b}})}]{Steane1996codes}%
  \BibitemOpen
  \bibfield  {author} {\bibinfo {author} {\bibfnamefont {A.~M.}\ \bibnamefont {Steane}},\ }\bibfield  {title} {\bibinfo {title} {Error correcting codes in quantum theory},\ }\href {https://doi.org/10.1103/PhysRevLett.77.793} {\bibfield  {journal} {\bibinfo  {journal} {Phys. Rev. Lett.}\ }\textbf {\bibinfo {volume} {77}},\ \bibinfo {pages} {793} (\bibinfo {year} {1996}{\natexlab{b}})}\BibitemShut {NoStop}%
\bibitem [{\citenamefont {Steane}(1996{\natexlab{c}})}]{Steane1996}%
  \BibitemOpen
  \bibfield  {author} {\bibinfo {author} {\bibfnamefont {A.~M.}\ \bibnamefont {Steane}},\ }\bibfield  {title} {\bibinfo {title} {Simple quantum error-correcting codes},\ }\href {https://doi.org/10.1103/PhysRevA.54.4741} {\bibfield  {journal} {\bibinfo  {journal} {Phys. Rev. A}\ }\textbf {\bibinfo {volume} {54}},\ \bibinfo {pages} {4741} (\bibinfo {year} {1996}{\natexlab{c}})}\BibitemShut {NoStop}%
\bibitem [{\citenamefont {Yamasaki}\ and\ \citenamefont {Koashi}(2024)}]{Yamasaki2024}%
  \BibitemOpen
  \bibfield  {author} {\bibinfo {author} {\bibfnamefont {H.}~\bibnamefont {Yamasaki}}\ and\ \bibinfo {author} {\bibfnamefont {M.}~\bibnamefont {Koashi}},\ }\bibfield  {title} {\bibinfo {title} {Time-efficient constant-space-overhead fault-tolerant quantum computation},\ }\href {https://doi.org/10.1038/s41567-023-02325-8} {\bibfield  {journal} {\bibinfo  {journal} {Nature Physics}\ }\textbf {\bibinfo {volume} {20}},\ \bibinfo {pages} {247} (\bibinfo {year} {2024})}\BibitemShut {NoStop}%
\bibitem [{\citenamefont {Tillich}\ and\ \citenamefont {Zémor}(2014)}]{Tillich2014}%
  \BibitemOpen
  \bibfield  {author} {\bibinfo {author} {\bibfnamefont {J.-P.}\ \bibnamefont {Tillich}}\ and\ \bibinfo {author} {\bibfnamefont {G.}~\bibnamefont {Zémor}},\ }\bibfield  {title} {\bibinfo {title} {Quantum ldpc codes with positive rate and minimum distance proportional to the square root of the blocklength},\ }\href {https://doi.org/10.1109/TIT.2013.2292061} {\bibfield  {journal} {\bibinfo  {journal} {IEEE Transactions on Information Theory}\ }\textbf {\bibinfo {volume} {60}},\ \bibinfo {pages} {1193} (\bibinfo {year} {2014})}\BibitemShut {NoStop}%
\bibitem [{\citenamefont {Leverrier}\ \emph {et~al.}(2015)\citenamefont {Leverrier}, \citenamefont {Tillich},\ and\ \citenamefont {Zémor}}]{Leverrier2015}%
  \BibitemOpen
  \bibfield  {author} {\bibinfo {author} {\bibfnamefont {A.}~\bibnamefont {Leverrier}}, \bibinfo {author} {\bibfnamefont {J.-P.}\ \bibnamefont {Tillich}},\ and\ \bibinfo {author} {\bibfnamefont {G.}~\bibnamefont {Zémor}},\ }\bibfield  {title} {\bibinfo {title} {Quantum expander codes},\ }in\ \href {https://doi.org/10.1109/FOCS.2015.55} {\emph {\bibinfo {booktitle} {2015 IEEE 56th Annual Symposium on Foundations of Computer Science}}}\ (\bibinfo {year} {2015})\ pp.\ \bibinfo {pages} {810--824}\BibitemShut {NoStop}%
\bibitem [{\citenamefont {Grassl}\ \emph {et~al.}(1997)\citenamefont {Grassl}, \citenamefont {Beth},\ and\ \citenamefont {Pellizzari}}]{Grassl1997}%
  \BibitemOpen
  \bibfield  {author} {\bibinfo {author} {\bibfnamefont {M.}~\bibnamefont {Grassl}}, \bibinfo {author} {\bibfnamefont {T.}~\bibnamefont {Beth}},\ and\ \bibinfo {author} {\bibfnamefont {T.}~\bibnamefont {Pellizzari}},\ }\bibfield  {title} {\bibinfo {title} {Codes for the quantum erasure channel},\ }\href {https://doi.org/10.1103/PhysRevA.56.33} {\bibfield  {journal} {\bibinfo  {journal} {Phys. Rev. A}\ }\textbf {\bibinfo {volume} {56}},\ \bibinfo {pages} {33} (\bibinfo {year} {1997})}\BibitemShut {NoStop}%
\bibitem [{\citenamefont {Kikura}\ \emph {et~al.}(2025{\natexlab{a}})\citenamefont {Kikura}, \citenamefont {Inoue}, \citenamefont {Yamasaki}, \citenamefont {Goban},\ and\ \citenamefont {Sunami}}]{kikura2025taming}%
  \BibitemOpen
  \bibfield  {author} {\bibinfo {author} {\bibfnamefont {S.}~\bibnamefont {Kikura}}, \bibinfo {author} {\bibfnamefont {R.}~\bibnamefont {Inoue}}, \bibinfo {author} {\bibfnamefont {H.}~\bibnamefont {Yamasaki}}, \bibinfo {author} {\bibfnamefont {A.}~\bibnamefont {Goban}},\ and\ \bibinfo {author} {\bibfnamefont {S.}~\bibnamefont {Sunami}},\ }\bibfield  {title} {\bibinfo {title} {Taming the recoil effect in cavity-assisted quantum interconnects},\ }\href {https://doi.org/10.1103/njh8-q7gb} {\bibfield  {journal} {\bibinfo  {journal} {PRX Quantum}\ }\textbf {\bibinfo {volume} {6}},\ \bibinfo {pages} {040351} (\bibinfo {year} {2025}{\natexlab{a}})}\BibitemShut {NoStop}%
\bibitem [{\citenamefont {Wu}\ \emph {et~al.}(2022)\citenamefont {Wu}, \citenamefont {Kolkowitz}, \citenamefont {Puri},\ and\ \citenamefont {Thompson}}]{Wu2022erasure}%
  \BibitemOpen
  \bibfield  {author} {\bibinfo {author} {\bibfnamefont {Y.}~\bibnamefont {Wu}}, \bibinfo {author} {\bibfnamefont {S.}~\bibnamefont {Kolkowitz}}, \bibinfo {author} {\bibfnamefont {S.}~\bibnamefont {Puri}},\ and\ \bibinfo {author} {\bibfnamefont {J.~D.}\ \bibnamefont {Thompson}},\ }\bibfield  {title} {\bibinfo {title} {Erasure conversion for fault-tolerant quantum computing in alkaline earth {Rydberg} atom arrays},\ }\href {https://doi.org/10.1038/s41467-022-32094-6} {\bibfield  {journal} {\bibinfo  {journal} {Nature Communications}\ }\textbf {\bibinfo {volume} {13}},\ \bibinfo {pages} {4657} (\bibinfo {year} {2022})}\BibitemShut {NoStop}%
\bibitem [{\citenamefont {Kikura}\ \emph {et~al.}(2025{\natexlab{b}})\citenamefont {Kikura}, \citenamefont {Tanji}, \citenamefont {Goban},\ and\ \citenamefont {Sunami}}]{kikura2025passive}%
  \BibitemOpen
  \bibfield  {author} {\bibinfo {author} {\bibfnamefont {S.}~\bibnamefont {Kikura}}, \bibinfo {author} {\bibfnamefont {K.}~\bibnamefont {Tanji}}, \bibinfo {author} {\bibfnamefont {A.}~\bibnamefont {Goban}},\ and\ \bibinfo {author} {\bibfnamefont {S.}~\bibnamefont {Sunami}},\ }\href {https://arxiv.org/abs/2507.01229} {\bibinfo {title} {Passive quantum interconnects: High-fidelity quantum networking at higher rates with less overhead}} (\bibinfo {year} {2025}{\natexlab{b}}),\ \Eprint {https://arxiv.org/abs/2507.01229} {arXiv:2507.01229 [quant-ph]} \BibitemShut {NoStop}%
\bibitem [{\citenamefont {Bonilla~Ataides}\ \emph {et~al.}(2021)\citenamefont {Bonilla~Ataides}, \citenamefont {Tuckett}, \citenamefont {Bartlett}, \citenamefont {Flammia},\ and\ \citenamefont {Brown}}]{bonilla2021xzzx}%
  \BibitemOpen
  \bibfield  {author} {\bibinfo {author} {\bibfnamefont {J.~P.}\ \bibnamefont {Bonilla~Ataides}}, \bibinfo {author} {\bibfnamefont {D.~K.}\ \bibnamefont {Tuckett}}, \bibinfo {author} {\bibfnamefont {S.~D.}\ \bibnamefont {Bartlett}}, \bibinfo {author} {\bibfnamefont {S.~T.}\ \bibnamefont {Flammia}},\ and\ \bibinfo {author} {\bibfnamefont {B.~J.}\ \bibnamefont {Brown}},\ }\bibfield  {title} {\bibinfo {title} {The xzzx surface code},\ }\href@noop {} {\bibfield  {journal} {\bibinfo  {journal} {Nature communications}\ }\textbf {\bibinfo {volume} {12}},\ \bibinfo {pages} {2172} (\bibinfo {year} {2021})}\BibitemShut {NoStop}%
\bibitem [{\citenamefont {Tuckett}\ \emph {et~al.}(2018)\citenamefont {Tuckett}, \citenamefont {Bartlett},\ and\ \citenamefont {Flammia}}]{Tuckett2018}%
  \BibitemOpen
  \bibfield  {author} {\bibinfo {author} {\bibfnamefont {D.~K.}\ \bibnamefont {Tuckett}}, \bibinfo {author} {\bibfnamefont {S.~D.}\ \bibnamefont {Bartlett}},\ and\ \bibinfo {author} {\bibfnamefont {S.~T.}\ \bibnamefont {Flammia}},\ }\bibfield  {title} {\bibinfo {title} {Ultrahigh error threshold for surface codes with biased noise},\ }\href {https://doi.org/10.1103/PhysRevLett.120.050505} {\bibfield  {journal} {\bibinfo  {journal} {Phys. Rev. Lett.}\ }\textbf {\bibinfo {volume} {120}},\ \bibinfo {pages} {050505} (\bibinfo {year} {2018})}\BibitemShut {NoStop}%
\bibitem [{\citenamefont {Tiurev}\ \emph {et~al.}(2023)\citenamefont {Tiurev}, \citenamefont {Derks}, \citenamefont {Roffe}, \citenamefont {Eisert},\ and\ \citenamefont {Reiner}}]{Tiurev2023correctingnon}%
  \BibitemOpen
  \bibfield  {author} {\bibinfo {author} {\bibfnamefont {K.}~\bibnamefont {Tiurev}}, \bibinfo {author} {\bibfnamefont {P.-J. H.~S.}\ \bibnamefont {Derks}}, \bibinfo {author} {\bibfnamefont {J.}~\bibnamefont {Roffe}}, \bibinfo {author} {\bibfnamefont {J.}~\bibnamefont {Eisert}},\ and\ \bibinfo {author} {\bibfnamefont {J.-M.}\ \bibnamefont {Reiner}},\ }\bibfield  {title} {\bibinfo {title} {Correcting non-independent and non-identically distributed errors with surface codes},\ }\href {https://doi.org/10.22331/q-2023-09-26-1123} {\bibfield  {journal} {\bibinfo  {journal} {{Quantum}}\ }\textbf {\bibinfo {volume} {7}},\ \bibinfo {pages} {1123} (\bibinfo {year} {2023})}\BibitemShut {NoStop}%
\bibitem [{\citenamefont {Khosravani}\ \emph {et~al.}(2026)\citenamefont {Khosravani}, \citenamefont {Escobar-Arrieta}, \citenamefont {Brown},\ and\ \citenamefont {Gutierrez}}]{khosravani2026heterogeneous}%
  \BibitemOpen
  \bibfield  {author} {\bibinfo {author} {\bibfnamefont {O.}~\bibnamefont {Khosravani}}, \bibinfo {author} {\bibfnamefont {G.}~\bibnamefont {Escobar-Arrieta}}, \bibinfo {author} {\bibfnamefont {K.~R.}\ \bibnamefont {Brown}},\ and\ \bibinfo {author} {\bibfnamefont {M.}~\bibnamefont {Gutierrez}},\ }\href {https://arxiv.org/abs/2603.06817} {\bibinfo {title} {Heterogeneous quantum error-correcting codes}} (\bibinfo {year} {2026}),\ \Eprint {https://arxiv.org/abs/2603.06817} {arXiv:2603.06817 [quant-ph]} \BibitemShut {NoStop}%
\bibitem [{\citenamefont {Zhou}\ \emph {et~al.}(2025)\citenamefont {Zhou}, \citenamefont {Zhao}, \citenamefont {Cain}, \citenamefont {Bluvstein}, \citenamefont {Maskara}, \citenamefont {Duckering}, \citenamefont {Hu}, \citenamefont {Wang}, \citenamefont {Kubica},\ and\ \citenamefont {Lukin}}]{zhou2025nature}%
  \BibitemOpen
  \bibfield  {author} {\bibinfo {author} {\bibfnamefont {H.}~\bibnamefont {Zhou}}, \bibinfo {author} {\bibfnamefont {C.}~\bibnamefont {Zhao}}, \bibinfo {author} {\bibfnamefont {M.}~\bibnamefont {Cain}}, \bibinfo {author} {\bibfnamefont {D.}~\bibnamefont {Bluvstein}}, \bibinfo {author} {\bibfnamefont {N.}~\bibnamefont {Maskara}}, \bibinfo {author} {\bibfnamefont {C.}~\bibnamefont {Duckering}}, \bibinfo {author} {\bibfnamefont {H.-Y.}\ \bibnamefont {Hu}}, \bibinfo {author} {\bibfnamefont {S.-T.}\ \bibnamefont {Wang}}, \bibinfo {author} {\bibfnamefont {A.}~\bibnamefont {Kubica}},\ and\ \bibinfo {author} {\bibfnamefont {M.~D.}\ \bibnamefont {Lukin}},\ }\bibfield  {title} {\bibinfo {title} {Low-overhead transversal fault tolerance for universal quantum computation},\ }\href {https://doi.org/10.1038/s41586-025-09543-5} {\bibfield  {journal} {\bibinfo  {journal} {Nature}\ }\textbf {\bibinfo {volume} {646}},\ \bibinfo {pages} {303} (\bibinfo {year} {2025})}\BibitemShut {NoStop}%
\bibitem [{\citenamefont {Sunami}\ \emph {et~al.}(2025{\natexlab{b}})\citenamefont {Sunami}, \citenamefont {Goban},\ and\ \citenamefont {Yamasaki}}]{sunami2025transversal}%
  \BibitemOpen
  \bibfield  {author} {\bibinfo {author} {\bibfnamefont {S.}~\bibnamefont {Sunami}}, \bibinfo {author} {\bibfnamefont {A.}~\bibnamefont {Goban}},\ and\ \bibinfo {author} {\bibfnamefont {H.}~\bibnamefont {Yamasaki}},\ }\href {https://arxiv.org/abs/2506.18979} {\bibinfo {title} {Transversal surface-code game powered by neutral atoms}} (\bibinfo {year} {2025}{\natexlab{b}}),\ \Eprint {https://arxiv.org/abs/2506.18979} {arXiv:2506.18979 [quant-ph]} \BibitemShut {NoStop}%
\bibitem [{\citenamefont {Hirano}\ \emph {et~al.}(2024{\natexlab{b}})\citenamefont {Hirano}, \citenamefont {Suzuki},\ and\ \citenamefont {Fujii}}]{hirano2024magicpool}%
  \BibitemOpen
  \bibfield  {author} {\bibinfo {author} {\bibfnamefont {Y.}~\bibnamefont {Hirano}}, \bibinfo {author} {\bibfnamefont {Y.}~\bibnamefont {Suzuki}},\ and\ \bibinfo {author} {\bibfnamefont {K.}~\bibnamefont {Fujii}},\ }\href {https://arxiv.org/abs/2407.07394} {\bibinfo {title} {Magicpool: Dealing with magic state distillation failures on large-scale fault-tolerant quantum computer}} (\bibinfo {year} {2024}{\natexlab{b}}),\ \Eprint {https://arxiv.org/abs/2407.07394} {arXiv:2407.07394 [quant-ph]} \BibitemShut {NoStop}%
\bibitem [{\citenamefont {Gidney}(2021)}]{gidney2021stim}%
  \BibitemOpen
  \bibfield  {author} {\bibinfo {author} {\bibfnamefont {C.}~\bibnamefont {Gidney}},\ }\bibfield  {title} {\bibinfo {title} {Stim: a fast stabilizer circuit simulator},\ }\href {https://doi.org/10.22331/q-2021-07-06-497} {\bibfield  {journal} {\bibinfo  {journal} {{Quantum}}\ }\textbf {\bibinfo {volume} {5}},\ \bibinfo {pages} {497} (\bibinfo {year} {2021})}\BibitemShut {NoStop}%
\bibitem [{\citenamefont {Haug}\ \emph {et~al.}(2025)\citenamefont {Haug}, \citenamefont {Hillmann}, \citenamefont {Kockum},\ and\ \citenamefont {Laer}}]{haug2025latticesurgerybellmeasurements}%
  \BibitemOpen
  \bibfield  {author} {\bibinfo {author} {\bibfnamefont {T.~H.}\ \bibnamefont {Haug}}, \bibinfo {author} {\bibfnamefont {T.}~\bibnamefont {Hillmann}}, \bibinfo {author} {\bibfnamefont {A.~F.}\ \bibnamefont {Kockum}},\ and\ \bibinfo {author} {\bibfnamefont {R.~V.}\ \bibnamefont {Laer}},\ }\href {https://arxiv.org/abs/2510.13541} {\bibinfo {title} {Lattice surgery with bell measurements: Modular fault-tolerant quantum computation at low entanglement cost}} (\bibinfo {year} {2025}),\ \Eprint {https://arxiv.org/abs/2510.13541} {arXiv:2510.13541 [quant-ph]} \BibitemShut {NoStop}%
\bibitem [{\citenamefont {Gottesman}(1996)}]{Gottesman1996class}%
  \BibitemOpen
  \bibfield  {author} {\bibinfo {author} {\bibfnamefont {D.}~\bibnamefont {Gottesman}},\ }\bibfield  {title} {\bibinfo {title} {Class of quantum error-correcting codes saturating the quantum hamming bound},\ }\href {https://doi.org/10.1103/PhysRevA.54.1862} {\bibfield  {journal} {\bibinfo  {journal} {Phys. Rev. A}\ }\textbf {\bibinfo {volume} {54}},\ \bibinfo {pages} {1862} (\bibinfo {year} {1996})}\BibitemShut {NoStop}%
\bibitem [{\citenamefont {Laflamme}\ \emph {et~al.}(1996)\citenamefont {Laflamme}, \citenamefont {Miquel}, \citenamefont {Paz},\ and\ \citenamefont {Zurek}}]{laflamme1996perfectquantumerrorcorrection}%
  \BibitemOpen
  \bibfield  {author} {\bibinfo {author} {\bibfnamefont {R.}~\bibnamefont {Laflamme}}, \bibinfo {author} {\bibfnamefont {C.}~\bibnamefont {Miquel}}, \bibinfo {author} {\bibfnamefont {J.~P.}\ \bibnamefont {Paz}},\ and\ \bibinfo {author} {\bibfnamefont {W.~H.}\ \bibnamefont {Zurek}},\ }\href {https://arxiv.org/abs/quant-ph/9602019} {\bibinfo {title} {Perfect quantum error correction code}} (\bibinfo {year} {1996}),\ \Eprint {https://arxiv.org/abs/quant-ph/9602019} {arXiv:quant-ph/9602019 [quant-ph]} \BibitemShut {NoStop}%
\bibitem [{\citenamefont {Gong}\ \emph {et~al.}(2021)\citenamefont {Gong}, \citenamefont {Yuan}, \citenamefont {Wang}, \citenamefont {Wu}, \citenamefont {Zhao}, \citenamefont {Zha}, \citenamefont {Li}, \citenamefont {Zhang}, \citenamefont {Zhao}, \citenamefont {Liu}, \citenamefont {Liang}, \citenamefont {Lin}, \citenamefont {Xu}, \citenamefont {Deng}, \citenamefont {Rong}, \citenamefont {Lu}, \citenamefont {Benjamin}, \citenamefont {Peng}, \citenamefont {Ma}, \citenamefont {Chen}, \citenamefont {Zhu},\ and\ \citenamefont {Pan}}]{10.1093/nsr/nwab011}%
  \BibitemOpen
  \bibfield  {author} {\bibinfo {author} {\bibfnamefont {M.}~\bibnamefont {Gong}}, \bibinfo {author} {\bibfnamefont {X.}~\bibnamefont {Yuan}}, \bibinfo {author} {\bibfnamefont {S.}~\bibnamefont {Wang}}, \bibinfo {author} {\bibfnamefont {Y.}~\bibnamefont {Wu}}, \bibinfo {author} {\bibfnamefont {Y.}~\bibnamefont {Zhao}}, \bibinfo {author} {\bibfnamefont {C.}~\bibnamefont {Zha}}, \bibinfo {author} {\bibfnamefont {S.}~\bibnamefont {Li}}, \bibinfo {author} {\bibfnamefont {Z.}~\bibnamefont {Zhang}}, \bibinfo {author} {\bibfnamefont {Q.}~\bibnamefont {Zhao}}, \bibinfo {author} {\bibfnamefont {Y.}~\bibnamefont {Liu}}, \bibinfo {author} {\bibfnamefont {F.}~\bibnamefont {Liang}}, \bibinfo {author} {\bibfnamefont {J.}~\bibnamefont {Lin}}, \bibinfo {author} {\bibfnamefont {Y.}~\bibnamefont {Xu}}, \bibinfo {author} {\bibfnamefont {H.}~\bibnamefont {Deng}}, \bibinfo {author} {\bibfnamefont {H.}~\bibnamefont {Rong}}, \bibinfo {author} {\bibfnamefont {H.}~\bibnamefont {Lu}}, \bibinfo {author} {\bibfnamefont {S.~C.}\
  \bibnamefont {Benjamin}}, \bibinfo {author} {\bibfnamefont {C.-Z.}\ \bibnamefont {Peng}}, \bibinfo {author} {\bibfnamefont {X.}~\bibnamefont {Ma}}, \bibinfo {author} {\bibfnamefont {Y.-A.}\ \bibnamefont {Chen}}, \bibinfo {author} {\bibfnamefont {X.}~\bibnamefont {Zhu}},\ and\ \bibinfo {author} {\bibfnamefont {J.-W.}\ \bibnamefont {Pan}},\ }\bibfield  {title} {\bibinfo {title} {Experimental exploration of five-qubit quantum error-correcting code with superconducting qubits},\ }\href {https://doi.org/10.1093/nsr/nwab011} {\bibfield  {journal} {\bibinfo  {journal} {National Science Review}\ }\textbf {\bibinfo {volume} {9}},\ \bibinfo {pages} {nwab011} (\bibinfo {year} {2021})}\BibitemShut {NoStop}%
\bibitem [{\citenamefont {Kliuchnikov}\ and\ \citenamefont {Maslov}(2013)}]{PhysRevA.88.052307}%
  \BibitemOpen
  \bibfield  {author} {\bibinfo {author} {\bibfnamefont {V.}~\bibnamefont {Kliuchnikov}}\ and\ \bibinfo {author} {\bibfnamefont {D.}~\bibnamefont {Maslov}},\ }\bibfield  {title} {\bibinfo {title} {Optimization of clifford circuits},\ }\href {https://doi.org/10.1103/PhysRevA.88.052307} {\bibfield  {journal} {\bibinfo  {journal} {Phys. Rev. A}\ }\textbf {\bibinfo {volume} {88}},\ \bibinfo {pages} {052307} (\bibinfo {year} {2013})}\BibitemShut {NoStop}%
\end{thebibliography}
\end{document}